\documentclass[amsmath,amssymb,aps,prb,twocolumn,floatfix,nofootinbib,letterpaper,superscriptaddress]{revtex4-2}

\newcommand{\pcol}[2]{\parbox[c]{#1}{\vspace{3pt}\raggedright #2\vspace{3pt}}}

\bibpunct{[}{]}{,}{n}{}{}

\usepackage{tabularx}
\usepackage{diagbox}
\usepackage{wasysym}
\usepackage[scr=boondox]{mathalpha}
\usepackage{makecell}
\usepackage{ragged2e} 

\usepackage[normalem]{ulem}
\usepackage{footnotebackref}
\usepackage{array}
\usepackage[shortlabels]{enumitem}

\usepackage{./MyStyle/myStyle}

\def\ii{\mathrm{i}}
\def\psf{\mathsf{p}}
\def\PP{\mathrm{p}}
\def\Mbs{\vec{m}}
\def\CCtwoSymbol{\Upsilon}

\def\SWAP{\mathrm{SWAP}}

\def\Area{\mathrm{A}_{\text{XY}}}

\def\diag{\mathrm{diag}}

\def\triv{$\ZZ_1$}

\def\dOR{\bar{\dd}}
\def\RL{{R}_{\LL}}
\def\RLp{{R}_{\LL'}}
\def\H{\mathcal{H}}

\def\OR{\mathtt{or}}

\def\OR{\mathrm{or}}

\def\Z{\ZZ}

\def\Cmo{\tilde{C}_{M_{\text{o}}}}
\def\MO{M_{\text{o}}}
\def\so{\mathscr{S}_{\text{o}}}
\def\sko{\mathscr{S}_{k,\mathrm{o}}}
\def\OO{\mathrm{o}}
\def\LL{l}

\def\Gwp{G_{\mathrm{space}}}
\def\Gpt{G_{\mathrm{pt}}}

\def\singletName{{singlet covering\,}}
\def\ee{\mathrm{e}}

\def\Hmc{\mathcal{H}}
\def\U{\mathrm{U}}
\def\SO{\mathrm{SO}}
\def\Z{\mathbb{Z}}
\def\ZZ{\mathbb{Z}}

\begin{document}

\title{Invariants for (2+1)D bosonic crystalline topological insulators \texorpdfstring{\\}{} for all 17 wallpaper groups} 

\author{Vladimir Calvera }\email{vcalvera@umn.edu}- 
\affiliation{Department of Physics, Stanford University, Stanford, California 94305, USA}
\affiliation{School of Physics and Astronomy and William I. Fine Theoretical Physics Institute,
University of Minnesota, Minneapolis, MN 55455, USA}
\author{Naren Manjunath}
\affiliation{Perimeter Institute for Theoretical Physics, 31 Caroline St N, Waterloo, ON N2L 2Y5, Canada}
\author{Maissam Barkeshli }
\affiliation{Department of Physics and Joint Quantum Institute, University of Maryland, College Park, Maryland 20742, USA}
\begin{abstract}
We study bosonic symmetry-protected topological (SPT) phases in (2+1) dimensions with symmetry $G = G_{\text{space}}\times K$, where $G_{\text{space}}$ is a general wallpaper group and $K=\text{U}(1),\mathbb{Z}_N, \text{SO}(3)$ is an internal symmetry. In each case we propose a set of many-body invariants that can detect all the different phases predicted from real space constructions and group cohomology classifications. They are obtained by applying partial rotations and reflections to a given ground state, combined with suitable operations in $K$. The reflection symmetry invariants that we introduce include `double partial reflections', `weak partial reflections' and their `relative' or `twisted' versions which also depend on $K$. We verify our proposal through exact calculations on ground states constructed using real space constructions.
We demonstrate our method in detail for the groups p4m and p4g, and in the case of p4m also derive a topological effective action involving gauge fields for orientation-reversing symmetries. Our results provide a concrete method to fully characterize (2+1)D crystalline topological invariants in bosonic SPT ground states. 
\end{abstract}

\maketitle 
\newpage

\tableofcontents

\section{Introduction}

\begin{table*}
    \centering
    \begin{tabular}{c|c|c||l|l|l}
    \multicolumn{6}{c}{\textbf{Group cohomology classification of crystalline SPTs: $\Hmc^3(\Gwp \times K,\U(1)^{\OR})$}} \\  \hline  
        $\#$ & $\Gwp$ & $\Hmc^3(\Gwp,\U(1)^{\OR})$ &  \multicolumn{1}{c|}{$K=\U(1)$} & \multicolumn{1}{c|}{$ K=\Z_N$} & \multicolumn{1}{c}{$K=\SO(3)$}  \\ \hline \hline 
         1&  p1 &
          $\ZZ_1$& 
         ${\ZZ}$ $ \times \red{\ZZ }$& 
         {$\Z_N$}$\times\red{\ZZ_N}$ &
         $\red{\ZZ }$\\
         2 &p2   & 
         $\Z_2^4$ & 
         {$\Z\times\Z_2^3$} $\times\red{\ZZ }$ & 
         {$\Z_N \times \Z_{(2,N)}^3$}$\times\red{\ZZ_N}$ & 
         {$\Z_2^3$}$\times\red{\ZZ }$\\
         3 &pm & 
         $\Z_2^2$ & 
         {$\Z\times\Z_2$} & 
         {$\Z_N \times \Z_{(2,N)}^3$}$\times\red{\ZZ_{(N,2)}}$ & 
         {$\Z_2^2$}\\
         4 & pg  & $\ZZ_1$
          & 
         {$\Z$}& 
         {$\Z_N$}$\times\red{\ZZ_{(N,2)}}$ & 
         {$\Z_2$}\\
         5 & cm  &
         $\Z_2$ &
         {$\Z$} &
         {$\Z_N \times \Z_{(2,N)}$ }$\times\red{\ZZ_{(N,2)}}$& 
         {$\Z_2$} \\
         6 & pmm  &
         $\Z_2^8$ &
         {$\Z\times \Z_2^3$} &
         {$\Z_N \times \Z_{(2,N)}^7$}$\times\red{\ZZ_{(N,2)}}$& 
         {$\Z_2^4$}\\
         7 & pmg&
         $\Z_2^3$&
         {$\Z\times\Z_2^2$} &
         {$\Z_N \times \Z_{(2,N)}^3$} $\times\red{\ZZ_{(N,2)}}$&
         {$\Z_2^3$}\\
         8 & pgg& 
         $\Z_2^2$ &
         {$\Z\times\Z_2$} &
         {$\Z_N \times \Z_{(2,N)}$} $\times\red{\ZZ_{(N,2)}}$&
         {$\Z_2^2$}\\
         9 & cmm&
         $\Z_2^5$ & 
         {$\Z\times\Z_2^2$} & 
         {$\Z_N \times \Z_{(2,N)}^4$} $\times\red{\ZZ_{(N,2)}}$& 
        {$\Z_2^3$}\\
         10 & p4 & 
         $\Z_4^2\times\Z_2$ & 
         {$\Z\times\Z_4\times\Z_2$} $\times\red{\ZZ }$& 
         {$\Z_N \times \Z_{(2,N)}\times\Z_{(4,N)}$} $\times\red{\ZZ_N}$& 
         {$\Z_2^2$}$\times\red{\ZZ }$\\
         11 & p4m & 
         $\Z_2^6$ & 
         {$\Z\times\Z_4\times\Z_2$} &
         {$\Z_N \times \Z_{(2,N)}^4\times\Z_{(4,N)}$} $\times\red{\ZZ_{(N,2)}}$& 
         {$\Z_2^3$}\\
         12 & p4g &
         $\Z_4\times\Z_2^2$ & 
         {$\Z\times\Z_4$} & 
         {$\Z_N \times \Z_{(2,N)}\times\Z_{(4,N)}$} $\times\red{\ZZ_{(N,2)}}$&
         {$\Z_2^2$}\\
         13 & p3  & 
         $\Z_3^3$ & 
         {$\Z\times\Z_3^2$ $\times\red{\ZZ}$} & 
         {$\Z_N \times\Z_{(3,N)}^2$}$\times\red{\ZZ_N}$ & 
         {$\red{\ZZ }$}\\
         14 & p3m1 & 
         $\Z_2$ & 
         {$\Z\times\Z_3^2$} & 
         {$\Z_N \times\Z_{(3,N)}^2\times \Z_{(2,N)}$}$\times\red{\ZZ_{(N,2)}}$ &
         {$\Z_2$}\\
         15 & p31m & 
         $\Z_6$ & 
         {$\Z\times\Z_3$}& 
         {$\Z_N \times \times\Z_{(3,N)}\times \Z_{(2,N)}$}$\times\red{\ZZ_{(N,2)}}$ & 
         {$\Z_2$}\\
         16 & p6 & 
         $\Z_6^2$ & 
         {$\Z\times\Z_6$$\times\red{\ZZ }$}& 
         {$\Z_N\times\Z_{(6,N)}$}$\times\red{\ZZ_N}$ & 
         {$\Z_2$}$\times\red{\ZZ}$\\
         17 & p6m & $\Z_2^4$ & 
         {$\Z\times\Z_6$} & 
         {$\Z_N \times\Z_{(6,N)}\times \Z_{(2,N)}^2$} $\times\red{\ZZ_{(N,2)}}$& 
         {$\Z_2^2$}
    \end{tabular}
    \caption{
    {The group cohomology classification of (2+1)D bosonic SPTs with only $\Gwp$ wallpaper group symmetry is given in the third column. To obtain the classification for $G = \Gwp \times K$, with $K = \U(1),\ZZ_N, \SO(3)$, we take the direct product between the third column and the desired $K$ column. The groups in black on the third, fourth and fifth column are protected by $\Gwp$ and corresponding $K$, while the \red{red} invariants are protected solely by $K$ (we have accounted for a possible reduction by the presence of reflections).
    We denote the greatest common divisor between $N$ and $k$ by $(N,k)$.}
    }
    \label{tab:classif-full}
\end{table*}

\begin{table*}[t]
    \centering
    \begin{tabular}{c|l|c|c|c|l|l}
            \multicolumn{7}{c}{\textbf{Real-space invariants for crystalline SPTs}} \\  \hline 
    Invariant & Symbol & Type & {Point group} & $K$ & Definition & 
    {Quantization}
    \\ \hline\hline
        Partial rotation & $\Theta_{\OO}$ & A1 & $C_{M_{\OO}}$ & - & Eq.~\eqref{eq:ThetaO} & 
         $\Z_{\MO} \left(^*\right)$ 
         \\
        \hline
         {Partial double reflection} &
         $\Sigma_{\OO,\LL}$ & A2 & 
         $ D_{2}$& -
         & Eq.~\eqref{eq:SigmaOL_defn} & $\Z_2$ \\
         \hline
         {Partial weak reflection} & $\Lambda_{\LL}$ & A3 & 
         $\ZZ \times D_{1}$& - 
         & Eq.~\eqref{eq:SigmaL_defn} & $\Z_2$ \\
         \hline
         \multirow{3}{*}{Discrete shift}&
         $\so^{\U(1)}$
          & B1 & 
          $C_{\MO}$& $\U(1)$ & Eq.~\eqref{eq:ThetaOU1} & $\ZZ_{M_{\OO}}$
         \\
        & $\so^{\ZZ_N}$
         & C1 & 
         $C_{\MO}$
         & $\ZZ_N$ & Eq.~\eqref{eq:ThetaOZN} & $\ZZ_{(M_{\OO},N)}$
         \\
        & $\so^{\SO(3)}$ & D1 & 
         $C_{\MO}$
         & $\SO(3)$ & Eq.~\eqref{eq:ThetaOSO3} & $\ZZ_{(M_{\OO},2)}$
         \\ \hline
         \multirow{2}{*}{\makecell{{Twisted relative} \\{partial reflection}}}& 
        $\CCtwoSymbol_{\LL}^{\ZZ_N}$ 
        & C4 & 
        $D_{1}$ & $\ZZ_{N}$
        & Eq.~\eqref{eq:QRS_defn} & $\Z_{(2,N)}$ \\
        & $\CCtwoSymbol_{\LL}^{\SO(3)}$
        & D4 &$D_{1}$ & $\SO(3)$ & Eq.~\eqref{eq:QRXZ_defn} & $\Z_2$ \\\hline 
        \multirow{2}{*}{\makecell{{Relative partial}\\ {double reflection} }} & $\tilde{\Sigma}_{\OO,\LL}^{\ZZ_N}$ & C2 & $D_{2}$ & $\ZZ_N$ & Eq.~\eqref{eq:SigmaS_defn} & $\ZZ_{(N,2)}$\\
        & $\tilde{\Sigma}_{\OO,\LL}^{\SO(3)}$ & D2 & $D_{2}$ & $\SO(3)$ & Eq.~\eqref{eq:SigmalXZ_defn} & $\ZZ_{2}$ \\ 
        \hline 
        \multirow{2}{*}{\makecell{{Relative partial}\\ {weak reflection}}}& $\tilde{\Lambda}_{\LL}^{\U(1)}$ & B3 & $D_{1}$ & $\U(1)$ & Eq.~\eqref{eq:DressedWeakPartialReflection}& $\ZZ_2$\\
        & $\tilde{\Lambda}_{\LL}^{\ZZ_N}$ & C3 & $ D_{1}$ & $\ZZ_N$ & Eq.~\eqref{eq:DressedWeakPartialReflection} & $\ZZ_{(N,2)}$
    \end{tabular}
    \caption{\label{tab:invts_def}
    \justifying{Summary of SPT invariant types with symmetry 
$G_{\text{space}} \times K$
    , for $G_{\text{space}}$ a point group and $K = \U(1), \Z_N, \SO(3)$}. 
    Type-A invariants are protected solely by
    $G_{\text{space}}$ while the other invariants also require $K$. Here $\OO$ denotes a rotation center of order $\MO$, $\LL$ denotes a reflection axis. 
    $C_{M_{\OO}}$ is an $M$-fold rotation around $\OO$, $D_{1}$ is reflection about line $\LL$, 
    {$D_{2}$ is the dihedral group generated by a two-fold rotation around $\OO$ and a reflection along $\LL$}, and $\Z$ is the group of translations parallel to line $\LL$. $(a,b)$ denotes the greatest common divisor of $a$ and $b$, and $\ZZ_1 = \{e\}$ is the trivial group. {These invariants together with the filling form a complete set. However, they are not all independent: there are non-trivial relations between different types of invariants, and also between the same type of invariant for different positions `$\OO$' or lines `$\LL$'.
    $(^*)$ The partial rotation invariant $\Theta_{\OO}$ satisfies $2\Theta_{\OO} = 0 \mod \MO$ if there exists a reflection line passing through $\OO$. 
    To the best of our knowledge, only partial rotation and discrete shift have appeared before in the literature.}
    }
\end{table*}

\begin{table}[h]
    \centering
    \begin{tabular}{c|c}
    \multicolumn{2}{c}{\textbf{List of notations}} \\  \hline 
        Symbol & Meaning  \\ \hline
         $\Gwp$ & \pcol{0.36\textwidth}{2d wallpaper group} \\
         $\Gpt$ & \pcol{0.36\textwidth}{Point group} \\
         $K$ & \pcol{0.36\textwidth}{Internal symmetry ($K = \U(1), \Z_N, \SO(3)$)} \\
         $\OO$ & \pcol{0.36\textwidth}{Origin of rotations in $\Gpt$} \\
         $\MO$ & \pcol{0.36\textwidth}{Order of rotations about $\OO$} \\
         $\alpha, \beta, \gamma, \delta$ & \pcol{0.36\textwidth}{Maximal Wyckoff positions} \\
         $\LL$ & \pcol{0.36\textwidth}{Reflection axis} \\
         $\lambda, \mu, \nu, \kappa$ & \pcol{0.36\textwidth}{Unit cell reflection axes} \\ \hline
        $\zero$ & \pcol{0.36\textwidth}{Identity group element} \\
         $\rbf_{\LL}$ & \pcol{0.36\textwidth}{Reflection about line $\LL$} \\
         $\hbf_{\OO}$ & \pcol{0.36\textwidth}{Elementary rotation about $\OO$} \\
         ${\bf x},{\bf y}$ & \pcol{0.36\textwidth}{Elementary translation} \\
        $\Sbf$ & \pcol{0.36\textwidth}{$\Z_N$ internal symmetry generator} \\
        $\Xbf,\Zbf$ & \pcol{0.36\textwidth}{Generators of $\ZZ_2\times \ZZ_2 \subset \SO(3)$} \\ \hline
        $\Cmo$ & \pcol{0.36\textwidth}{$\MO$-fold rotation operator about $\OO$} \\
        $R_{\LL}$ & \pcol{0.36\textwidth}{Reflection operator along line $\LL$} \\
        $T_{\vec{a}}$ & \pcol{0.36\textwidth}{Operator for translations by $\vec{a}$} \\
        $U_{\gbf}$ & \pcol{0.36\textwidth}{Operator for internal symmetry $\gbf \in K$} \\
        $\tilde{C}_{k,\OO}({\bf g})$ & \pcol{0.36\textwidth}{$2\pi/k$ rotation about $\OO$ dressed with $U_{\bf g}$, see Eq.~\eqref{eq:CKODef}} \\ \hline
        $D$ & \pcol{0.36\textwidth}{Region in which partial symmetry acts} \\ \hline
        $\sigma$ & \pcol{0.36\textwidth}{Reflection gauge field (here and below, see Footnote~\ref{fn:origin})} \\
        $\omega$ & \pcol{0.36\textwidth}{Rotation gauge field} \\
        $\vec{R}$ & \pcol{0.36\textwidth}{Translation gauge field}
    \end{tabular}
    \caption{List of notations used in this paper. }
    \label{tab:list_of_notations}
\end{table}

\begin{table*}[t]
    \centering
    \begin{tabular}{l|l||w{c}{0.085\textwidth}|c||c|c||c|c}
    \multicolumn{8}{c}{\textbf{Independent set of pure crystalline invariants for each wallpaper group}} \\  \hline 
        \multicolumn{2}{c||}{} & \multicolumn{2}{c||}{Type-A1:$\Theta_{\OO}$} & \multicolumn{2}{c||}{Type-A2:$\Sigma_{\OO,\LL}$} & \multicolumn{2}{c}{Type-A3:$\Lambda_{\LL}$}  \\ \hline
        $\#$ & $G_{\text{space}}$ & Class. & Invariants & Class. & Invariants & Class. & Invariants  \\ \hline\hline
        1 & p1 &\triv  &  -& \triv& -&\triv &- \\ 
        2&p2 & $\Z_2^4$ & $\Theta_{\alpha}, \Theta_{\beta}, \Theta_{\gamma}, \Theta_{\delta}$ & \triv&- & \triv& -\\ 
        3 &pm &\triv  & - &\triv & -& $\Z_2^2$ & $\Sigma_{\lambda},\Sigma_{\mu}$ \\ 
        4 &pg & \triv& - & \triv& - & \triv& - \\ 
        5 &cm & \triv & - & \triv& - & $\Z_2$ & $\Sigma_{\lambda}$\\ 
        6 &pmm & $\Z_2^4$&  $\Theta_{\alpha},\Theta_{\beta},\Theta_{\gamma},\Theta_{\delta}$& $\Z_2^4$ &$\Sigma_{\alpha,\lambda},\Sigma_{\delta,\lambda},\Sigma_{\beta,\mu},\Sigma_{\gamma,\mu}$&\triv&-\\ 
        7 &pmg & $\Z_2^2$& $\Theta_{\alpha}, \Theta_{\beta}$  & \triv& - &$\Z_2$ & $\Sigma_{\lambda}$\\ 
        8 &pgg & $\Z_2^2$& $\Theta_{\alpha},\Theta_{\beta}$ & \triv & - &\triv & -\\ 
        9&cmm & $\Z_2^3$ & $\Theta_{\alpha},\Theta_{\beta},\Theta_{\gamma}$ & $\Z_2^2$ & $\Sigma_{\alpha,\lambda}, \Sigma_{\alpha,\mu}$ &\triv  & -\\ 
        10&p4 & $\Z_4^2\times\Z_2$& $\Theta_{\alpha},\Theta_{\beta},\Theta_{\gamma}$ & \triv &- &\triv &-\\ 
        11&p4m &$\Z_2^3 $  & $\frac{\Theta_{\alpha}}{2},\frac{\Theta_{\beta}}{2},\Theta_{\gamma}$ & $\Z_2^3$ & $\Sigma_{\alpha,\lambda},\Sigma_{\gamma,\lambda},\Sigma_{\beta,\mu}$ & \triv &-\\ 
        12&p4g &$\Z_4\times\Z_2 $  & $\Theta_{\alpha},\Theta_{\beta}$ & $\Z_2$ & $\Sigma_{\beta}$ & \triv & -\\ 
        13&p3 & $\Z_3^3$& $\Theta_{\alpha}, \Theta_{\beta}, \Theta_{\gamma}$ & \triv & -  &\triv  & -\\ 
        14&p3m1 &\triv & -& \triv& -& $\Z_2$& $\Sigma_{\lambda}$\\ 
        15&p31m & $\Z_3$& $\Theta_{\beta}$ & \triv& -& $\Z_2$& $\Sigma_{\lambda}$\\ 
        16&p6 & $\Z_6\times\Z_3\times\Z_2$& $\Theta_{\alpha}, \Theta_{\beta}, \Theta_{\gamma}$ &\triv  & -&\triv &-\\ 
        17&p6m & $\Z_2^2$& $\frac{\Theta_{\alpha}}{3},\Theta_{\gamma}$ &$\Z_2^2$ &$\Sigma_{\alpha,\mu},\Sigma_{\gamma,\mu}$ & \triv &  -
    \end{tabular}
    \caption{An independent set of bosonic SPT invariants associated only to wallpaper group symmetry. The product of the terms in each row equals the pure $\Gwp$ classification listed in Table~\ref{tab:classif-full}. 
    We follow the unit cell conventions in App.~\ref{app:UnitCellConventions}; For concreteness, when `$\OO$' or `$\LL$' are degenerate, the invariant is evaluated using `$\OO_1$' or `$\LL_1$', respectively. The invariants constitute a generating set for the respective abelian groups, listed under the `Class.' column,  associated with type-A1, A2, and A3. The notation $\frac{1}{(\MO/2)}\Theta_{\OO}$ in the rows 11 and 17 arises because, in these cases, $\Theta_{\OO} = (\MO/2) k \mod \MO$, indicating that $k$ serves as the generator of a \(\mathbb{Z}_2\) subgroup in the classification.
    }
    \label{tab:pure_invts}
\end{table*}

\begin{table*}[t]
    \centering
    \begin{tabular}{r|l||c|w{c}{0.07\textwidth}||c|w{c}{0.07\textwidth}||c|w{c}{0.07\textwidth}}
    \multicolumn{8}{c}{\textbf{Independent set of mixed invariants between each wallpaper group and $\U(1)$ or $\ZZ_N$}} \\  \hline 
         & & 
        \multicolumn{2}{c||}{$K=\U(1)$}& 
        \multicolumn{4}{c}{$K=\ZZ_N$}
        \\ 
        \hline
        $\#$ & $G_{\text{space}}$  & B1: $\mathscr{S}_{\OO}^{\U(1)}$ &
        \multicolumn{1}{c||}{$\OO$ }
        & C1: $\mathscr{S}_{\OO}^{\ZZ_N}$ & 
        \multicolumn{1}{c||}{$\OO$}
        & C4: $\CCtwoSymbol_{\LL}^{\ZZ_N}$ & 
        \multicolumn{1}{c}{$\LL$}
        \\ \hline\hline
         1 &p1 
         &\triv &- &\triv &-  & \triv & -\\
         2&p2  & $\Z_2^3$ & $\alpha, \beta, \gamma$ & $\Z_{(2,N)}^3$ & $\alpha, \beta, \gamma$ & \triv  &-   \\
         4&pg  &\triv &- &\triv &-  & \triv & -\\
         5&cm &\triv &- &\triv &-  & $\Z_{(2,N)}$ & $\lambda$ \\
         6&pmm & $\Z_2^3$ & $\alpha,\beta,\gamma$ &$\Z_{(2,N)}^3$ & $\alpha,\beta,\gamma$ & $\Z_{(2,N)}^4$ & $\lambda,\mu,\nu,\kappa$\\
         7&pmg & $\Z_2^2$ & $\alpha,\beta$ &$\Z_{(2,N)}^2$ & $\alpha,\beta$ & $\Z_{(2,N)}$ & $\lambda$\\
         8&pgg & $\Z_2$ & $\alpha$ & $\Z_{(2,N)}$ & $\alpha$ & \triv & - \\
         9&cmm & $\Z_2^2$ & $\alpha,\gamma$ &$\Z_{(2,N)}^2$ & $\alpha,\gamma$ &$\Z_{(2,N)}^2$ & $\lambda,\mu $\\
         10&p4 & $\Z_4 \times \Z_2$ & $\alpha,\gamma$ &$\Z_{(4,N)}\times\Z_{(2,N)}$ & $\alpha,\gamma$ &  \triv & -  \\
         11&p4m & $\Z_4\times\Z_2$ & $\alpha,\gamma$ & $\Z_{(4,N)}\times\Z_{(2,N)}$ & $\alpha,\gamma$ &$\Z_{(2,N)}^3$ & $\lambda,\mu,\nu$ \\
         12&p4g & $\Z_4$ & $\alpha$ &$\Z_{(4,N)}$ & $\alpha$ & $\Z_{(2,N)}$ & $\lambda$\\
         13&p3 & $\Z_3^2$ & $\alpha,\beta$ &$\Z_{(3,N)}^2$ & $\alpha,\beta$ & \triv & -  \\
         14&p3m1 & $\Z_3^2$& $\alpha,\beta$ &$\Z_{(3,N)}^2$ & $\alpha,\beta$   & $\Z_{(2,N)}$ & $\lambda$ \\
         15&p31m & $\Z_3$ & $\alpha$ &$\Z_{(3,N)}$ & $\alpha$ & $\Z_{(2,N)}$ & $\lambda$ \\
         16&p6 & $\Z_6$ & $\alpha$& $\Z_{(6,N)}$& $\alpha$ & \triv & - \\
         17&p6m & $\Z_6$ & $\alpha$ & $\Z_{(6,N)}$ & $\alpha$ & $\Z_{(2,N)}^2$ & $\lambda, \mu$  \\
         \multicolumn{8}{c}{} 
         \vspace{-1em}
         \\
         \hline
                  & & 
        \multicolumn{2}{c||}{$K=\U(1)$}& 
        \multicolumn{4}{c}{$K=\ZZ_N$}
        \\ 
         \hline
         $\#$ & $G_{\text{space}}$  & B3: $\tilde{\Lambda}_{\LL}^{\U(1)}$ &
        \multicolumn{1}{c||}{$\LL$ }
        & C3: $\tilde{\Lambda}_{\LL}^{\ZZ_N}$ & 
        \multicolumn{1}{c||}{$\LL$}
        & C4: $\CCtwoSymbol_{\LL}^{\ZZ_N}$ & 
        \multicolumn{1}{c}{$\LL$}
        \\ \hline\hline
        3&pm 
         &$\Z_2$ & $\lambda$
         &$\Z_{(2,N)}$& $\lambda$
         &$\Z_{(2,N)}^2$ & $\lambda, \mu$ \\
    \end{tabular}
    \caption{{For each type of invariant, we list the rotation center (`o') or reflection line (`l') about which the invariant needs to be evaluated.} The unit cell notation is from App.~\ref{app:UnitCellConventions}. For $K = \U(1),\Z_N$, there is an additional filling invariant valued in $\Z, \Z_N$ respectively; {the type-B1 and C1 invariants around the rotation centers not included in the table can be used to determine the filling invariant modulo some integer (see Sec.~\ref{sec:overall}).} The product of terms in each row, together with the filling, equals the mixed classification listed in Table~\ref{tab:classif-full}. {The group pm is an exception that is handled with type B3 and C3 invariants instead of type B1 and C1, respectively (Sec.~\ref{sec:InvariantsB4C4}).}
    }
    \label{tab:mixed_invts_1}
\end{table*}

\begin{table}[t]
    
    \begin{tabular}{
    w{c}{0.004\textwidth}|w{l}{0.035\textwidth}||w{c}{0.025\textwidth}|w{c}{0.03\textwidth}||w{c}{0.03\textwidth}|w{c}{0.065\textwidth}}
     \multicolumn{6}{c}{\makecell{\textbf{Independent set of mixed invariants} \\\textbf{between each wallpaper group and $\SO(3)$}} } \\  \hline 
        $\#$ &
        $G_{\text{space}}$  & 
        D1
        & 
        $\OO$ 
        & 
        D4
        & 
        $\LL$
        \\ \hline\hline
        1 & p1 & \triv & - &\triv & -\\
        2 & p2   & $\Z_2^3$ & $\alpha,\beta,\gamma$ & \triv & - \\
        3 & pm &\triv & -  & $\Z_2^2$ & $\lambda,\mu$\\
        4 & pg  &\triv  &-  & $\Z_2$ & $\lambda$\\
         5 &cm & \triv & - & $\Z_2$ & $\lambda$ \\
         6  &pmm  & \triv & - & $\Z_2^4$ & $\lambda, \mu,\nu,\kappa$\\
         7 &pmg & $\Z_2^2$& $\alpha,\beta$ & $\Z_2$ & $\lambda$\\
         8 &pgg & $\Z_2^2$ & $\alpha,\beta$ & \triv & -\\
         9 &cmm & $\Z_2^2$ & $\alpha,\gamma$ & $\Z_2$ & $\lambda$\\
         10 &p4 & $\Z_2^2$ & $\alpha,\beta$ & \triv &-\\
         11 &p4m & \triv &- & $\Z_2^3$& $\lambda,\mu,\nu$\\
         12 &p4g & $\Z_2$ & $\alpha$ &  $\Z_2$ & $\lambda$\\
         13 &p3 & \triv & -   & \triv & -\\
         14 &p3m1 & \triv &   - & $\Z_2$ & $\lambda$\\
         15 &p31m & \triv  & -  & $\Z_2$ & $\lambda$\\
         16 &p6 & $\Z_2$ & $\alpha$ & \triv & - \\
         17 &p6m  & \triv & - & $\Z_2^2$ & $\lambda,\mu$
    \end{tabular}
\caption{
{An independent set of mixed bosonic SPT invariants between an internal $\SO(3)$ symmetry and $G_{\text{space}}$. These are of type D1 ($\mathscr{S}^{\SO(3)}_{\OO}$) and type D4 ($\CCtwoSymbol_{\LL}^{\SO(3)}$). The unit cell notation is from App.~\ref{app:UnitCellConventions}. The product of invariants on each row is equal to $\H^1(\Gwp,\Z_2)$, or equivalently the mixed-$\SO(3)$ classification in} Table~\ref{tab:classif-full}.
}
    \label{tab:mixed_invts_2}
\end{table}

The characterization and classification of topological phases with crystalline symmetries has seen remarkable progress over the last several years (for a partial list of references, see Refs.~\cite{hasan2010,fu2011topological,Benalcazar2014,ando2015topological,watanabe2015filling,watanabe2016filling,schindler2018higher,khalaf2018higher,Benalcazar2019HOTI,bernevig2013topological,Chiu2016review,Kruthoff2017TIBandComb,Bradlyn2017tqc,Po2017symmind,watanabe2018structure,khalaf2018symmetry,tang2019comprehensive,Cano_2021,Elcoro2021tqc,herzogarbeitman2022interacting,Essin2014spect,Essin2013SF,YangPRL2015,hermele2016,zaletel2017,song2017,Huang2017lowerDimCrysSPT,Song2020RealSpaceRecupeTopCrystallineState,Thorngren2018,manjunath2021cgt,Manjunath2020fqh,Miert2018dislocationCharge,Li2020disc,Liu2019ShiftIns,zhang2022fractional,zhang2022pol,zhang2023complete,manjunath2022mzm,manjunath2023characterization}). It is now well understood how to both classify and characterize free fermion phases with crystalline symmetries in (2+1) dimensions based on their topological band structure \cite{bernevig2013topological,Chiu2016review,Kruthoff2017TIBandComb,Bradlyn2017tqc,Po2017symmind,watanabe2018structure,khalaf2018symmetry,tang2019comprehensive,Cano_2021,Elcoro2021tqc}. Beyond free fermions, one can consider symmetry-protected topological (SPT) states, which can be adiabatically connected to a trivial product state through a finite-depth circuit that breaks symmetry, but not through one that preserves symmetry \cite{Chen2013SPTGroupCohomology,Senthil2015SPT,gu2014,kapustin2014SPTbeyond,kapustin2015fSPT,Wang2020fSPT}. We can also consider invertible topological states, which have the property that any invertible state $\ket{\Psi}$ has an inverse, denoted $\ket{\Psi^{-1}}$, such that $\ket{\Psi} \otimes \ket{\Psi^{-1}}$ can be adiabatically connected to a trivial product state \cite{barkeshli2021invertible,aasen2021characterization}\footnote{Note that SPT states are always invertible, but in (2+1) dimensions the converse is true if and only if the invertible state has vanishing chiral central charge, that is, it does not have gapless chiral edge states when defined on open boundaries.}. In these cases, where the system can have arbitrarily strong interactions, our understanding is less complete than for free fermions. It is now quite well understood how to \textit{classify} crystalline topological states in these cases using a combination of physical constructions \cite{song2017,Huang2017lowerDimCrysSPT,Song2020RealSpaceRecupeTopCrystallineState} and mathematical techniques based on topological quantum field theory (TQFT) and higher category theory \cite{Barkeshli2019,freed2016,Thorngren2018,manjunath2021cgt,Manjunath2020fqh,bulmashSymmFrac,barkeshli2021invertible,aasen2021characterization,manjunath2022mzm}. However, we still do not fully understand how to extract a complete set of crystalline topological invariants given a microscopic lattice model or ground state wave function.

Previous work on this question has proceeded in two broad directions. The first is to measure the invariants using the response of the system to inserting symmetry defects, specifically fluxes of internal symmetries and lattice defects such as disclinations and dislocations. This approach has been explored in detail in systems with orientation-preserving crystalline symmetries \cite{Benalcazar2014,Miert2018dislocationCharge,Li2020disc,Liu2019ShiftIns,Song_2020monopole,You2020hoe,zhang2022fractional,zhang2022pol,barkeshli2025disclinations}. While defects have also been studied theoretically in the orientation-reversing case \cite{barkeshli2020reflection,Barkeshli2020Anomaly}, they are not as well understood. 

A second approach is to apply \textit{partial} symmetry operations, that is, to measure the expectation value of the ground state with respect to a symmetry operator which is restricted to act only on a subregion \cite{shiozaki2017MBIfSPTs,zhang2023complete,manjunath2023characterization,kobayashi2025crystalline}. Ref.  \cite{zhang2023complete,manjunath2023characterization} showed that partial rotations can completely characterize invertible fermionic states with $\Gwp$ wallpaper group and $\U(1)$ charge conservation symmetries, where $\Gwp$ is orientation-preserving. Indeed, if we additionally know the charge per unit cell (filling), the Chern number and the chiral central charge $c_-$ of the system, this characterization was shown to be complete. While partial reflections have analogously been shown to characterize certain topological invariants associated with reflection symmetries \cite{shiozaki2017MBIfSPTs}, there is currently no systematic procedure to obtain a full set of invariants for each wallpaper group $\Gwp$ using partial symmetry operations.

The goal of this paper is to obtain a complete characterization of crystalline topological invariants based on partial symmetry operations for bosonic SPT states with symmetry $G = \Gwp \times K$ where $\Gwp$ is a \textit{general} wallpaper group in 2$d$ ($d$ denotes the space dimension), and $K = \U(1), \Z_N$ or $\SO(3)$ is an internal unitary symmetry group. These symmetries determine a rich classification of SPT phases; detecting them requires new types of invariants, which we develop in this paper. Although we focus on bosonic SPT states in this work, we expect that our results should generalize to invertible fermionic states, with some modifications. Note that all the invariants we propose are expected to be well-defined for SPT states with arbitrarily strong interactions.

We use the classification of crystalline SPTs based on the `crystalline equivalence principle' (CEP), which states that we should treat spatial symmetries as on-site symmetries, with the only caveat that space-time orientation-reversing symmetries become anti-unitary symmetries \cite{Thorngren2018}. The classification for on-site symmetries is obtained using the group cohomology framework \cite{Chen2013SPTGroupCohomology}. The CEP has been extensively checked for bosonic SPTs by matching its predictions to independent real-space classifications \cite{song2017,Song2020RealSpaceRecupeTopCrystallineState}. 

For each pair $(\Gwp,K)$, the known classification is comprised of three groups of topological invariants: 1) pure crystalline invariants; 2) pure internal invariants; and 3) mixed invariants. This is summarized in Table~\ref{tab:classif-full}. The `pure' invariants are those invariants that are protected only by $\Gwp$ (crystalline) or $K$ (internal). The mixed invariants are protected by both $K$ and $\Gwp$. Note that the allowed values of pure internal invariants may be constrained in the presence of reflection symmetry; for example, the Hall conductance is forced to be zero if there are reflections. 

Assuming that the pure $K$ invariants are known, our main result is that partial symmetries are able to detect the pure crystalline and mixed invariants, except for the $\U(1)$ filling and $\Z_N$ filling (charge per unit cell). We confirm this by constructing exactly solvable ground states for each type of invariant and analytically demonstrating that the quantities defined in Table~\ref{tab:invts_def} take the expected nontrivial values. In particular, several of our schemes that detect invariants for reflection symmetries have not appeared in previous work.

We obtain our main result as follows. First, we define a general set of partial symmetry expectation values that are summarized in Table~\ref{tab:invts_def}; these quantities are defined with respect to a specific point or line within a real-space unit cell, and can be applied to different $\Gwp$. For convenience we refer to them as Type-A, Type-B, and so on; this notation will be further explained below in Sec.~\ref{sec:invts_def}. 
Using them, we propose a complete set of crystalline invariants
\footnote{In this draft, we use the term `invariant' to refer to two different objects: (1) the quantized coefficients appearing in a topological field theory; and (2) the quantized numbers extracted from expectation values of partial symmetry operations, both of which can be related to each other. }
for each $\Gwp$ in Tables~\ref{tab:pure_invts} (pure crystalline) and~\ref{tab:mixed_invts_1},~\ref{tab:mixed_invts_2} (mixed between $\Gwp$ and $K$). We do so by repeatedly evaluating selected invariants from Table~\ref{tab:invts_def} at different locations in the real-space unit cell. Our procedure reproduces the mathematical classification of SPT states based on group cohomology, which is summarized in Table~\ref{tab:classif-full}. Note that our methods are equally applicable to symmorphic and non-symmorphic lattices, and can therefore handle glide symmetries as well. 

The formulas cited in Table~\ref{tab:invts_def} make the assumption that all pure $K$ SPT invariants are trivial. In general, some of them need to be modified when the $K$ invariant is non-trivial, but to simplify our formulas we do not consider this most general case. The given formulas correctly predict the difference between the crystalline invariants of two states which share the same $K$ invariant. 
We will briefly discuss the case with nontrivial $K$ invariants in Sec.~\ref{sec:Disc-InternalSPTs}. 
Furthermore, note that type 1 and 2 invariants require only the density matrix for a single ground state on a open disk. This adds to a growing body of work devoted to extracting topological invariants from a single ground state wave function \cite{levin2006,kitaev2006topological,shiozaki2017MBIfSPTs,dehghani2021,cian2021,cian2022extracting,Kim2022ccc,fan2022,zhang2023complete}. We elaborate on this in Sec.~\ref{sec:Disc-singleWfn}.

To further illustrate our approach, we consider in detail the wallpaper groups p4m (\# 11) and p4g (\# 12). p4m is the symmetry group of the square lattice that is a symmorphic group that is an extension of p4 (\# 10). The p4g group is also an extension of p4 but is nonsymmorphic.
We show how the invariants in Table~\ref{tab:invts_def} reproduce the group cohomology classification. In the p4m case, we also construct a topological effective action involving gauge fields for the wallpaper group symmetry, and relate the field theory coefficients to the above partial symmetry invariants. Our method of deriving the action can be generalized to all $\Gwp$. Although we do not explore this here, these effective actions may be useful to motivate alternative characterizations based on symmetry defects. 

\subsection{Prior work}\label{sec:PriorWork}

\subsubsection{Real space constructions/invariants:} 
A number of previous works have developed real-space constructions of bosonic SPT ground states with the symmetries of interest in this paper. These include \cite{song2017,Huang2017lowerDimCrysSPT} (for $G = \Gwp$), and \cite{Song2020RealSpaceRecupeTopCrystallineState} (for $G = \Gwp \times K$). 

In particular, Ref.~\cite{Huang2017lowerDimCrysSPT} studied SPT phases protected by $G=\Gwp$. They argued that 2d bosonic crystalline SPTs can be built from 0-dimensional blocks, which can be thought of as the bosonic version of  `atomic insulators (AI)'. Additionally, \cite{Huang2017lowerDimCrysSPT} argued that their classification can be understood in terms of `point group SPT invariants' and  related `weak invariants', which are defined in terms of the atomic limit. The main result of \cite{Huang2017lowerDimCrysSPT} relevant to us is summarized in their Table III. 

Ref.~\cite{Song2020RealSpaceRecupeTopCrystallineState} extended the above construction to $G = \Gwp \times K$ SPT phases with arbitrary $K$ \footnote{They actually consider a more general setting where $G$ is not necessarily a direct product of the wallpaper group and an internal symmetry group. }. They argued that their method reproduces the classification using the crystalline equivalence principle (CEP) of Ref.~\cite{Thorngren2018} in the sense that both methods give the same abelian group of distinct SPT phases. In particular, Table I in their Supplemental Material gives the classification for various $K$ ($G_0$ in their notation) in terms of $d$-dimensional blocks (column labeled by $E_{d,\infty}^{d}$). 

Note that while these past works focused on computing various cohomology classifications of SPT phases, our goal is rather to explicitly give the ground state expectation values which detect these SPT phases.

\subsubsection{Fermions} Fermionic analogs of the partial rotation invariants $\Theta_{\OO}$ in Table~\ref{tab:invts_def} have been studied numerically in Refs. \cite{shiozaki2017MBIfSPTs,zhang2023complete,herzogarbeitman2022interacting}; these works have established that partial rotations are indeed a viable method of extracting rotational SPT invariants in microscopic models. Analogous numerical studies have not yet been carried out for the other invariants {relevant to reflection symmetry}, particularly those that depend on both $K$ and $\Gwp$. 

\subsection{Organization of paper}

The rest of the paper is organized as follows. In Sec.~\ref{sec:MainRes} we give the basic definition and properties of each invariant in Table~\ref{tab:invts_def}. In Secs.~\ref{sec:p4m},~\ref{sec:p4g} we consider the groups p4m and p4g respectively, and illustrate how to use these invariants to obtain a full characterization. In Sec.~\ref{sec:GroupCoh} we consider the 17 wallpaper groups in general and discuss how the invariants we define fully capture their group cohomology classification. Then in Sec.~\ref{sec:Discussion} we conclude and discuss future directions. Some background details and several technical derivations have been placed in the appendices. A summary of the notation we use is given in Table~\ref{tab:list_of_notations}. A brief summary of crystallographic notation is given in App.~\ref{app:CrystallographyConcepts}.

\section{Main results}\label{sec:MainRes}

\subsection{Conceptual origin of invariants}\label{sec:assumptions} 

Before getting into the specifics of each invariant, let us explain some general ideas that are useful in deriving them. These ideas also help us verify their robustness as topological invariants to some extent. 

First, the classification of SPT phases using group cohomology, combined with the crystalline equivalence principle, suggests a connection between SPT phases and topological quantum field theories (TQFTs). TQFT invariants for internal symmetries can be obtained by computing partition functions on manifolds with specific background gauge field configurations, but for an arbitrary crystalline symmetry it is not clear what manifold we should consider even after applying the CEP. This was only known previously for the partial rotation invariants in (2+1)D that we will introduce below \cite{shiozaki2017MBIfSPTs,turzillo2025}. In this paper we give a TQFT interpretation for most of the remaining invariants involving reflection symmetry. Altogether, we now have an understanding for almost every invariant we propose based on TQFT partition functions, except for the `weak' invariants in Table~\ref{tab:invts_def} that depend on system size and are therefore different from the other cases.

Another perspective comes from studying the entanglement spectrum of an SPT state, certain features of which are expected to be universal within the SPT phase. The low-lying ground state entanglement spectrum in many examples coincides with that of a (1+1)D conformal field theory (CFT). We start by assuming that the density matrix $\rho_D$ of the SPT state within a suitably chosen region $D$ equals $\rho_{CFT}$, the density matrix of the CFT living on the boundary of $D$. Then the real space invariant can be calculated in terms of correlation functions in the CFT.  This CFT calculation has been done for partial rotation invariants (Type A1, B1, C1, D1) in Ref.~\cite{zhang2023complete} (similar calculations were done in Ref.~\cite{shiozaki2017MBIfSPTs}). For bosonic SPT states, the expected answer from TQFT is contained in the leading term of the CFT correlation function, while the remaining terms become negligible for sufficiently large $|\partial D|$. However this calculation has not been done for the remaining invariants.

A third, independent motivation comes from performing explicit evaluations in simple lattice models. In this paper we construct multiple exactly solvable ground states in spin models motivated by known real-space constructions. For these states, the SPT invariants have a clear interpretation in terms of symmetry charges {or one-dimensional SPT states} localized at specific points or lines in the real-space unit cell. 
We verify that each partial symmetry invariant does take nontrivial values in at least one of the states we construct, is related to the real-space invariants in a simple way, and obeys the quantization conditions predicted by group cohomology.

The new invariants we propose in this paper were motivated by a combination of the above ideas. The fact that we can provide intuition for these invariants both from TQFT and microscopic real-space calculations gives us confidence that our partial symmetry invariants are indeed robust signatures of a topological phase. 

{
\subsection{Definition of invariants}\label{sec:invts_def}

Consider a bosonic SPT state $\ket{\Psi}$ on a torus or an open disk.  We organize the different invariants into the following classes:
\begin{enumerate}
    \item Type-A refers to pure crystalline invariants. There are three sub-classes, which we denote by A1, A2, A3. 
    \item The remaining invariants are all mixed invariants between $\Gwp$ and $K$. We refer to these invariants as Type-B, C or D for $K = \U(1),\Z_N,\SO(3)$ respectively.
    \item Invariants of type A1, B1, C1 and D1 can all be detected by performing suitable partial rotations on a given ground state.
    \item The remaining invariants can all be detected by performing suitable partial reflections on a ground state, {possibly on tori with twisted boundary conditions or different system sizes. } 
    \item `Weak' invariants are partly protected by a translation symmetry, and must be computed by taking ratios of expectation values for different system sizes.
    \item The invariants which depend on both $K$ and $\Gwp$ come in two types:
    \begin{enumerate}
        \item The `relative' invariants are defined as the difference between two quantities extracted from partial symmetry expectation values defined using the same space group symmetry but different internal symmetries; they can be computed on an open patch with arbitrary boundary conditions. 
        \item The `twisted' invariants need to be defined on a torus with specific boundary conditions along one direction.  
    \end{enumerate} 
\end{enumerate}
Below we give some general intuition behind deriving the different invariants, before giving their explicit definition and quantization rules. In Sec.~\ref{sec:overall}, we explain with examples how suitable combinations of these invariants can be used to characterize a given SPT state. Note that there are several relations between the different invariants, some of which we will point out as we go along. In Sec.~\ref{sec:GroupCoh}, we give a more mathematical discussion of the precise group cohomology classes measured by each invariant. The analysis there allows us to conclude that the above invariants fully distinguish all bosonic SPT invariants predicted by group cohomology, except for filling invariants which must be obtained separately.
 
\subsubsection{Type A1:  {Partial rotation}}
 The results in this section appeared previously in Ref.~\cite{zhang2023complete} in the context of invertible fermionic states. First we define $\Cmo|_D$ to be the restriction of the rotation operator $\Cmo$ to some symmetric open region $D$ centered at $\OO$. 

As in the case of invertible fermionic states, the expectation value of the partial rotation behaves as
\begin{align}\label{ThetaOExplicit}
     \bra{\Psi} \Cmo|_D \ket{\Psi} &= e^{- \gamma |\partial D| + i \frac{2\pi}{\MO} \Theta_{\OO}} (1 + O(e^{- \epsilon |\partial D|})),
\end{align}
$\gamma$ sets the amplitude of the expectation value, while $\epsilon$ is some positive number that captures subleading contributions. As $\tilde{C}_{\MO}$ is a symmetry, $\ket{\Psi}$ and $\tilde{C}_{\MO}|_{D}\ket{\Psi}$ will look locally the same away from the boundary of $D$, except for a possible phase. The exponential decay in $\partial D$ comes from correlations between points close to $\partial{D}$. 

We summarize the expression in Eq.~\ref{ThetaOExplicit} as
\begin{align}\label{eq:ThetaO}
    \arg\mel{\Psi}{\tilde{C}_{\MO}|_D}{\Psi} \to  \frac{2\pi}{M_{\OO}} \Theta_{\OO} \mod 2\pi.
\end{align}
$\arg \, z$ stands for the argument of the complex number $z$ and the symbol `$\to$' means that the quantization is obtained for a large enough region $D$. While we expect that a similar scaling of the expectation values in Eq.~\ref{ThetaOExplicit} applies to all the invariants we present in this paper, we do not have such explicit formulas for them. Therefore we use the $\rightarrow$ symbol in later equations to indicate that suitable choices of large regions are required to get the correct results numerically. 

We argue that $\Theta_{\OO}$ is a many-body topological invariant for bosonic states and is quantized modulo $\MO$. This can be shown by assuming that the ground state's density matrix $\rho_D$ within region $D$ is equivalent to $\rho_{\text{CFT}}$, the density matrix of the conformal field theory (CFT) on the boundary $\partial D$. This relationship was first proposed in the context of fractional quantum Hall states \cite{Haldane2008entanglement} and has since been applied to various gapped topological states, as discussed for example in \cite{Qi2012entanglement}. In this case, the expectation value on the left-hand side can be simplified using CFT techniques to an expression involving only the $G$-crossed modular data associated to defects of the rotational symmetry, which depends on the invariant $\Theta_{\OO}$. This calculation was carried out for invertible fermionic states in Ref.~\cite{zhang2023complete}, and can be adapted to bosonic SPT states. The result is that in bosonic SPT phases, $\Theta_{\OO}$ is defined mod $\MO$, is a many-body topological invariant (because it depends only on TQFT data), and it fully characterizes the SPT invariant for pure $\MO$-fold point group rotations about $\OO$. Furthermore, $\Theta_{\OO}$ is quantized to integer values when $c_- = 0$.

The quantization of the rotation invariants is different between bosonic SPTs and invertible fermionic states. In the fermionic case, the partial rotation expectation value can depend on the size of $D$. Consider a four-fold rotation center. Enlarging $D$ adds new states in orbits of 4, which are permuted under the rotation. In the fermionic case, the new state inside $D$ differs by a factor $\ket{\psi} = c^{\dagger}_1c^{\dagger}_2c^{\dagger}_3c^{\dagger}_4 \ket{0}$, where $\ket{0}$ is the vaccumm. Due to the anticommutation relations of fermionic operators, $\bra{\psi}\tilde{C}_{4,\OO} \ket{\psi} =- 1$, if $\tilde{C}_{4,\OO}^4 = +1$. Therefore, the topological invariant associated to partial rotations with $\tilde{C}_{4,\OO}^4 = +1$ is defined mod 2 and not mod 4 \cite{zhang2023complete}. On the other hand, the fact that bosonic operators commute implies that the analogous partial rotation expectation value in the bosonic SPT case is independent of the size of $D$, and remains quantized mod 4.  

In the presence of a reflection $R_{\LL}$ along a line $\LL$ that goes through $\OO$ ($\LL \ni \OO$), the SPT state must be an eigenstate of the reflection. Therefore $\mel{\Psi}{O}{\Psi} = \mel{\Psi}{R^{\dagger}_{\LL}O R_{\LL}}{\Psi}$ for any operator $O$. In the case $O = \Cmo|_D$, we obtain the constraint $\mel{\Psi}{\Cmo|_D}{\Psi} = \left[\mel{\Psi}{\Cmo|_D}{\Psi}\right]^\dagger$, thus $2\Theta_{\OO}/\MO = 0 \mod 1 $. As reflections also forces a vanishing chiral central charge ($c_- = 0 $), $\Theta_\OO$ is constrained to take certain values: 1) If $\MO$ is even, $\Theta_\OO \in \{ 0, \MO/2\}$; 2) If $\MO$ is odd, $\Theta_\OO = 0$.   We remark that any reflection lines which do not intersect $\OO$ will not affect the quantization of $\Theta_{\OO}$. This is relevant to understanding certain $\Z_4$ rotation invariants that appear in the classification of SPT phases with $\Gwp=\text{p4g}$.

Eq.~\eqref{eq:ThetaO} along with the quantization conditions on $\Theta_{\OO}$ can be analytically verified in different exactly solvable models. First, we consider fixed-point wavefunctions for general bosonic SPT states with $\Gwp$ symmetry, which were previously constructed in Refs.~\cite{song2017,Huang2017lowerDimCrysSPT}. It is straightforward to apply partial rotation operations to these wavefunctions and analytically verify Eq.~\eqref{eq:ThetaO}. We can also consider exactly solvable models for bosonic SPTs based on stacking Affleck-Kennedy-Lieb-Tasaki model (AKLT) chains; these are not in a fixed point limit but are nonetheless analytically trackable. We compute $\Theta_{\OO}$ for these models in App.~\ref{app:singletCover:PartialRotation}. 

\subsubsection{Type B1,C1,D1: Discrete shift}

Next we define a set of mixed invariants between the rotation point group at $\OO$ and the internal symmetry $K$. When $K = \U(1)$, this invariant is called the `discrete shift', and has been studied in several recent works \cite{Liu2019ShiftIns,Li2020disc,zhang2022fractional,zhang2022pol}. 

For each $K$, we define a `dressed' rotation operator $\tilde{C}_{k,\OO}({\bf g})$ which corresponds to a $2\pi/k$ rotation about $\OO$ composed with an element ${\bf g} \in K$, such that ${\bf g}^k$ is the identity in $K$:
\begin{equation}\label{eq:CKODef}
    \tilde{C}_{k,\OO}({\bf g}) := \Cmo^{\frac{\MO}{k}} \times U_{\bf g}.
\end{equation}
These operators are of order $k$, and are used to define $\Theta_{k,\OO}(\gbf)$ as
\begin{equation}\label{eq:ThetaKODef}
    \frac{k}{2\pi} \arg\bra{\psi} \tilde{C}_{k,\OO}({\bf g})|_D \ket{\psi} \rightarrow \Theta_{k,\OO}({\bf g}) \mod k.
\end{equation}

Similar CFT arguments to those of Ref.~\cite{zhang2023complete} show that $\Theta_{k,\OO}({\bf g})$ measures a many-body topological invariant defined modulo $k$. When $c_- = 0$, $\Theta_{k,\OO}({\bf g})$ is integer valued and gets contributions from the pure internal invariant ($\ZZ_k$ Hall conductance $\sigma_{\gbf}$), the pure crystalline invariant ($\Theta_{\OO}$), and the mixed invariant ($\sko(\gbf)$). We can thus extract $\sko(\gbf)$ if we know $\sigma_{\gbf}$. In this work, we will assume that $\sigma_{\gbf} =0$ and thus (from the CFT calculation in \cite{zhang2023complete}):
\begin{equation}\label{eq:ThetakO:ThetaOSko}
    \Theta_{k,\OO}({\bf g}) = \Theta_{\OO} + \sko(\gbf) \mod k .
\end{equation}
Note that the expressions we obtain can be used to detect difference in crystalline invariants (either mixed or pure) between two SPTs with the same $\sigma_{\bf g}$. The case with non-trivial $\sigma_{\gbf}$ is discussed further in Sec.~\ref{sec:Discussion}.

In a real space picture, $\sko(\gbf)$ measures the $\gbf$ charge localized at $\OO$. However, even if $\gbf$ is an element of a larger Abelian group, the charge is defined at most modulo $\MO$ because one can always move charge away from $\OO$ in an $\MO$-fold symmetric way. Furthermore, whenever there is an internal symmetry element, $\kbf$, such that $\kbf^{-1}\gbf \kbf =\gbf^{-1}$, we expect that $\sko(\gbf)  = -\sko(\gbf) \mod k$ from the same argument used for the Class A1 invariants. 

We now identify a set of invariants that are complete for each $K \in \{\U(1),\ZZ_N, \SO(3)\}$. When $K = \U(1)$, we take $k = \MO$ and $U_{\bf g} = e^{\frac{2\pi i}{\MO} \hat{N}}$ where $\hat{N}$ is the boson number operator, and $\Theta_{\OO}^{\U(1)} := \Theta_{\MO,\OO}({\bf g})$ for this ${\bf g}$. Following Eq.~\ref{eq:ThetakO:ThetaOSko}, we can extract the mixed invariant $\so^{\U(1)}$ as
\begin{equation}\label{eq:ThetaOU1}
    \so^{\U(1)} := \Theta_{\OO}^{\U(1)}-\Theta_{\OO}  \mod \MO. 
\end{equation}
Here $\Theta_{\OO}$ is the pure rotation SPT invariant discussed above, while $\so^{\U(1)}$ is the mixed SPT invariant. Note that spatial reflections do not constrain the value of $\so^{\U(1)}$. 

When $K = \Z_N$, we take $k = (\MO,N)$ and $\gbf = \Sbf^{N/(N,k)}$ where $\Sbf$ is the generator of $\ZZ_N$, and $\Theta_{\OO}^{\Z_N} := \Theta_{k,\OO}({\bf g})$ for this $k,{\bf g}$. Following Eq.~\ref{eq:ThetakO:ThetaOSko}, we can extract the mixed invariant $\so^{\Z_N}$ as
\begin{equation}\label{eq:ThetaOZN}
    \so^{\Z_N} := \Theta_{\OO}^{\Z_N}-\Theta_{\OO}  \mod (\MO,N). 
\end{equation}

When $K = \SO(3)$, we can take $k=\MO$ and ${\bf g}$ to be a $2\pi/\MO$ spin rotation about any axis. We define $\Theta_{\OO}^{\SO(3)} := \Theta_{\MO,\OO}({\bf g})$. Applying the above CFT arguments to the $\Z_{\MO}$ subgroup of $\SO(3)$, we can define the mixed invariant
\begin{equation}\label{eq:ThetaOSO3}
    {\so^{\SO(3)}:=\frac{(\MO,2)}{\MO} \left(\Theta_{\OO}^{\SO(3)}- \Theta_{\OO} \right) \mod (\MO,2).} 
\end{equation}

Note that $\so^K$ fixes the $K$ charge at disclinations centred at $\OO$. In particular, when the disclination angle is $\Omega$, this invariant contributes an excess $K$ charge at a disclination given by
\begin{align}
    Q &=
\frac{\Omega}{2\pi} \times \begin{cases}
       \so^{\U(1)} 
       \mod 1 & ,K = \U(1);\\
       \so^{\Z_N}  \mod \frac{(\MO,N)}{\MO} & ,K = \Z_N;\\
\frac{\MO }{(\MO,2)} \so^{\SO(3)} \mod 1 & ,K = \SO(3).
    \end{cases}
\end{align}
Furthermore, differences $\so^K - \mathscr{S}_{\OO'}^K$ are related to a polarization of $K$ charge; this gives an alternative many-body definition of polarization in 2d systems with rotational symmetry. See Ref.~\cite{zhang2022pol} for a more complete discussion of this point.

Finally note that the filling is partially determined by partial rotation invariants. Class A1 invariants determine the $\U(1)$-filling and $\Z_N$-filling modulo $M$ and $(M,N)$, respectively ({c.f.} Sec.~\ref{sec:p4mxU1}).

\subsubsection{Type A2: Partial double reflection}
\label{sec:InvariantsA2}

Next we discuss a different class of pure crystalline invariants that explicitly depends on reflection symmetries. Suppose there is a high symmetry point $\OO$ with at least a $D_2 \cong \Z_2 \times \Z_2$ site symmetry group generated by
two reflections $\rbf_{\LL},\rbf_{\LL'}$ about perpendicular lines $\LL,\LL'$ passing through $\OO$.
In this case, we can define an invariant $\Sigma_{\OO,\LL}$, which is protected by the pair of reflections\footnote{Note that a origin $\OO$ and a line $\LL$ uniquely fix the second reflection axis $\LL'$, which is why we do not include $\LL'$ in the label of the invariant.}. 

\begin{figure}
    \centering
    \includegraphics[width=0.5\textwidth]{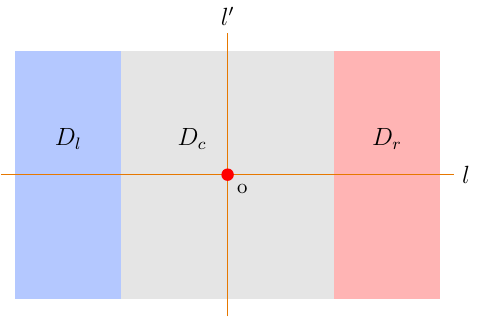}
    \caption{Definition of the decomposition of the disk $D$ into three regions $D_l$, $D_c$ and $D_r$ used in the definition of $\Sigma_{\OO,\LL}$. }
    \label{fig:DefinitionDForPartialReflection}
\end{figure}

We choose a $D_2$ invariant region $D$ that we decompose into three contiguous regions $D_l, D_c$ and $D_r$ as shown in Fig.~\ref{fig:DefinitionDForPartialReflection}. We choose the decomposition such that: 1) $D_c$ is symmetric under $\rbf_{\LL}$ and $\rbf_{\LL'}$; 2) $D_l$ and $D_r$ are symmetric under $\rbf_{\LL}$; and 3) $D_l$ and $D_r$ are mapped to each other by $\rbf_{\LL'}$. We define $\Sigma_{\OO,\LL}$ by
\begin{align}\label{eq:SigmaOL_defn}
   \frac{2}{2\pi}\arg \mel{\Psi}{\RL|_{D_{c}}\RLp|_{D_l\cup D_r}}{\Psi} \to   \Sigma_{\OO,\LL} \mod 2 .
\end{align}
We refer to this expectation value as `partial double reflection'. 

In Appendix~\ref{app:ReflectionInvaraint}, we show that the above expectation value simulates the partition function of the underlying TQFT on a space-time manifold $\Mmc^3 \cong \RP^2_{\rbf_{\LL}} \times S^1_{\rbf_{\LL}\rbf_{\LL'}}$:
\begin{equation}
 (-1)^{\Sigma_{\OO,\LL}} = \mathcal{Z}(\RP^2_{\rbf_{\LL}} \times S^1_{\rbf_{\LL}\rbf_{\LL'}}).
\end{equation}
Here $\RP^2_{\rbf_{\LL}}$ (resp. $ S^1_{\rbf_{\LL}\rbf_{\LL'}}$) is the manifold $\RP^2$ (resp. $S^1$) with $\rbf_{\LL}$ ($\rbf_{\LL}\rbf_{\LL'}$) holonomy along the non-trivial 1-cycle (where $\rbf_{\LL}\rbf_{\LL'}$ is treated as an internal symmetry).

It turns out that the partition function $\mathcal{Z}(\RP^2_{\rbf_{\LL}} \times S^1_{\rbf_{\LL}\rbf_{\LL'}})$, together with the partition function $\Zmc(\RP^3_{\rbf_{\LL}\rbf_{\LL'}})$ (which is related to $
\Theta_{\OO}$) are enough to detect all the SPTs protected by $D_2=\ZZ_2^{\rbf_{\LL}} \times \ZZ_2^{\rbf_{\LL}\rbf_{\LL'}}$.\footnote{The superscript in $\ZZ_2^{\rbf_{\LL}}$ denotes that the $\mathbb{Z}_2$ generator is a reflection $\rbf_{\LL}$.}

We verified that the partial double reflection indeed detects the topological invariant for an explicit non-trivial example. In App.~\ref{app:singletCover:def}, we constructed a state by placing singlets on the bonds of the square lattice, referred to as `\singletName' in the following. This state can also be constructed by starting with AKLT states on every axis on the square lattice and removing the projector to the spin $S=1$ sector on every lattice site. In App.~\ref{app:singletCover:2Reflection} we evaluated the partial double reflection and found a non-trivial value as expected from the fact that the AKLT state belongs to a non-trivial SPT protected by reflection symmetry \cite{Pollmann2010EntanglementSpectrumSPT1d,pollmann2012symmetry}.

To put the above invariant in context, recall that there is a unique non-trivial SPT in (1+1)D protected by reflection symmetry. This SPT is detected by evaluating a partial reflection, which simulates the partition function on $\RP^2$ with $\rbf$ flux along the non-trivial 1-cycle ($\Zmc(\RP^2_{\rbf})$)\cite{pollmann2012symmetry, cho2015}. According to the crystalline equivalence principle, SPTs protected by reflection are in one-to-one correspondence with SPTs protected by time-reversal symmetry. Therefore, there is a unique non-trivial reflection SPT in $(1+1)D$ but no non-trivial SPT in $(2+1)D$ \cite{Chen2013SPTGroupCohomology} with a single reflection. An alternative perspective to understand the lack of non-trivial reflection SPTs in $(n+1)D$, is to use the folding trick \cite{Lake2016FoldingTrick}, which roughly says that we can understand reflection SPTs protected in $(n+1)D$ by restricting to the $(n-1)d$ reflection hyperplane and treating the reflection as an on-site $\ZZ_2$ symmetry. Therefore, a non-trivial $(2+1)D$ reflection SPT corresponds to placing a $(1+1)D$ SPT protected by $\ZZ_2$ on the reflection axis. However, it is also known that there is no non-trivial $\ZZ_2$ SPT in $(1+1)D$ \cite{Chen2013SPTGroupCohomology}. 

The Type A1 and A2 invariants are not all independent. For example, since $\rbf_{\LL} \rbf_{\LL'}$ is a $C_2$ rotation around $\OO$, the quantity $\Sigma_{\OO,\LL} + \Sigma_{\OO,\LL'}$ should depend on $\Theta_{\OO}$. Indeed, when $M_{\OO}=2$ , we can show the relation 
\begin{equation}
    \Sigma_{\OO,\LL} + \Sigma_{\OO,\LL'} = \Theta_{\OO} \mod 2
\end{equation}
for states that admit an atomic limit (See App.~\ref{app:RelationBetweenTypeAInvariants}). In Table~\ref{tab:pure_invts}, we have presented one independent set of invariants, with the convention that we first list all possible $\Theta_{\OO}$, followed by the remaining independent choices of $\Sigma_{\OO,\LL}$.

\subsubsection{Type A3: Weak partial reflection}\label{sec:InvariantsA3}

When the unit cell contains reflection lines but no $C_2$-symmetric points lying on them (wallpaper groups \textit{pm}, \textit{pg}, \textit{cm}, \textit{p31m}, and \textit{p3m1}), the type A2 invariant cannot be used. In this case, we introduce an alternative `Type A3' invariant $\Lambda_{\LL}$, protected by the combination of $D_1$ reflection symmetry about the line $\LL$, generated by the operator $\RL$, and $\Z$ translation symmetry along $\LL$. {Because translation symmetry is essential for its definition, we refer to this as a \emph{weak partial reflection}. This invariant differs from A1 and A2 in that it requires evaluating the ground state on systems of different sizes.}

Consider the ground state on an $L_1 \times L_2$ torus, with $L_1$ and $L_2$ much larger than the correlation length. Here $L_1$ is the length along the direction of $\LL$. Denote the ground state by $\ket{\Psi(L_1,L_2)}$. Let $D$ be a region that is invariant under $\rbf_{\LL}$ and fully contains $\LL$.   We define the invariant as 
\begin{widetext}
    \begin{equation}\label{eq:SigmaL_defn}
\frac{1}{\pi}\arg\frac{\mel{\Psi(L_1+1,L_2)}{\RL |_D}{\Psi(L_1+1,L_2)}}{\mel{{\Psi(L_1,L_2)}}{\RL |_D}{{\Psi(L_1,L_2)}} } \to \Lambda_{\LL} \mod 2.
\end{equation}
\end{widetext}
This equation becomes exact in the $L_1,L_2\to \infty$ limit.  $\Lambda_{\LL}$ is related to $\Sigma_{\OO,\LL}$, when they can both be defined (see App.~\ref{app:RelationBetweenTypeAInvariants} for the precise relation). 

In a fixed-point limit where the degrees of freedom are all localized at specific points in the unit cell, this invariant measures the reflection eigenvalue of these degrees of freedom per unit length along the line $\LL$. 

\begin{figure}[h!]
    \centering
    \includegraphics[width=0.45\textwidth]{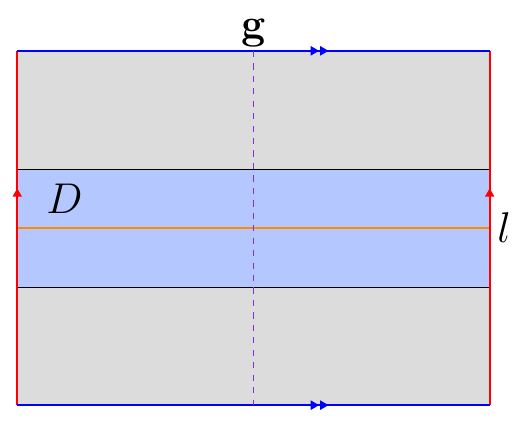}
    \caption{Setup to evaluate $\CCtwoSymbol_{\LL}(\gbf;\jbf)$. The red and blue lines denote the identification of sides of the rectangular space into a torus. $D$ is the region in light blue. The dashed purple line denotes boundary conditions twisted by the group element $\gbf$. The reflection line $\LL$ is shown in orange. }
    \label{fig:DefinitionDForQInvariant}
\end{figure}

\subsubsection{Type C4,D4: Twisted relative partial reflection}

Given a reflection $\rbf_{\LL}$ and pair of group elements $\jbf,\gbf \in K$ such that $\jbf^2 = \zero$ and $\gbf \jbf=\jbf\gbf$, we can define new invariants $\CCtwoSymbol_{\LL}(\gbf;\jbf)$.

Let $\ket{\Psi_{\gbf}}$ be the ground state on a torus with twisted boundary conditions by $\gbf$ around a loop that intercepts the reflection line $\LL$ once (see Fig.~\ref{fig:DefinitionDForQInvariant}). Let $D$ be a large region containing $\LL$ that is $\rbf_{\LL}$ invariant. $\CCtwoSymbol_{\LL}(\gbf; \jbf)$ is defined as 
\begin{equation}
    \frac{1}{\pi} \arg \frac{\mel{\Psi_{\gbf}}{(\RL U_{\jbf})|_{D}}{\Psi_{\gbf}}}{\mel{\Psi_{\gbf}}{(R_{\LL})|_{D}}{\Psi_{\gbf}}} \to \CCtwoSymbol_{\LL}(\gbf; \jbf) \mod 2.
\end{equation}

We first argue that $\mel{\Psi_{\gbf}}{\RL |_{D}}{\Psi_{\gbf}}$ simulates the partition function on $S^1\times \RP^2$ with $\gbf$ holonomy along $S^1$. The spatial manifold where $\ket{\Psi_{\gbf}}$ lives is $S^1_x\times S^1_y$, where $S^1_{x/y}$ represents the circle along the horizontal/vertical direction. The twisted boundary conditions are understood as $\gbf$ flux along $S^1_x$. For a fixed vertical cut of the spatial region, the above corresponds to the evaluation of partial reflection for a (1+1)D SPT state. This effectively "simulates" the manifold $\RP^2$ \cite{shiozaki2017MBIfSPTs} for each point in $S^1_x$, thus simulating $S^1\times \RP^2$ in total. We define $\CCtwoSymbol_{\LL}(\gbf;\jbf)$ using of a ratio of expectation values in order to get rid of spurious bulk contributions that can depend on the size of the torus.

When $K=\ZZ_{N}$, we define the type C4 invariant as
\begin{equation}\label{eq:QRS_defn}
    \CCtwoSymbol^{\ZZ_N}_{\LL} := \CCtwoSymbol_{\LL}(\Sbf; \bf{0}),
\end{equation}
where $\Sbf$ is a generator of $\ZZ_N$. 

When $K=\SO(3)$, we define the type D4 invariant as 
\begin{equation}\label{eq:QRXZ_defn}
    \CCtwoSymbol^{\SO(3)}_{\LL} := \CCtwoSymbol_{\LL}(\Zbf; \Xbf) - \CCtwoSymbol_{\LL}(\Zbf; \bf{0})
\end{equation}
where $\Zbf$ and $\Xbf$ are the elements in $\SO(3)$ corresponding to $\pi$-rotations around the z and x axes, respectively.

Note that we don't define a type B4 invariant with $K = \U(1)$ because the group cohomology calculation tells us that this invariant should be trivial. Furthermore, the type C4 invariant for a $\ZZ_{N}$ is if $\ZZ_{N}$ is a subgroup of a $\ZZ_{2N}$ symmetry.

In Appendix~\ref{app:singletCover:TwistedReflection} we construct a state by stacking AKLT chains, and explicitly verify that the type-C4 and D4 invariants for this case take the expected non-trivial values.

\subsubsection{Type C2,D2: Relative partial double reflection}\label{sec:TypeC3D3}

The above type C4 and D4 invariants require the use of multiple ground states on a torus with twisted boundary conditions along different cycles. 
It may be more desirable to have an alternative scheme using a single ground state wave function {on a disk}, as was the case for all the partial rotation invariants. We now present such a scheme by modifying the partial double reflection operators appearing in the type A2 case. Because the type C4 and D4 invariants appearing in Tables~\ref{tab:mixed_invts_1},~\ref{tab:mixed_invts_2} form a complete set, the invariants presented in this section can be expressed in terms of them. We expect that, when they exist, type C2 and D2 invariants give the same information as type C4 and D4 invariants, respectively, but we have not checked this.

Pick regions $D_l$, $D_c$ and $D_r$ as in Fig.~\ref{fig:DefinitionDForPartialReflection}, as well as group elements $\jbf,\kbf \in K$ such that $\jbf^2=\bf{0}$ and $\kbf\jbf=\jbf\kbf$. We define the operator
\begin{equation}\label{eq:dressedPartialDoubleReflection}
    \Rmc_{\OO,\LL}(\kbf,\jbf) = {
    \left(\RLp\right)|_{D_l\cup D_r}\left(\RL U_{\jbf}\right)|_{D_c}
    \left(U_{\kbf}\right)|_{D_l}
    \left(U_{\kbf}^{\dagger}\right)|_{D_r}
    },
\end{equation}
and the invariant ${\Sigma}_{\OO,\LL}(\kbf,\jbf)$ as
\begin{equation}
    \frac{1}{\pi }\arg \mel{\Psi}{\Rmc_{\OO,\LL}(\kbf,\jbf)}{\Psi} \to \Sigma_{\OO,\LL}(\kbf,\jbf)  \mod 2. 
\end{equation}
The expectation value of $\Rmc_{\OO,\LL}(\kbf,\jbf)$ simulates the partition function on the same manifold as Type A2 but with different holonomies. In App.~\ref{app:DressedPartialDoubleReflection} we argue that 
\begin{equation}
    (-1)^{\Sigma_{\OO,\LL}(\kbf,\jbf)} = \Zmc(\RP^2_{\rbf_{\LL} \jbf} \times S^1_{\jbf \kbf\hbf_{\OO} }).
\end{equation}

As in the case of rotations, we define a quantity that measures the mixed SPT invariant by appropriately subtracting pure invariants. 

When $K=\ZZ_{N}$, we define the type C2 invariant as
\begin{equation}\label{eq:SigmaS_defn}
    \tilde\Sigma^{\ZZ_N}_{\OO,\LL} := \Sigma_{\OO,\LL}(\Sbf, \zero) - \Sigma_{\OO,\LL}(\zero,\zero)
\end{equation}
where $\Sbf$ is a generator of $\ZZ_N$.

When $K=\SO(3)$, we define the type D2 invariant as 
\begin{equation}\label{eq:SigmalXZ_defn}
    \tilde\Sigma^{\SO(3)}_{\OO,\LL} := \Sigma_{\OO,\LL}(\Zbf, \Xbf) - \Sigma_{\OO,\LL}(\zero, \Xbf)
\end{equation}
where $\Zbf$ and $\Xbf$ are the elements in $\SO(3)$ corresponding to $\pi$-rotations around the z and x axes, respectively.

In App.~\ref{app:singletCover:2Reflection}, we evaluate the invariants for the \singletName state and explicitly verify that the type-C2 and D2 invariants for this case take the desired non-trivial values.

\subsubsection{Type B3, C3: Relative partial weak reflection}\label{sec:InvariantsB4C4}

For certain wallpaper groups, there exist $\mathbb{Z}_2$ invariants corresponding to the charge mod 2 per unit length along a reflection axis for $K=\U(1), \Z_N$. Except for the wallpaper group \textit{pm}, these invariants can be detected using the previously defined constructions.\footnote{One might expect that the same issue would appear for the wallpaper groups pg and cm, which also lack rotational symmetries. We find that there are no mixed invariants at all with $K = \U(1)$, and the sole mixed $K=\Z_N$ invariant for the group cm can be detected by type C4 invariants.}

To address the above exception, we propose to use a relative version of type-A3 invariants to detect these states. In other words, let $\Lambda_{\LL}(\jbf)$ be $\Lambda_{\LL}$ evaluated with $R_\LL \to R_\LL U_{\jbf}$.
We define the relative weak partial reflection (type B3 and C3) as
\begin{equation}\label{eq:DressedWeakPartialReflection}
   \tilde{\Lambda}_{\LL}^{K}:={\Lambda_{\LL}(\pi) - \Lambda_{\LL}}  \mod 2; 
\end{equation}
where $\pi$ is the order two element in $K = \U(1), \ZZ_{N}$.

\subsection{Overall classification}\label{sec:overall}

A central result of this work is that all the SPT invariants which depend on $\Gwp$ can be obtained by evaluating the invariants in Table~\ref{tab:invts_def} at suitable locations in the real-space unit cell of $\Gwp$. The invariants which depend only on $\Gwp$ are given in Table~\ref{tab:pure_invts}, while those that depend on both $\Gwp$ and $K$ are given in Table~\ref{tab:mixed_invts_1},~\ref{tab:mixed_invts_2}.

For a given $\Gwp$, each row of the table gives one independent set of invariants. This set need not be unique, as mentioned at different points in the previous section. 

\subsubsection{How to read the tables: an example}
For a concrete example of how to read the tables, consider the group pmm (\# 6). The unit cell for this group is shown in Fig.~\ref{fig:UnitCell_pmm_6}. There are four high symmetry points denoted $\OO = \alpha, \beta, \gamma, \delta$; these are order two rotation centers. Each point also lies on two mutually perpendicular reflection axes, which we denote by $l = \lambda, \mu,\nu,\kappa$. Now the classification of pure crystalline invariants is given by $\H^4(\text{pmm},\Z^{\OR}) \cong \Z_2^8$. As listed in Table~\ref{tab:pure_invts}, one independent set is given by $\Theta_{\OO}$ for each $\OO$ (that is, four type-A1 invariants), along with four invariants of type A2. Note that an alternative but equivalent set is given by the 8 different invariants of type A2 (all possible choices of $\Sigma_{\OO,l}$ where $\OO$ lies on $l$).

For pmm, the classification of mixed invariants when $K = \U(1),\Z_N,\SO(3)$ is given by $\Z\times \Z_2^3, \Z_N \times \Z_{(2,N)}^7, \Z_2^4$ respectively. First let $K = \U(1)$. The $\Z$ invariant corresponds to the filling (charge per unit cell) $\nu$. This invariant is not listed in the table, as it is common to each $\Gwp$ and also cannot be fully determined by partial point group operations. We assume that the filling is either already specified or can be calculated separately. But the remaining $\Z_2^3$ classification can be obtained by evaluating $\so^{\U(1)}$ at any three high symmetry points. Note that there are only three independent mixed invariants because $\sum_{\OO} \so^{\U(1)} = \nu \mod 2$. There are no mixed invariants that depend on reflection symmetries.

When $K = \Z_N$, we have a $\Z_N$ analog of the filling as well as three independent $\so^{\Z_N}$ invariants, when $N$ is even. In this case, there are four additional $\Z_{(2,N)}$ invariants of type C4, which can be measured by $\CCtwoSymbol_{\LL}^{\Z_N}$ where $\LL$ runs over the four reflection lines. 

Finally, when $K = \SO(3)$, the mixed SPT invariants can be detected by evaluating the type-D4 invariant $\CCtwoSymbol_{\LL}^{\SO(3)}$ on the four reflection lines. Note that we could alternatively measure three independent type-D1 invariants along with a single type-D4 invariant. (See App.~\ref{app:RelationBetweenTypeDInvariants:pmm}, which gives relations between invariants of type D1 and D2)

\begin{figure}
    \centering
    \includegraphics[width=0.35\textwidth]{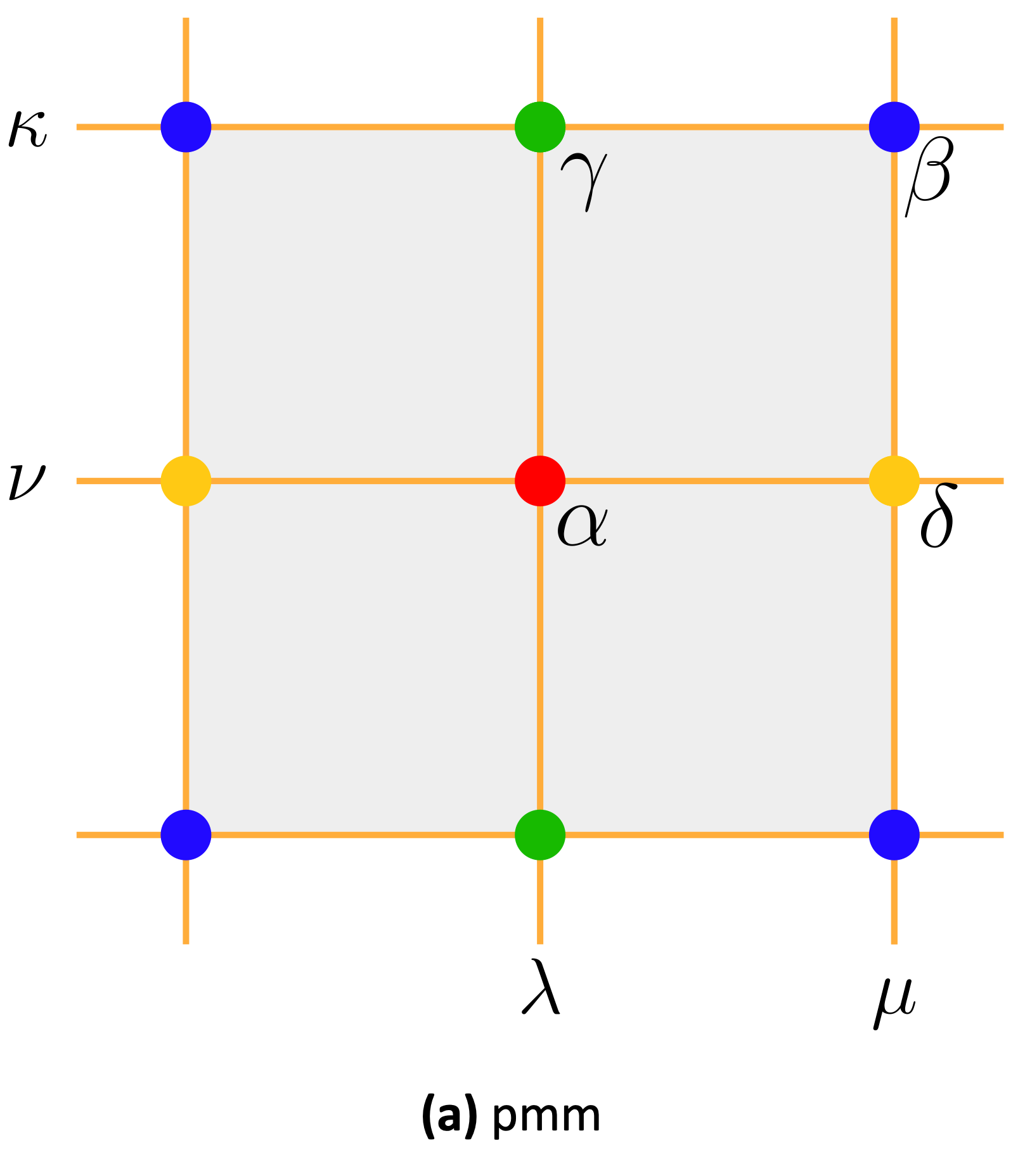}
    \caption{Unit cell for space group pmm.}
    \label{fig:UnitCell_pmm_6}
\end{figure}

\section{\texorpdfstring{$\text{{p4m}}$}{p4m}}\label{sec:p4m}

In this section we focus on the symmetries of the square lattice { $\mathrm{p4m}=\ZZ^2\rtimes(\ZZ_4\rtimes\ZZ_2)$}. 
 We explain the specific SPT invariants that arise in detail, and derive a topological effective action describing the response. 
 As in previous sections we assume that: 1) the chiral central charge $c_-=0$; and that 2) the topological invariants for internal symmetries are trivial. If these assumptions are not satisfied, the invariants may satisfy other quantization conditions \cite{zhang2023complete,turzillo2025}.

In this section and the next, we will repeatedly refer to a set of real-space constructions which provide representative ground states for a large class of bosonic SPT phases and which give a simple way to understand the response properties of invariants appearing in a topological field theory \cite{song2017,Huang2017lowerDimCrysSPT,Song2020RealSpaceRecupeTopCrystallineState}. In these constructions, a crystalline SPT wave function is constructed as a tensor product of lower-dimensional states defined at the high symmetry points and lines of the real-space unit cell. These constructions generally assume a `Wannier limit' in which the degrees of freedom are supported at such high symmetry regions. The various SPT invariants can then be understood in terms of the symmetry eigenvalues or quantum numbers of the localized degrees of freedom; the precise values are often obvious from the construction. {When the degrees of freedom are localized on single points, we refer to the such states as atomic insulators (AI) \cite{Huang2017lowerDimCrysSPT}.}
Based on prior numerical studies on some of the invariants in this paper, for example Refs.~\cite{zhang2022fractional,zhang2022pol,zhang2023complete}, it is reasonable to believe that many predictions motivated by this construction should also hold away from the Wannier limit.

\subsection{Conventions}

\subsubsection{Unit cell}

The wallpaper group \text{p4m} has 6 Wyckoff positions with a non-trivial site-group (Fig.~\ref{fig:p4m+p4g_UnitCell}). There are three maximal Wyckoff positions, $\alpha,\beta,\gamma$, that also appear in the space group \text{p4}. $\alpha$ and $\beta$ have site-groups isomorphic to $D_4 \cong \ZZ_4\rtimes\ZZ_2$, while $\gamma$ is two-fold degenerate and has a site-group isomorphic to $D_2 \cong \ZZ_2\times\ZZ_2$. Compared to the wallpaper group p4$=\ZZ^2\rtimes\ZZ_4$, there are three new Wyckoff positions. These positions lie on the reflection lines $\lambda,\mu,\nu$.
Each of these positions have site group $D_1 \cong \ZZ_2$. There is also the generic Wyckoff position with a trivial site-group. 

\begin{figure*}
    \centering
    \includegraphics[width=0.8\textwidth]{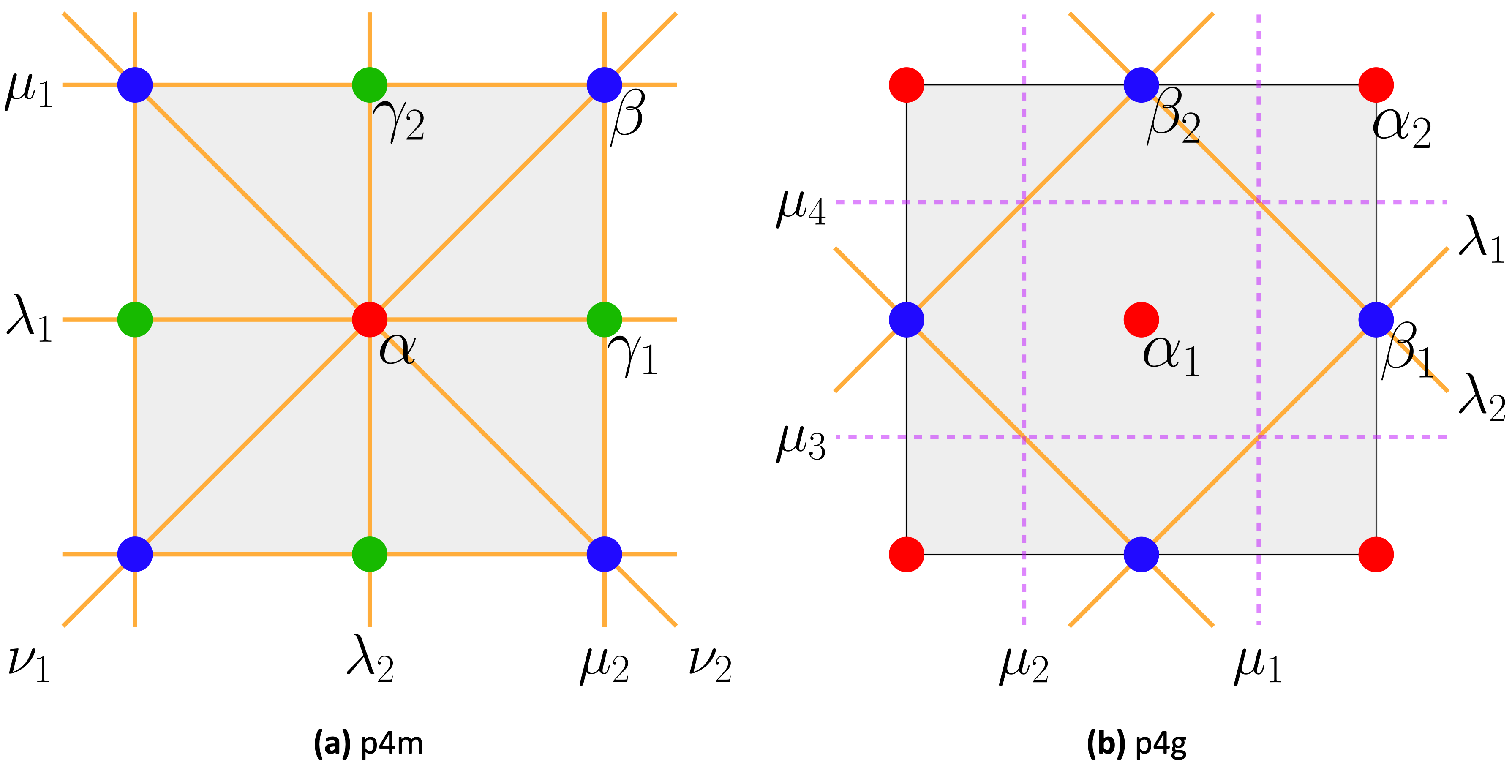}
    \caption{Unit cell conventions for \textbf{(a)} p4m and \textbf{(b)} p4g. 
    {Maximal Wyckoff positions are labeled by early Greek letters ($\alpha,\beta,\gamma)$. Orange and dashed purple lines correspond to reflection and glide axes, respectively. These lines are labeled by mid-range Greek letters ($\lambda,\mu,\nu$). 
    We use labels with the same subscripts to denote 
    positions or lines that are related by a point group symmetry. } }
    \label{fig:p4m+p4g_UnitCell}
\end{figure*}

\subsubsection{Definition of gauge fields}

We fix an origin $\OO$ such that $M_{\OO} =4$, that is, $\OO$ either belongs to $\alpha$ or $\beta$, and is contained in the horizontal line $\LL = \lambda_1$ or $\mu_1$ respectively. Furthermore, we parametrize the group elements $\gbf\in \text{p4m}$ as $\xbf^{t_{x,\gbf}}\ybf^{t_{y,\gbf}}\hbf_{\OO}^{n_{\gbf}} \rbf_{\LL}^{r_{\gbf}}$ where $\xbf (\ybf)$ is translation by one unit-cell in the x(y) direction, $\rbf_{\LL}$ is the reflection about $\LL$ as defined above, and $\hbf_{\OO}$ is a counter-clockwise rotation by 90 degrees around $\OO$. $t_{x,\gbf}, t_{y,\gbf}, n_{\gbf}$ and $r_{\gbf}$ are integers with the redundancies $n_{\gbf} \sim n_{\gbf}+ 4$ and $r_{\gbf} \sim r_{\gbf}+ 2$, which follow from $\hbf^4_{\OO}=\rbf_{\LL}^2=1$. Furthermore, $\hbf_{\OO} \xbf = \ybf \hbf_{\OO} $, 
$\hbf_{\OO} \ybf = \xbf^{-1} \hbf_{\OO} $,
$\hbf_{\OO} \rbf_{\LL} = \rbf_{\LL} \hbf_{\OO}^3 $,  $\rbf_{\LL} \xbf =  \xbf \rbf_{\LL}$, and 
$\rbf_{\LL} \ybf =  \ybf^{-1} \rbf_{\LL}$.

Per the crystalline equivalence principle, we define the topological effective action for the group p4m in terms of a background gauge field for an internal symmetry isomorphic to p4m. Therefore we consider a closed 3-manifold $\mathcal{M}^3$ with a triangulation and define a flat gauge field on the links of the triangulation as ${B} = (\vec{R},\omega,\sigma)$.\footnote{\label{fn:origin}Note that the components of the gauge field $B $ depend on a choice of origin $\OO$ and a reflection axis $\LL$. For ease of notation, we will suppress these dependencies while writing the gauge fields, and only make them explicit in the field theory coefficients. 
} The three components are gauge fields for translations, rotations and reflections respectively.
The basic quantization of the three components is 
\begin{align}
    \frac{1}{2\pi}\vec{R}_{} &\in \Z^2 & \nonumber \\
    \omega &\in \frac{2\pi}{4} \Z; & \omega\sim \omega+ 2\pi \nonumber \\
    \sigma &\in 
     \Z; & \sigma \sim \sigma + 2
\end{align} 
Recall that the reflection gauge field should be identified with the first Stiefel-Whitney class of the space-time manifold \cite{kapustin2015fSPT}. 

\subsection{Pure crystalline invariants}
The pure crystalline SPT invariants are classified by $\Hmc^4(\text{p4m},\ZZ^{\OR}) = \ZZ_2^6$. We will explain how to understand them in terms of SPTs protected only by different site symmetry groups. Recall that the site groups of the maximal Wyckoff positions $\OO$ are given by the dihedral group $D_{M_{\OO}}$ with $M_{\OO}$ being the order of rotations about $\OO$. $\MO$ equals $2$ or $4$ for p4m.

\paragraph{Single WP:} The effective Lagrangian density for $D_{2n}$-SPTs is (see App.~\ref{app:Cohomology:D2n} for a derivation):
\begin{equation}\label{eq:topActionD2n}
    \Lmc_{D_{2n}} = k_{1,\OO} n \omega \frac{\dOR\omega}{2\pi} + k_{2,\OO,\LL}\pi \sigma  \frac{\dOR\omega}{2\pi}.
\end{equation}
with $k_{1,\OO},k_{2,\OO,\LL} \in \ZZ_2$. $\dOR$ is the differential twisted by orientation. $k_{1,\OO}$ and $k_{2,\OO,\LL}$ depend on the rotation center $\OO$ and reflection line $\LL$ used to define the $D_{2n}$ gauge fields. {In the case of a single WP, $\OO$ is fixed uniquely, but we will need the subscript in the discussion below.} 

Recall that the partial symmetry operators evaluate the partition functions on certain manifolds, and these can be evaluated explicitly for the topological action in Eq.~\ref{eq:topActionD2n} \cite{NatInvariants,shiozaki2017MBIfSPTs}:
\begin{equation}
    \begin{split}
        e^{ \frac{2\pi i }{2n}\Theta_{\OO} }  &= \Zmc(L(2n,1)_{\hbf_{\OO}^n}) = (-1)^{k_{1,\OO}}\\
        (-1)^{\Sigma_{\OO,\LL}}  &= \Zmc(\RP^2_{\rbf_{\LL}}\times S^1_{\hbf_{\OO}^n}) = (-1)^{k_{2,\OO,\LL}}.
    \end{split}
\end{equation}
Here $L(p,1)_{\gbf}$ is a lens space with $\gbf$ holonomy along its non-contractible cycle. This implies
\begin{equation}\label{eq:relation:k and RSI}
    k_{1,\OO} = \frac{\Theta_{\OO}}{n} \mod 2; \quad k_{2,\OO,\LL}  = \Sigma_{\OO,\LL} \mod 2.
\end{equation}

Now that we have related the real-space invariants to coefficients of the topological action, we can easily calculate the real-space invariants for the atomic insulators previously alluded to. For each site group $G_{\OO}$, an AI is fully determined by how the localized degrees of freedom at $\OO$ transform as a one-dimensional irreducible representation (1d-irrep) of $G_{\OO}$. For $G_{\OO} = D_{2n}$, there are only 4 1d-irreps that are specified by their eigenvalues under the generators $\hbf_{\OO}$ ($2n$-fold rotation) and $\rbf_{\LL}$ (reflection). These are denoted as $\lambda_{\hbf_{\OO}},\lambda_{\rbf_{\LL}} \in \{+1,-1\}$, where $\lambda_{\gbf}$ is the $\gbf$-eigenvalue of the irrep. It is straightforward to evaluate the real space invariants for the AIs in terms of their eigenvalues:
\begin{equation}
    \begin{split}
        e^{\tfrac{i \pi}{n} \Theta_{\OO}}|_{\rm{AI}}= \lambda_{\hbf_{\OO}};  \\
        (-1)^{\Sigma_{\OO,\LL} }|_{\rm{AI}}= \lambda_{\rbf_{\LL}}.
    \end{split} 
\end{equation}
Combining the above equation and Eq.~\ref{eq:relation:k and RSI}, we find that a state described by Eq.~\ref{eq:topActionD2n} is in the same phase as an atomic insulator of localized degrees of freedom with angular momentum $n k_{1,\OO} \mod 2n$, and even(odd) parity under $\rbf_{\LL}$ when $k_{2,\OO,\LL} = 0 (1)$.

\paragraph{p4m:} Mathematically, the most general effective action has the Lagrangian
\begin{equation}
   \mathcal{L} = B^* \varphi_3
\end{equation}
where $[\varphi_3]$ is an element of $\H^3(\text{p4m},\U(1)^{\OR})$, where we have identified $\U(1)$ with the real numbers modulo $2\pi$. $B^*$ denotes the pullback operation using $B$.\footnote{{In reality, $G$ gauge fields are constructed by pulling back cochains along the classifying map $f_B:\mathcal{M}\to BG$, where $BG$ is the classifying space of principal $G$-bundles and $\mathcal{M}$ is the space-time manifold. Accordingly, the notation $B^*$ is short-hand for the pullback $f_B^*$. }} That is, $B^*\varphi_3$ is a cocycle defined on the 3-simplices of the triangulation of $\mathcal{M}^3$, and the cocycle condition ensures that the associated partition function $e^{ i \int_{\mathcal{M}^3} \mathcal{L}}$ is invariant under retriangulations. Therefore, to get a general form for $\mathcal{L}$ we first need to find a general expression for $\varphi_3$.

The generators of $\Hmc^3(\text{p4m},\U(1)^{\OR})$ can be obtained as cup products of generators of $\Hmc^2(\text{p4m},\ZZ^{\OR}) \cong \ZZ_4\times\ZZ_2\times\ZZ$ with the generators of $\Hmc^1(\text{p4m},\U(1)) \cong \ZZ_2^3$.\footnote{See Sec.~\ref{sec:GroupCoh} for a sketch of the argument, and App.~\ref{app:Coh:p4m} for more explicit calculations for p4m. }
Roughly speaking, the elements of the former group correspond to symmetry fluxes, while those of the latter correspond to symmetry charges, and the cup product implements flux-charge attachment. 

Any element $[\Xi_1]\in\Hmc^1(\text{p4m},\U(1) )$ can be written as 
\begin{equation}
    B^*\Xi_1 =  q_{1,\OO}2\omega + q_{2,\OO,\LL} \pi \sigma + q_{3,\OO} \Mbs \cdot \vec{R} \in 2\pi \RR /\ZZ,
\end{equation}
with $q_{1,\OO},q_{2,\OO,\LL},q_{3}\in\ZZ_2$ and $\Mbs = [1/2,1/2]$. The quantities $2q_{1,\OO}, q_{2,\OO,\LL},$ and $q_{3}$ correspond to the charge under $\hbf_{\OO}, \rbf_{\LL}, \xbf$, respectively.

Similarly, an element $[\Xi_2]\in \Hmc^2(\text{p4m},\ZZ^{\OR})$ can be written as 
\begin{equation}
    B^*\Xi_2 = 
    j_{1,\OO} \frac{\dOR\omega}{2\pi} + 
    j_{2,\OO} \frac{\dOR(\vec{R}\cdot\Mbs)}{2\pi} + 
    j_{3} \Area,
\end{equation}
where $\Area$ is the "area form" which reduces to $\Area = \frac{1}{2\pi} R_x \wedge R_y$ in the absence of rotation or reflection fluxes \cite{manjunath2021cgt}. The coefficients are quantized as $(j_{1,\OO},j_{2},j_{3}) \in\ZZ_4 \times\ZZ_2 \times \ZZ$. 
 
Taking the cup product of both types of terms, and eliminating the redundant terms, we find the response action to be
\begin{widetext}
    \begin{equation}\label{eq:actionp4m}
    \begin{split}
    \Lmc_{\text{p4m}} &= k_{1,\alpha}  2\omega\frac{\dOR\omega}{2\pi}
    + k_{2,\alpha,\lambda_1}  \pi \sigma \frac{\dOR\omega}{2\pi} 
    + k_{3,\alpha}  (\vec{R}\cdot \Mbs) \frac{\dOR\omega}{2\pi}+ k_{4,\alpha,\lambda_1}  \pi \sigma\frac{\dOR\vec{R}\cdot \Mbs}{2\pi}
    + k_{5}  2\omega \Area
    + k_{6}  \pi\sigma \Area.
    \end{split}
\end{equation}
\end{widetext}
A physical interpretation of each field theory coefficient will be given below. However, following the general strategy of this paper, we will first relate the coefficients to the invariants $\Theta_{\OO},\Sigma_{\OO,\LL}$ defined previously, for different $\OO$ and $\LL$. 
We start by restricting the space group $\Gwp$ to site groups $G_{\OO'}$ for different $\OO'$. This restriction induces a map at the cohomology level $\Hmc^3(\Gwp,\U(1)^{\OR}) \to \Hmc^3(G_{\OO'},\U(1)^{\OR})$ that, after pulling back by $B$, allows us to express $k_{1,\OO'}$ and $k_{2,\OO',\LL'}$ in terms of the $k$-invariants appearing in Eq.~\ref{eq:actionp4m}:
\begin{equation}\label{eq:Lp4m_with_k1_k2}
    \begin{split}
        k_{1,\alpha} &= k_{1,\alpha} \\
        k_{2,\alpha,\lambda_1} &= k_{2,\alpha,\lambda_1} \\
        k_{1,\beta} &= k_{1,\alpha} + k_{3,\alpha} + k_{5}\\
        k_{2,\beta,\mu_1} &= k_{2,\alpha,\lambda_1} + k_{3,\alpha} + k_{6}\\
        k_{1,\gamma_1} &= k_{3,\alpha} \\
        k_{2,\gamma_1,\lambda_1} &= k_{2,\alpha,\lambda_1} + k_{4,\alpha,\lambda_1} \\
        k_{1,\gamma_2} &= k_{3,\alpha} \\
        k_{2,\gamma_2,\mu_1} &= k_{2,\alpha,\lambda_1} + k_{3,\alpha} + k_{4,\alpha,\lambda_1}.
    \end{split}
\end{equation}
Then we solve for $k_{j,\alpha}$:
\begin{equation}\label{eq:p4m_KinTermsOfLocal}
    \begin{split}
        k_{1,\alpha} &= k_{1,\alpha} \\
        k_{2,\alpha,\lambda_1} &= k_{2,\alpha,\lambda_1} \\
        k_{3,\alpha} &= k_{1,\gamma_1} \\
        k_{4,\alpha,\lambda_1} &= k_{2,\gamma_1,\lambda_1} + k_{2,\alpha,\lambda_1} \\
        k_{5} &= k_{1,\alpha} + k_{1,\gamma_1} + k_{1,\beta} \\
        k_{6} &= k_{2,\alpha,\lambda_1} + k_{2,\gamma_1,\lambda_1} + k_{2,\gamma_2,\mu_1} + k_{2,\beta,\mu_1} \\
        & = k_{2,\alpha,\lambda_1} + k_{2,\beta,\mu_1} + k_{1,\gamma_1}
    \end{split}.
\end{equation}
Given that $k_{1,\OO}, k_{2,\OO,\LL}$ can be directly obtained from $\Theta_{\OO}, \Sigma_{\OO,\LL}$, we can extract all the field theory coefficients from an SPT wavefunction in terms of the real space invariants $\Theta_{\OO}$ and $\Sigma_{\OO,\LL}$. This is summarized in Table~\ref{tab:pure_invts}. { See Table~\ref{tab:p4m_relations} for a list of all the relations between SPT invariants for p4m, and App.~\ref{app:RelationBetweenTypeAInvariants} for a general discussion of the relations between Type A invariants.}

We now give a direct interpretation of the coefficients in Eq.~\ref{eq:p4m_KinTermsOfLocal} that is motivated by the field theory:
\begin{enumerate}
    \item \underline{$k_{1,\alpha}$}: $\ell_{\alpha}=2k_{1,\alpha}$ is the $C_{M_{\alpha}}$ angular momentum of the ground state on a subregion. It has been well studied in the absence of reflection symmetry \cite{zhang2023complete}.
    \item \underline{$k_{2,\alpha,\lambda_1}$}: As mentioned previously, in the Wannier limit real-space construction, $p_{\alpha,\lambda_1}= k_{2,\alpha,\lambda_1} \mod 2$ is the parity under the reflection $\rbs_{\l_1}$ of the degrees of freedom localized at $\alpha$. Although we have not numerically checked it, we expect this interpretation to hold away from the Wannier limit, analogous to how fractional $\U(1)$ charges are found to be localized at crystalline defects in topological insulators or Chern insulators away from the Wannier limit \cite{zhang2022fractional,zhang2022pol}. A second interpretation suggested by the field theory is that it determines the angular momentum of defects of the reflection symmetry, but establishing this requires a further study of reflection symmetry defects which we do not pursue here.
    \item \underline{$k_{3,\alpha}$}: The quantity $\vec{\mathscr{P}}_{\alpha,s} :={k_{3,\alpha}(1/2,1/2)}$ can be understood as an angular momentum polarization, consistent with the fact that it can be expressed as a difference between $k_{1\OO}$ at different $\OO$. (See \cite{zhang2022pol,manjunath2023characterization} for a discussion on relating angular momentum polarization to other field theory coefficients.)
    
    \item \underline{$k_{4,\alpha,\lambda_1}$}: Analogously, $\vec{\mathscr{P}}_{\alpha,\lambda_1,p} := {k_{4,\alpha,\lambda_1}(1/2,1/2)}$ is a {\bf polarization of the reflection eigenvalue} $k_{2,\OO,\LL}$, or equivalently a difference between $k_{2,\OO,\LL}$ for two different choices of $\OO$ measured with respect to a single reflection axis $\LL$.

    \item \underline{$k_{5}$}: The quantity $k_{5} := \nu_s$ is interpreted as an `angular momentum per unit-cell' \cite{manjunath2021cgt,manjunath2023characterization}, and can be thought of as a generalized filling invariant. It is origin-independent, assuming we only consider the origins $\alpha,\beta$ which have the maximal site group $D_4$.
    
    \item \underline{$k_{6}$}:
    Finally, the quantity $ k_{6} := \nu_r$ measures a weighted sum of reflection eigenvalues for different points about a horizontal axis, and is also origin-independent. It can be thought of as a measure of the {\bf total reflection eigenvalue per unit cell}.
\end{enumerate}
We emphasize that $k_{4,\alpha},k_{6}$ and the partial reflection invariants which extract them have not appeared previously in the literature.

\subsection{Mixed invariants}
Next we discuss the invariants that are protected by both \text{p4m} and $K$.  The full effective action capturing these mixed invariants is given in Eq.~\ref{eq:FullMixedACtion}. The relations between the invariants that we have derived are summarized in Table~\ref{tab:p4m_relations}.

\begin{table}[t]
    \centering
    \begin{tabular}{c|c}
        Invariants being related & Relation \\ \hline \hline
        \pcol{0.25\textwidth}{ All coefficients of Eq.~\eqref{eq:actionp4m} in terms of $k_{1\OO},k_{2,\OO,\LL}$}&Eq.~\eqref{eq:p4m_KinTermsOfLocal} 
         \\ 
         \hline
        $\nu, \so^{\U(1)}$ & Eq.~\eqref{eq:nu in terms of shift}\\ 
         \hline
    $\bar{\nu}, \so^{\U(1)}$ {\color{blue}for p4g} & Eq.~\eqref{eq:p4g_nu-vs-shift}\\
     \hline
         $\overline{\mathscr{P}}_{\OO},\mathscr{S}_{\OO'}^{\U(1)}$ & Eq.~\eqref{eq:m2-from-m1}\\ 
         \hline 
         $\mathscr{S}_{\OO}^{\SO(3)},\CCtwoSymbol_{\LL}^{\SO(3)}$ & Eq.~\eqref{eq:RelationBetweenD1&D2} \\
    \end{tabular}
    \caption{Partial list of relations between the SPT invariants with symmetry $\text{p4m} \times K, K = \U(1), \SO(3)$. For comparison, one analogous relation for wallpaper group p4g is given in {\color{blue}blue}. Relations for $K=\ZZ_N$ can be obtained from those for $K=\U(1)$ after reducing modulo $N$, and changing superscripts to $\ZZ_N$.}
    \label{tab:p4m_relations}
\end{table}

\subsubsection{\texorpdfstring{$K=\U(1)$}{K=U1}: }\label{sec:p4mxU1} 
We begin with $K=\U(1)$. Note that the Hall conductance is forced to vanish because of reflection symmetry, therefore there are no pure $K$-SPT invariants. 

\paragraph{Single WP:} Let $A$ be a $\U(1)$ gauge field. The most general mixed $\U(1)\times D_{M_{\OO}}$ topological action is 
\begin{equation}\label{eq:p4m_actionCharge}
    \Lmc^{\text{mixed}}_{D_{M_{\OO}},\U(1)} = \mathscr{S}^{\U(1)}_{\OO} A \frac{\dOR \omega}{2\pi},
\end{equation}
with $\mathscr{S}^{\U(1)}_{\OO}\in \ZZ_{M_{\OO}}$. In the absence of reflections, the above action appeared in Refs.~\cite{zhang2022fractional,zhang2022pol} and defines the discrete shift. Although those works studied systems without reflection symmetry, the quantization of the shift is unaffected by reflections. $\mathscr{S}^{\U(1)}_{\OO}$ is extracted by a type-B1 invariant, which is a relative partial rotation. Previous real-space constructions in the Wannier limit interpret $\mathscr{S}^{\U(1)}_{\OO}$ as the $\U(1)$ charge localized at $\OO$. Note that $\mathscr{S}^{\U(1)}_{\OO}$ is only defined modulo reduction mod $M_{\OO}$  because one can move charge away from $\OO$ in multiples of $M_{\OO}$ symmetrically.

\paragraph{p4m:} According to the group cohomology calculation, the most general action is
\begin{equation}\label{eq:mixed:p4m-U1}
    \begin{split}
        \Lmc_{\text{p4m},\U(1)}^{\text{mixed}}  
        = & \mathscr{S}^{\U(1)}_{\OO}A\frac{\dOR\omega_{\OO}}{2\pi} + \overline{\mathscr{P}}_{\OO}A \frac{\dOR \vec{R}\cdot \Mbs}{2\pi}\\ &\quad + \nu A \Area,
    \end{split}
\end{equation}
where $\mathscr{S}^{\U(1)}_{\OO},\overline{\mathscr{P}}_{\OO},\nu \in \ZZ_4\times\ZZ_2\times\ZZ$ and $\Mbs = (1/2,1/2)$. Note that $\vec{\mathscr{P}}_{\OO} := \frac{\overline{\mathscr{P}}_{ \OO}}{2}(1,1)$ is the charge polarization, and $\nu$ is the filling; these quantities were studied previously in Refs.~\cite{zhang2022pol,manjunath2023characterization}. 

We can measure $\overline{\mathscr{P}}_{\OO}$ by reexpressing it in terms of $\mathscr{S}^{\U(1)}_{\OO'}$ for different $\OO'$, see Eq.~\eqref{eq:m2-from-m1} below. The filling $\nu$ needs to be computed separately, for example by finding the $\U(1)$ charge of the ground state for different system sizes. However, $\nu$ can be partially determined if we know $\mathscr{S}^{\U(1)}_{\OO'}$ at each $\OO'$. In the Wannier limit, we can calculate $\nu$ as the sum of charge at high symmetry positions (counted with multiplicity) and the charge away from them: 
\begin{equation}
    \nu = Q_\alpha +  Q_{\beta} + 2Q_{\gamma} + Q_{\text{away}}.
\end{equation}
However, because of rotation symmetry any charge assigned to $Q_{\text{away}}$ always appears in multiples of $4$. Using $Q_{\OO} = \mathscr{S}^{\U(1)}_{\OO} \mod M_{\OO}$, we obtain 
\begin{equation}\label{eq:nu in terms of shift}
    \nu = \mathscr{S}^{\U(1)}_{\alpha} + \mathscr{S}^{\U(1)}_{\beta} + 2\mathscr{S}^{\U(1)}_{\gamma} \mod 4. 
\end{equation}
This implies that one of the $\ZZ_4$ factors is redundant assuming $\nu$ is known. 

We now establish the relation between $\overline{\mathscr{P}}{\OO}$ and $\mathscr{S}^{\U(1)}{\OO'}$. From the results in Table~\ref{tab:Resp4m}, and noting that upon restriction to $\text{p1}$ the generator $\Area$ remains while all others vanish, we find
\begin{equation}
    \begin{aligned}
        \mathscr{S}^{\U(1)}_{ \alpha} &= \mathscr{S}^{\U(1)}_{ \alpha}&\mod 4 \\
        \mathscr{S}^{\U(1)}_{ \beta} &= \mathscr{S}^{\U(1)}_{ \alpha} + 2 \overline{\mathscr{P}}_{ \alpha} + \nu &\mod 4 \\
        \mathscr{S}^{\U(1)}_{ \gamma} &= \mathscr{S}^{\U(1)}_{ \alpha} + \overline{\mathscr{P}}_{ \alpha} &\mod 2.
    \end{aligned}
\end{equation}
This implies that
\begin{equation}\label{eq:m2-from-m1}
    \begin{split}
        \mathscr{S}^{\U(1)}_{ \alpha } &= \mathscr{S}^{\U(1)}_{ \alpha}\\
        \overline{\mathscr{P}}_{ \alpha } &= \mathscr{S}^{\U(1)}_{ \gamma} - \mathscr{S}^{\U(1)}_{ \alpha} \mod 2.
    \end{split}
\end{equation}

\subsubsection{\texorpdfstring{$K=\ZZ_N$}{K=ZN}:}
Next, we consider $K = \ZZ_N$. In what follows, we assume that the pure $\ZZ_N$-SPT invariant is trivial in the ground state; we discuss the consequences of relaxing this assumption in Sec.~\ref{sec:Discussion}.

\paragraph{Single WP:} mixed $D_{1}\times\ZZ_N$ SPTs are classified by $\ZZ_{(2,N)}$.\footnote{See also Ref.~\cite{Yoshida2015BSPTsReflection} where an effective boundary field theory approach was used.} 
The corresponding action is
\begin{equation}\label{eq:ActionMixedTorD1ZN}
    \Lmc_{{D}_{1},\ZZ_N}^{\text{mixed}} = {t}_{1,\LL}  \frac{N}{(N,2)} A\sigma\sigma,
\end{equation}
where $\sigma$ is the reflection gauge field, $A$ is the $\ZZ_N$ gauge field (taking values in $\tfrac{2\pi}{N}\ZZ$),  and $t_{1,\LL} \in \ZZ_{2}$.

Since $\rbf_{\LL}$ acts as a $\ZZ_2$ on-site symmetry on $\LL$, the mixed SPT can be understood as decorations of $\LL$ with (1+1)D $\ZZ_2\times \ZZ_{N}$ SPTs. An example of a non-trivial decoration is the AKLT chain. In this case, we can evaluate the type C4 invariant (Eq.~\ref{eq:QRS_defn}) $\CCtwoSymbol_{\LL}^{\ZZ_N}= 1 \mod 2$ explicitly. Therefore, $\CCtwoSymbol_{\LL}^{\ZZ_N} = t_{1,\LL}$.

In the presence of $D_{2n}$ symmetry around $\OO$, the mixed terms are 
\begin{equation}\label{eq:ActionMixedTorD2nZN}
    \begin{split}
        \Lmc_{{D}_{2n},\ZZ_N}^{\text{mixed}} 
        =&   \frac{N}{(N,2)} A \sigma\left[t_{1,\OO,\LL}\sigma  + t_{2,\OO}   {\frac{n\omega_{\OO}}{2\pi}}\right]  \\ 
        &\quad+ \mathscr{S}^{\Z_N}_{\OO} A \frac{\dOR{\omega}}{2\pi} ,
    \end{split}
\end{equation}
where we have added a subscript $\OO$ to the coefficients for later convenience.
$t_{2,\OO}$ measures the difference of $t_{1,\OO,\LL} - t_{1,\OO,\LL'}$ for $\rbf_{\LL'} =\hbf_{\OO}\rbf_{\LL}$, which can be shown by restricting the action to the two $D_{1}\times \Z_{N}$ subgroups. $\mathscr{S}^{\Z_N}_{\OO}$ is the $\ZZ_N$ version of shift and can be detected by $\Theta_{\OO}^{\ZZ_{2n}}$. 

Note that when $N$ is odd, the only allowed term is $\mathscr{S}^{\Z_N}_{\OO}  A \frac{\bar{\dd}\omega}{2\pi}$ with $\mathscr{S}^{\Z_N}_{\OO} $ is defined modulo $(N,n)$.

\paragraph{p4m:} The mixed SPT states are classified by
\begin{equation}
    \begin{split}
        \Hmc^2(\text{p4m},\ZZ_N^{\OR}) \cong & 
    (\Hmc^2(\text{p4m},\ZZ^{\OR})\otimes \ZZ_N ) \\ 
    &\quad \oplus \Tor[\Hmc^3(\text{p4m},\ZZ^{\OR}),\ZZ_N ].
    \end{split}
\end{equation}
The first term corresponds to the $\ZZ_N$ version of the $\text{p4m}\times \U(1)$ invariants. The tensor product means that we need to reduce the various invariants modulo $N$, which is expected as charge is now defined modulo $N$: 
\begin{equation}
    \Hmc^2(\text{p4m},\ZZ^{\OR})\otimes \ZZ_N  = \ZZ_{(N,4)} \times \ZZ_{(N,2)} \times \ZZ_{N}.
\end{equation}
An effective action capturing these invariants is the same as Eq.~\ref{eq:mixed:p4m-U1} (now with $\Z_N$ superscripts):
\begin{equation}\label{eq:mixed:p4m-Zn}
    \begin{split}
        \Lmc_{\text{p4m},\Z_N}^{\text{mixed}} 
        = & \mathscr{S}^{\Z_N}_{\OO}A\frac{\dOR\omega_{\OO}}{2\pi} + \overline{\mathscr{P}}_{\OO}^{\Z_N}A \frac{\dOR \vec{R}\cdot \Mbs}{2\pi}\\ &\quad + \nu^{\Z_N} A \Area,
    \end{split}
\end{equation}
The coefficients have the quantization $(\mathscr{S}^{\Z_N}_{\OO},\overline{\mathscr{P}}_{\OO}^{\Z_N},\nu^{\Z_N})\in \ZZ_{(N,4)} \times \ZZ_{(N,2)} \times \ZZ_{N}$.

Now consider the second piece. The group cohomology calculation shows that $\Hmc^3(\text{p4m},\ZZ^{\OR}) = \Z_{2}^3$ and is trivial upon restriction to p4. Evaluating the torsion, we get
\begin{equation}\label{eq:mixed_p4m_Zn_Tor}
    \Tor[\Hmc^3(\text{p4m},\ZZ^{\OR}),\ZZ_N ] = \ZZ_{(2,N)} \times \ZZ_{(2,N)} \times \ZZ_{(2,N)}.
\end{equation}
The topological action describing these `torsion' mixed SPTs is
\begin{equation}\label{eq:ActionMixedTorp4mZN}
      \Lmc_{\text{p4m},\ZZ_N}^{\text{Tor}} = \frac{2N}{(N,2)} A{\sigma(t_{1,\OO,\LL}  \frac{\sigma}{2}+t_{2,\OO}\frac{2\omega}{2\pi} + t_{3} \frac{\vec{R}\cdot \Mbs}{2\pi} )},
\end{equation}
where $t_{1,\OO,\LL},t_{2,\OO},t_{3} \in \ZZ_2^3$.

In p4m, any reflection is in the conjugacy class of one of the following reflections: $\rbf_{\l_1}$, $\rbf_{\m_1}$ or $\rbf_{\n_1}$ (see Fig.~\ref{fig:p4m+p4g_UnitCell}). For $N$ even, there are 3 root phases according to the real space construction. Corresponding ideal states are constructed by placing $\ZZ_2\times \ZZ_N$ (1+1)D SPTs on each conjugacy class of reflection lines. By restricting the action in Eq.~\ref{eq:ActionMixedTorp4mZN} from $\text{p4m}$ to the three different $D_1$ subgroups generated by each of the three reflections, we find the following relation between the topological action and the invariants: 
\begin{equation}
    \begin{aligned}
        \CCtwoSymbol_{\lambda_1}^{\ZZ_N} &= t_{1, \a,\lambda_{1}} \quad &\mod 2; \\
        \CCtwoSymbol_{\nu_1}^{\ZZ_N} &= t_{1,\a,\lambda_1} +t_{2, \a}&\mod 2 ;\\
        \CCtwoSymbol_{\mu_1}^{\ZZ_N} &= t_{1, \a,\lambda_1}+ t_{3\,\,}&\mod 2 .
    \end{aligned}
\end{equation}
See App.~\ref{app:singletCover:2Reflection} for an explicit computation of the $\CCtwoSymbol_{\LL}^{\ZZ_2}$ invariant in an ideal SPT ground state with p4m symmetry.

\subsubsection{\texorpdfstring{$K=\SO(3)$}{K=SO3}:} 
We now consider spin rotation symmetry, $K= \SO(3)$. The spin Hall conductivity is forced to be trivial due to spatial reflections. 

\paragraph{single WP:} $D_1\times \SO(3)$ mixed SPTs are classified by $\ZZ_2$. The corresponding action is
\begin{equation}\label{eq:ActionMixedTorD1SO3}
    \Lmc_{{D}_{1},\SO(3)}^{\text{mixed}} = u_{\LL}  \pi{\ww_2\sigma},
\end{equation}
where $\ww_{2}$ (the Stiefel-Whitney class of the $\SO(3)$ bundle) is the pullback of the generator of $\Hmc^2(\SO(3),\ZZ_2)$ and $u_{\LL}\in\ZZ_2$. The $u_{\LL}=1$ case can be understood as placing an AKLT chain (the non-trivial $\SO(3)$ (1+1)D SPT phase) on $\LL$. We can detect this using the Type-C4 invariant $\CCtwoSymbol_{\LL}^{\SO(3)}$ (see Eq.~\eqref{eq:QRXZ_defn}).

The AKLT chain is characterized by fractionalized $\SO(3)$ spins in the presence of open boundary conditions. Consequently, a non-trivial $u_{\LL}$ (or $\CCtwoSymbol_{\LL}^{\SO(3)}$) signals the presence of fractionalized spins at the points where $\LL$ interesects the system boundary. 

For $D_{2n}$ symmetry, the classification is $\ZZ_{2}\times\ZZ_2$, with corresponding topological action    
\begin{equation}
\Lmc_{D_{2n},\SO(3)}^{\text{mixed}} =2\pi\ww_{2}(u_{1,\OO,\LL}  \frac{\sigma}{2}+u_{2,\OO}\frac{n\omega}{2\pi}),
\end{equation}
where $u_{1,\OO,\LL}, u_{2,\OO} \in \ZZ_2$. Analogously to the $\ZZ_N$ case: $u_{1,\OO,\LL}$ corresponds to the invariant $u_{1,\LL}$ for the $D_1$ subgroup generated by $\rbf_{\LL}$.

We can understand $ 2\pi \ww_{2} \frac{n\omega}{2\pi}$ by restricting $\SO(3)$ to a $\SO(2)$ ( $\cong \U(1)$) subgroup. Under this restriction, $\ww_2 \to \frac{\dd A}{2\pi} \mod 2$, where $A$ is the $\SO(2)$ gauge field. Integrating by parts then gives $2\pi \ww_{2} \frac{n\omega}{2\pi} \to  n A \frac{\dd\omega}{2\pi}$. Therefore, there is charge $n u_{2,\OO}$ localized at $\OO$, which implies that a disclination of angle $\pm \pi/n$ carries charge $u_{2,\OO}/2$ under $\SO(2)$. In other words, the $\SO(3)$ spin at a disclination core $S_{\OO}$ satisfies $S_{\OO} = \frac{u_{2,\OO}}{2} \mod 1$. From the action restricted to $\SO(2)$, together with the known relation between dressed partial rotations and the coefficients of the topological action, we obtain $u_{2,\OO} = \mathscr{S}_{\OO}^{\SO(3)}$.

\paragraph{p4m:}

The new mixed SPT invariants are classified by $\Hmc^1(\text{p4m}, \Hmc^2(\SO(3), \U(1))) = \ZZ_2 \times\ZZ_2\times \ZZ_2$. 
The topological action for these mixed invariants is
\begin{equation}\label{eq:p4m_mixed_SO3}
\begin{split}
\Lmc_{\text{p4m},\SO(3)}^{\text{mixed}} =&  
2\pi\ww_{2}(u_{1,\OO,\LL}  \frac{\sigma}{2}+u_{2,\OO}\frac{2\omega}{2\pi}+u_{3}\frac{\vec{R}\cdot\Mbs}{2\pi} ), 
\end{split}
\end{equation}
where $u_{1,\OO,\LL},u_{2,\OO},u_{3}\in\ZZ_2$. The real space construction and the folding trick suggest, that the root states are obtained by placing AKLT states on the conjugacy classes of the reflection lines $\lambda_1$, $\mu_1$, and $\nu_1$.

From the analysis of a single WP, we have $u_{1,\alpha,\l_1} = \CCtwoSymbol_{{\l_1}}^{\SO(3)}$ and $u_{2,\alpha} = \mathscr{S}_{\alpha}^{\SO(3)}$. We now derive the relation 
\begin{equation}\label{eq:RelationBetweenD1&D2}
    \mathscr{S}_{\alpha}^{\SO(3)} = \CCtwoSymbol_{\l_1}^{\SO(3)} + \CCtwoSymbol_{\n_1}^{\SO(3)}.
\end{equation}
One approach is to start from the action in Eq.~\ref{eq:p4m_mixed_SO3} and restrict to the two $D_1$ subgroups generated by $\rbf_{\l_1}$ and $\rbf_{\n_1}$, which allow us to relate the corresponding coefficients in the action (See App.~\ref{app:RelationBetweenTypeDInvariants:p4m} for details). 

Another approach is to observe that, according to the real-space construction, the three root states for mixed p4m-$\SO(3)$ SPTs can be obtained by placing Haldane chains (the nontrivial (1+1)D $\SO(3)$ SPT) along each of the conjugacy classes of reflection lines. In these idealized states, the only sources of fractional spins are the dangling ends of the Haldane chains. In particular, when constructing a $\pi/2$ disclination, one effectively removes half of the $\lambda_1$ and $\nu_1$ reflection lines. Since $\Upsilon^{\SO(3)}_{\LL} = 1$ only if a Haldane chain is placed on $\LL$, the spin at a disclination core is  $\frac{ \mathscr{S}_{\alpha}^{\SO(3)}}{2} = \frac{\Upsilon^{\SO(3)}_{\lambda_1}+\Upsilon^{\SO(3)}_{\nu_1}}{2} \mod 1$. 

Upon restricting p4m to p1, the mixed SPT invariants are classified by $u_{3,\OO}\in \ZZ_2$. \footnote{The mixed SPTs are classified by $\ZZ_2^2$ but the $C_4$ rotation imposes that the two indices to be equal, thus reducing the classification to $\ZZ_2$} The invariant associated with this $\ZZ_2$ is the translation-$\SO(3)$ Lieb-Shultz-Mattis anomaly of the edge theory -- the parity of fractional spins per unit cell on the edge theory. By restricting the action in Eq.~\ref{eq:p4m_mixed_SO3} to the site groups at $\alpha$ and $\beta$ (see App.~\ref{app:RelationBetweenTypeDInvariants:p4m}), we obtain 
\begin{equation}
    u_{3} = u_{1,\alpha,\l_1}+u_{1,\beta,\m_1} = \CCtwoSymbol_{\l_1}^{\SO(3)} + \CCtwoSymbol_{\m_1}^{\SO(3)}.
\end{equation}
We can interpret this as follows: if translations and reflection symmetries are present on the edge, the fractional $\SO(3)$ charge on the unit cell can determined by considering only the center ($\CCtwoSymbol_{\l_1}^{\SO(3)}$) and middle ($\CCtwoSymbol_{\m_1}^{\SO(3)}$) of the edge unit cell because any contribution away from these two points comes in pairs and thus cancels out. 

\subsubsection{Summary}
The full crystalline action for SPT with wallpaper group p4m and internal symmetry $K = \U(1),\Z_N, \SO(3)$ is
\begin{widetext}
    \begin{equation}\label{eq:FullMixedACtion}
    \begin{aligned}
        \Lmc_{\text{p4m},K} &= \Lmc_{\text{p4m}} + \Lmc_{\text{p4m},K}^{\text{mix}} \\
        \Lmc_{\text{p4m}} &= k_{1,\alpha}  2\omega\frac{\dOR\omega}{2\pi}
    + k_{2,\alpha,\lambda_1}  \pi \sigma \frac{\dOR\omega}{2\pi} 
    + k_{3,\alpha}  (\vec{R}\cdot \Mbs) \frac{\dOR\omega}{2\pi}+ k_{4,\alpha,\lambda_1}  \pi \sigma\frac{\dOR\vec{R}\cdot \Mbs}{2\pi}
    + k_{5}  2\omega \Area
    + k_{6}  \pi\sigma \Area. \\
    \Lmc_{\text{p4m},K}^{\text{mix}} &= \begin{cases}
          \mathscr{S}^{\U(1)}_{\OO}A\frac{\dOR\omega_{\OO}}{2\pi} + \overline{\mathscr{P}}_{\OO}A \frac{\dOR \vec{R}\cdot \Mbs}{2\pi} + \nu A \Area &; K = U(1) ;\\
          \mathscr{S}^{\Z_N}_{\OO}A\frac{\dOR\omega_{\OO}}{2\pi} + \overline{\mathscr{P}}_{\OO}^{\Z_N}A \frac{\dOR \vec{R}\cdot \Mbs}{2\pi}+ \nu^{\Z_N} A \Area + \frac{2N}{(N,2)} A{\sigma(t_{1,\OO,\LL}  \frac{\sigma}{2}+t_{2,\OO}\frac{2\omega}{2\pi} + t_{3} \frac{\vec{R}\cdot \Mbs}{2\pi} )} &; K=\ZZ_N\\
          
2\pi\ww_{2}(u_{1,\OO,\LL}  \frac{\sigma}{2}+u_{2,\OO}\frac{2\omega}{2\pi}+u_{3}\frac{\vec{R}\cdot\!\Mbs}{2\pi} ) &; K =\SO(3) \\
    \end{cases}
    \end{aligned} 
\end{equation}
For $K=\ZZ_N$ and $N$ even, we are allowed to have a term $\sigma_{\ZZ_N} A \frac{\bar{\dd} A}{2\pi}$, with $\sigma_{\ZZ_N}$ a multiple of $N/2$. This term likely modifies the relations between the real space invariants and the effective action coefficients presented above.
\end{widetext}

\section{\texorpdfstring{$\text{p4g}$}{p4g}}\label{sec:p4g}

Next, we study a non-symmorphic example, $\Gwp = \text{p4g}$, which is a non-trivial extension of $D_1$ by $\text{p4}$. In this extension, the generator of $D_1$ squares to a translation—{i.e.}, the generator of $D_1$ becomes a glide symmetry in $\text{p4g}$. We take the unit cell to be as in Fig.~\ref{fig:p4m+p4g_UnitCell}. Because of the glide symmetry, every WP is degenerate. Therefore the MWP contains only two orbits invariant under a rotation, $\alpha$ and $\beta$. We have $G_{\alpha_j}\cong C_4$ and $G_{\beta_j} \cong D_2$, where $j=1,2$. The extra glide transformation relates the two MWPs with $M_{\OO} = 4$ of {p4}. Moreover, every point in the unit cell is mapped to a distinct point under the glide, which has direct consequences for the filling invariants, as discussed below.

First we study the pure crystalline invariants. The group cohomology result is 
\begin{equation}
    \Hmc^3(\text{p4g}, \U(1)^{\OR}) = \ZZ_4 \times\ZZ_2^2.
\end{equation}
This can be understood as $\Hmc^3(\text{p4g}, \U(1)^{\OR}) \cong  \Hmc^3(G_{\alpha_1}, \U(1)^{\OR}) \times \Hmc^3(G_{\beta_1}, \U(1)^{\OR})$. In other words, there is a $\Z_4$ type-A1 invariant for $\alpha$ and a $\Z_2^2$ invariant (one $\Z_2$ from type A1 and A2 respectively) for $\beta$. Since there is no reflection line passing through $\alpha$, the $\Z_4$ invariant is not reduced by reflections, as it would have been in a symmorphic lattice.

Next we study the mixed invariants. We note the following group cohomology results {(obtained with GAP)}:
\begin{equation}
    \begin{split}
        \Hmc^1(\text{p4g}, \ZZ^{\OR}) & = \ZZ_2 \\
        \Hmc^2(\text{p4g}, \ZZ^{\OR}) & = \ZZ \times \ZZ_4 \\
        \Hmc^3(\text{p4g}, \ZZ^{\OR}) &  = \ZZ_2.\\
    \end{split}
\end{equation}

When $K=\U(1)$, the mixed phases are classified by $\Hmc^2(\text{p4g}, \H^1(\U(1),\U(1)^{\OR})) \cong \H^2(\text{p4g},\Z^{\OR})$. The $\ZZ_4$ factor corresponds to having an integer invariant $\so^{\U(1)}$ mod 4 at $\alpha_1$, and the $\ZZ$ factor is equal to the total charge in each fundamental domain. The fundamental domain is a subset of the unit cell that generates the full unit cell upon action of translations or glides. For symmorphic groups, the unit cell is the same as the fundamental domain, but this is not true in non-symmorphic groups. An important result is that the true integer invariant in the non-symmorphic case is the filling per fundamental domain $\bar{\nu}$, while in contrast the filling per unit cell $\nu = 2 \bar{\nu}$ is an even integer. 

Consider the AI limit. Let $Q_{\OO}$ be the charge localized at $\OO$. Note that $Q_{\OO} = \so^{\U(1)}$ mod $M_{\OO}$, $Q_{\alpha_1}=Q_{\alpha_2}$ and $Q_{\beta_1} = Q_{\beta_2}$. Due to the $C_4$ symmetry around $\alpha_1$, the charge away from 
high symmetry points $\alpha_1,\alpha_2, \beta_1, \beta_2$ appear in multiples of 4, {i.e.} $Q_{\text{away}} = 0 \mod 4$. We can calculate the total filling as $\nu = 2\bar{\nu} = 2Q_{\alpha_1} + 2Q_{\beta_1} + Q_{\text{away}} $. Taking a mod 4 reduction and dividing by 2, we find $\bar{\nu} = Q_{\alpha_1}+ Q_{\beta_1} \mod 2$. Therefore, $Q_{\beta_1}$ is determined from $Q_{\alpha_1}$ and $\bar{\nu}$, and thus
\begin{equation}\label{eq:p4g_nu-vs-shift}
    \bar{\nu} = \mathscr{S}_{\alpha_1}^{\U(1)} + \mathscr{S}_{\beta_1}^{\U(1)} \mod 2.
\end{equation}
We have thus shown that $\mathscr{S}_{\beta_1}^{\U(1)}$ is not an independent invariant. (The analogous relation for p4m is Eq.~\eqref{eq:nu in terms of shift}.) 

When $K=\ZZ_N$, the mixed phases are classified by two pieces: the first is $\Hmc^2(\text{p4g}, \ZZ^{\OR})\otimes\ZZ_N$ (charges modulo $N$), and the second is a new piece $\Tor[\Hmc^3(\text{p4g}, \ZZ^{\OR}),\ZZ_N] \cong \ZZ_{(N,2)}$. From our previous discussion of the group \text{p4m}, this piece can be detected by type-C1 invariants. In {p4g}, all reflections are conjugate to $\rbf_{\lambda_1}$. {Therefore, this new mixed SPT state can be constructed by placing mixed $\ZZ_2 \times \ZZ_N$ SPT states along the reflection axis $\lambda_1$.}

For $K=\SO(3)$, the mixed SPT states are classified by $\ZZ_2$. As above, they correspond to placing a (1+1)D $\SO(3)$-SPT state on $\lambda_1$. 

In principle, one could construct the effective actions for p4g  directly from the corresponding group cohomology cocycles, but we will leave this for future work. 
An interesting feature of these cocycles is that the area form\footnote{By `area form', we mean the generator of the $\ZZ$ factor in $\Hmc^2(\Gwp,\ZZ^{\text{or}})$. } of p4g restricts to \textit{twice} the area form of p4. This can be understood geometrically: the area form of p4g actually represents the fundamental domain which is half as large as the unit cell of p4 due to the glide symmetry.

\section{Group cohomology interpretation of invariants}\label{sec:GroupCoh}

The previous two sections studied two specific wallpaper groups at length. In this section, we consider the 17 wallpaper groups in general and discuss how the various invariants in this paper fit into the group cohomology classification of bosonic SPTs. We also argue why the invariants listed in the tables give a complete classification as per group cohomology (possibly up to specifying some filling invariants). All the cohomology groups we actually need to compute have integer coefficients and can be evaluated using the GAP program \cite{GAP4}. These groups have been tabulated in Tables~\ref{tab:GrCoho_untwisted},~\ref{tab:GrCoho_twisted}. A physicist's introduction to the group cohomology definitions and formulas that we use in this paper can be found in Refs.~\cite{Chen2013SPTGroupCohomology,barkeshli2021invertible}. 

Table~\ref{tab:pure_invts} lists the classification of Type A1, A2 and A3 invariants, which are pure crystalline invariants. Consider a high-symmetry point $\OO$. If $G_{\OO} = C_{\MO}$, the classification of $G_{\OO}$ SPTs is $\H^4(G_{\OO},\Z) \cong \Z_{\MO}$, and this is detected by the type A1 invariant. If $G_{\OO} = D_{2\MO}$, the classification of $G_{\OO}$ SPTs is $\H^4(G_{\OO},\Z^{\OR}) \cong\Z_{2}\times \Z_{(2,\MO)}$, and the two factors are detected by the type A1 and A2 invariants at $\OO$, respectively. Finally, suppose $\Gwp$ has a $\Z \rtimes \Z_2^{\bf r}$ subgroup generated by a translation and a reflection about an axis orthogonal to the translation direction. In this case $\H^4(\Z \rtimes \Z_2^{\bf r},\Z^{\OR}) \cong \Z_2$, and this is the class detected by the type-A3 invariant.     

For every wallpaper group $\Gwp$ we have shown that type A1, A2 and A3 invariants evaluated for different $\OO$ and $\LL$ measure enough information to fully determine the SPT class. An independent and complete set of invariants for each wallpaper group is shown in Table~\ref{tab:pure_invts}. To obtain this result we used the following steps: first we constructed candidate generators for $\Hmc^3(G,\U(1)^{\OR})$ by taking the cup product between generators of $\Hmc^1(G,\U(1))$ and $\Hmc^2(G,\ZZ^{\OR})$. To check that these classes are indeed generators, we evaluated the cohomology invariants associated to the type A invariants in Table~\ref{tab:pure_invts}, and found that the set of candidate generators are indeed independent and complete. 

Next, consider $G = \Gwp \times \U(1)$. Using the Kunneth formula, we have 
\begin{equation}
    \H^4(G, \Z^{\OR}) \cong \H^4(\Gwp,\Z^{\OR}) \times \H^2(\Gwp,
    \Z^{\OR}) .
\end{equation}
The second term classifies the different assignments of $\U(1)$ charge at points in the unit cell. This term always includes a $\Z$ factor, which gives the filling per fundamental domain (the usual filling $\nu$ for symmorphic wallpaper groups and $\nu/2$ for non-symmorphic wallpaper groups). The remaining factors can be determined using Type-B1 invariants. (Note that we can also recover partial information about $\nu$ from Type-B1 invariants.) 

For $G = \Gwp \times \Z_N$, the mixed invariants are classified by the following term in the Kunneth decomposition:
\begin{equation}
    \begin{split}
        &\H^2(\Gwp,
    \H^1(\Z_N,\U(1)^{\OR})) = \H^2(\Gwp,
    \Z_N^{\OR}) \\
    &= \H^2(\Gwp,\Z^{\OR}) \otimes \Z_N \\ &\quad \,\,\,\,\times \Tor[\H^3(\Gwp,\Z^{\OR}),\Z_N].
    \end{split}
\end{equation}
The second equality uses the Universal Coefficient Theorem. The first term is a $\Z_N$ analog of the mixed invariant that appeared above with $K = \U(1)$. This term always contains a $\Z_N$ factor which corresponds to the $\Z_N$ filling per unit cell. There are additional pieces, which are detected by the Type-C1 invariants. In all cases except for the group pm, these pieces can be interpreted as classifying the different assignments of $\Z_N$ charge at points in the unit cell. The special case of pm was addressed in Sec.~\ref{sec:InvariantsB4C4}.

The Tor term on the last line is nontrivial only in the presence of reflection symmetries. It classifies (1+1)D SPTs of $\Z_N \times \Z_2^{\rbf}$ symmetry that can be placed on the reflection axes in the unit cell. These correspond to the Type-C4 invariants in Table~\ref{tab:mixed_invts_1}.

Finally, for $G = \Gwp \times \SO(3)$, the mixed invariants are classified by
\begin{equation}
    \begin{split}
         &\H^1(\Gwp,
    H^2(\SO(3),\U(1)^{\OR})) = \H^1(\Gwp,
    \Z_2) \\
    &=    \Tor[\H^2(\Gwp,\Z^{\OR}),\Z_2] \\
    & \quad\,\,\,\,\times\H^1(\Gwp,\Z^{\OR}) \otimes \Z_2.
    \end{split}
\end{equation}
The Tor term on the last line captures two distinct types of invariants. One type consists of mixed invariants between $\SO(3)$ and $C_{\MO}$ subgroups of $\Gwp$; specifically, these invariants measure SPTs in which AKLT chains are placed on the boundaries of fundamental domains of the rotation symmetry. This subgroup can be detected by invariants of type D1. The second type classifies mixed invariants between $\SO(3)$ and reflection subgroups of $\Gwp$. Specifically, these invariants measure SPTs in which AKLT chains are placed on the reflection axes. The corresponding invariants are of type D2. In general, there is some freedom in attributing a given factor of this Tor classification to type D1 or D2, if both rotations and reflections are present. 

The final term containing the tensor product $\otimes$ is nontrivial only in the presence of reflections. In this case, the classification is always a single factor of $\Z_2$, and the state carrying this invariant is constructed by stacking AKLT chains along the reflection axis. This topological phase can also be detected by an invariant of type D2. 

Based on this discussion, we conclude that the new invariants we propose in this paper do capture the full mathematical classification of bosonic SPTs predicted by group cohomology.

\section{Discussion}\label{sec:Discussion}

We have considered bosonic SPT states in (2+1)D with symmetry $G = \Gwp \times K$ where $\Gwp$ is any of the 17 2d wallpaper groups and the internal symmetry $K = \U(1), \Z_N$ or $\SO(3)$. We have provided formulas to extract all the SPT invariants that depend on $\Gwp$, including pure crystalline invariants as well as mixed invariants between $\Gwp$ and $K$. It was known that all the new invariants involving reflection symmetry can be physically understood in terms of lower-dimensional states decorated on a reflection axis. That they can all be detected with expectation values of partial reflections, suitably combined with other operations in $G$, is the main new result of this paper.

Below we address some related issues. First we consider the case where the given state does have pure $K$ invariants. Next, we address to what extent the invariants presented here can be obtained from a single ground state wave function, and whether we might instead need a family of wave functions in some cases. Finally, we comment on future directions suggested by this work.  

\subsection{The case where pure \texorpdfstring{$K$}{K} SPT invariants are nontrivial}\label{sec:Disc-InternalSPTs}
So far in this paper, we have made the assumption that all topological invariants associated to $K$ symmetry alone are trivial. In this section we consider the case where the given state might have nontrivial $K$ SPT invariants, and to what extent our partial symmetry approaches can capture them.

When $K = \U(1)$ the only pure $K$ invariant is the bosonic Hall conductance $\sigma_H = 2C_{\U(1)} e_b^2/h$ where $e_b$ is the charge of an elementary boson. The Chern number $C_{\U(1)}$ is a $\Z$ invariant. When $K = \Z_N$ we have a $\Z_N$ analog of this, whose coefficient $C_{\Z_N}$ is defined mod $N$. Finally, for $K = \SO(3)$ there is a spin Hall conductance which is $\Z$ classified. Note that $C_K$ and $-C_K$ must necessarily be equal for any $\Gwp$ with orientation-reversing elements whenever $G = \Gwp \times K$ and $\Gwp$ is unitary. 
Interestingly, our approach turns out to give partial information about these invariants, as we now discuss.

The only way in which the pure $K$ invariants affect our previous results is to modify the formulas for the partial rotation invariants $\Theta_{\OO}^K$. (Formulas to isolate the pure $K$ invariants have been given in \cite{turzillo2025}.) Using CFT arguments, it was shown \cite{zhang2023complete} that the most general formula in this case is
\begin{equation}\label{eq:TildeThetaOK}
     \Theta_{\OO}^K-\Theta_{\OO} - C_{K} =\begin{cases}
        \so^{\U(1)}  \mod \MO & K = \U(1) \\
         \so^{\Z_N} \mod (\MO,N) & K = \Z_N \\
         \frac{\MO}{(\MO,2)}\so^{\SO(3)}  \mod \MO & K = \SO(3)
    \end{cases}  
\end{equation}
Thus for a given $\MO$-fold rotation about $\OO$, $\Theta_{\OO}^{\U(1)}$ contains information about $C_{\U(1)} \mod \MO$, while $\Theta_{\OO}^{\Z_N}$ contains information about $C_{\Z_N} \mod (\MO,N)$. In defining $\Theta_{\OO}^{\SO(3)}$, we consider an $\MO$-fold rotation instead of the 2-fold rotation used previously. Note that the pure $\SO(3)$ SPT invariant $C_{\SO(3)}$ manifests as a Hall conductance which must be an even multiple of the elementary BIQH conductance $C_{\U(1)} = 1$; therefore it only contributes when $\MO$ is even and $\MO \ge 4$. We can thus conclude that if $\M$ is the largest possible value of $\MO$ for the given lattice, partial rotations can at best determine $C_{\U(1)} \mod M$ and $C_{\Z_N} \mod (M,N)$. However, in order to extract even this information, we need to determine $\Theta_{\OO}$ and $\so^K$ separately for sufficiently many origins $\OO$. The precise extent to which this is possible depends on $\Gwp$.

\subsection{Requirement of a single ground state wave function}\label{sec:Disc-singleWfn}
In general, a partial symmetry invariant of the form $\bra{\psi} \hat{O} \ket{\psi}$ only requires a single ground state wave function $\ket{\psi}$, while an invariant involving an expression such as $\bra{\psi_{\gbf_1}} \hat{O} \ket{\psi_{\gbf_2}}$ requires a family of ground states with boundary conditions twisted by ${\gbf_1},{\gbf_2} \in K$. From our definitions we can see that all the invariants except for the `weak' invariants, and Types C4 and D4 can be obtained from a single wave function. For the weak invariants, we need to consider the ground state on different system sizes, while in the other two cases we need to consider twisted boundary conditions. Note that we can alternatively measure type C4 and D4 invariants using type C2 and D2 respectively, and these use a single wave function. Therefore the only case in which we still need multiple wave functions is the weak case. 

There are 5 wallpaper groups in which it is necessary to measure weak invariants: pm, cm, pmg, p3m1, and p31m. The common feature of these lattices is that they have a reflection axis which does not pass through any $C_2$ rotation center. Indeed, whenever such a point with $D_2$ symmetry exists, we can measure the same invariant differently: for example, we can replace the type A3 invariant with a type A2 invariant, which does not require multiple ground states. It is not clear to us whether there is an alternative way to measure the weak invariants using a single ground state wave function.

\subsection{Future directions}

A natural extension of this work is to consider invertible fermionic states with general wallpaper group symmetries. Real-space classifications of such states are already available, for various $K$ \cite{zhang2022realspace}. In this case, the full symmetry is given by a group extension of the `bosonic' symmetry group $G_b$ by fermion parity $\Z_2^f$. We expect that the main additional step is to properly define partial symmetry invariants for operations in $G_b$ that extend fermion parity in different ways. It would also be interesting to fully understand the relations between the different invariants studied here, since we have not derived all possible relations in this paper.

Another important direction is to numerically test the various predictions made here in ground states that are away from any ideal limit, and also to potentially simplify the formulas to make them more natural to implement in an experiment or a quantum simulation.

Finally, the invariants described in this paper are also relevant to symmetry-enriched topological (SET) phases, which harbor topologically degenerate ground states and anyonic excitations. Recent work has already shown the applicability of partial rotations in obtaining the symmetry fractionalization data and response invariants for fractional Chern insulators (FCI) \cite{kobayashi2024FCI}. We expect that the partial reflection invariants defined here will prove similarly useful in studying SET phases with reflection symmetry, such as quantum spin liquids.

\section{Acknowledgements}
We thank Dominic Else for sharing GAP code. This work was supported in part by NSF-BSF award DMR-2310312 at Stanford (VC). Research at Perimeter Institute is supported in part by the Government of Canada through the Department of Innovation, Science and Economic Development and by the Province of Ontario through the Ministry of Colleges and Universities. MB is supported by NSF DMR-2345644. 

\bibliography{References,refs}

\begin{thebibliography}{78}%
\makeatletter
\providecommand \@ifxundefined [1]{%
 \@ifx{#1\undefined}
}%
\providecommand \@ifnum [1]{%
 \ifnum #1\expandafter \@firstoftwo
 \else \expandafter \@secondoftwo
 \fi
}%
\providecommand \@ifx [1]{%
 \ifx #1\expandafter \@firstoftwo
 \else \expandafter \@secondoftwo
 \fi
}%
\providecommand \natexlab [1]{#1}%
\providecommand \enquote  [1]{``#1''}%
\providecommand \bibnamefont  [1]{#1}%
\providecommand \bibfnamefont [1]{#1}%
\providecommand \citenamefont [1]{#1}%
\providecommand \href@noop [0]{\@secondoftwo}%
\providecommand \href [0]{\begingroup \@sanitize@url \@href}%
\providecommand \@href[1]{\@@startlink{#1}\@@href}%
\providecommand \@@href[1]{\endgroup#1\@@endlink}%
\providecommand \@sanitize@url [0]{\catcode `\\12\catcode `\$12\catcode `\&12\catcode `\#12\catcode `\^12\catcode `\_12\catcode `\%12\relax}%
\providecommand \@@startlink[1]{}%
\providecommand \@@endlink[0]{}%
\providecommand \url  [0]{\begingroup\@sanitize@url \@url }%
\providecommand \@url [1]{\endgroup\@href {#1}{\urlprefix }}%
\providecommand \urlprefix  [0]{URL }%
\providecommand \Eprint [0]{\href }%
\providecommand \doibase [0]{https://doi.org/}%
\providecommand \selectlanguage [0]{\@gobble}%
\providecommand \bibinfo  [0]{\@secondoftwo}%
\providecommand \bibfield  [0]{\@secondoftwo}%
\providecommand \translation [1]{[#1]}%
\providecommand \BibitemOpen [0]{}%
\providecommand \bibitemStop [0]{}%
\providecommand \bibitemNoStop [0]{.\EOS\space}%
\providecommand \EOS [0]{\spacefactor3000\relax}%
\providecommand \BibitemShut  [1]{\csname bibitem#1\endcsname}%
\let\auto@bib@innerbib\@empty
\bibitem [{\citenamefont {Hasan}\ and\ \citenamefont {Kane}(2010)}]{hasan2010}%
  \BibitemOpen
  \bibfield  {author} {\bibinfo {author} {\bibfnamefont {M.~Z.}\ \bibnamefont {Hasan}}and\ \bibinfo {author} {\bibfnamefont {C.~L.}\ \bibnamefont {Kane}},\ }\bibfield  {title} {\bibinfo {title} {Colloquium: Topological insulators},\ }\href {https://doi.org/10.1103/RevModPhys.82.3045} {\bibfield  {journal} {\bibinfo  {journal} {Rev. Mod. Phys.}\ }\textbf {\bibinfo {volume} {82}},\ \bibinfo {pages} {3045} (\bibinfo {year} {2010})}\BibitemShut {NoStop}%
\bibitem [{\citenamefont {Fu}(2011)}]{fu2011topological}%
  \BibitemOpen
  \bibfield  {author} {\bibinfo {author} {\bibfnamefont {L.}~\bibnamefont {Fu}},\ }\bibfield  {title} {\bibinfo {title} {Topological crystalline insulators},\ }\href@noop {} {\bibfield  {journal} {\bibinfo  {journal} {Physical review letters}\ }\textbf {\bibinfo {volume} {106}},\ \bibinfo {pages} {106802} (\bibinfo {year} {2011})}\BibitemShut {NoStop}%
\bibitem [{\citenamefont {Benalcazar}\ \emph {et~al.}(2014)\citenamefont {Benalcazar}, \citenamefont {Teo},\ and\ \citenamefont {Hughes}}]{Benalcazar2014}%
  \BibitemOpen
  \bibfield  {author} {\bibinfo {author} {\bibfnamefont {W.~A.}\ \bibnamefont {Benalcazar}}, \bibinfo {author} {\bibfnamefont {J.~C.~Y.}\ \bibnamefont {Teo}}, and\ \bibinfo {author} {\bibfnamefont {T.~L.}\ \bibnamefont {Hughes}},\ }\bibfield  {title} {\bibinfo {title} {Classification of two-dimensional topological crystalline superconductors and majorana bound states at disclinations},\ }\href {https://doi.org/10.1103/PhysRevB.89.224503} {\bibfield  {journal} {\bibinfo  {journal} {Phys. Rev. B}\ }\textbf {\bibinfo {volume} {89}},\ \bibinfo {pages} {224503} (\bibinfo {year} {2014})}\BibitemShut {NoStop}%
\bibitem [{\citenamefont {Ando}\ and\ \citenamefont {Fu}(2015)}]{ando2015topological}%
  \BibitemOpen
  \bibfield  {author} {\bibinfo {author} {\bibfnamefont {Y.}~\bibnamefont {Ando}}and\ \bibinfo {author} {\bibfnamefont {L.}~\bibnamefont {Fu}},\ }\bibfield  {title} {\bibinfo {title} {Topological crystalline insulators and topological superconductors: From concepts to materials},\ }\href@noop {} {\bibfield  {journal} {\bibinfo  {journal} {Annu. Rev. Condens. Matter Phys.}\ }\textbf {\bibinfo {volume} {6}},\ \bibinfo {pages} {361} (\bibinfo {year} {2015})}\BibitemShut {NoStop}%
\bibitem [{\citenamefont {Watanabe}\ \emph {et~al.}(2015)\citenamefont {Watanabe}, \citenamefont {Po}, \citenamefont {Vishwanath},\ and\ \citenamefont {Zaletel}}]{watanabe2015filling}%
  \BibitemOpen
  \bibfield  {author} {\bibinfo {author} {\bibfnamefont {H.}~\bibnamefont {Watanabe}}, \bibinfo {author} {\bibfnamefont {H.~C.}\ \bibnamefont {Po}}, \bibinfo {author} {\bibfnamefont {A.}~\bibnamefont {Vishwanath}}, and\ \bibinfo {author} {\bibfnamefont {M.}~\bibnamefont {Zaletel}},\ }\bibfield  {title} {\bibinfo {title} {Filling constraints for spin-orbit coupled insulators in symmorphic and nonsymmorphic crystals},\ }\href@noop {} {\bibfield  {journal} {\bibinfo  {journal} {Proceedings of the National Academy of Sciences}\ }\textbf {\bibinfo {volume} {112}},\ \bibinfo {pages} {14551} (\bibinfo {year} {2015})}\BibitemShut {NoStop}%
\bibitem [{\citenamefont {Watanabe}\ \emph {et~al.}(2016)\citenamefont {Watanabe}, \citenamefont {Po}, \citenamefont {Zaletel},\ and\ \citenamefont {Vishwanath}}]{watanabe2016filling}%
  \BibitemOpen
  \bibfield  {author} {\bibinfo {author} {\bibfnamefont {H.}~\bibnamefont {Watanabe}}, \bibinfo {author} {\bibfnamefont {H.~C.}\ \bibnamefont {Po}}, \bibinfo {author} {\bibfnamefont {M.~P.}\ \bibnamefont {Zaletel}}, and\ \bibinfo {author} {\bibfnamefont {A.}~\bibnamefont {Vishwanath}},\ }\bibfield  {title} {\bibinfo {title} {Filling-enforced gaplessness in band structures of the 230 space groups},\ }\href@noop {} {\bibfield  {journal} {\bibinfo  {journal} {Physical review letters}\ }\textbf {\bibinfo {volume} {117}},\ \bibinfo {pages} {096404} (\bibinfo {year} {2016})}\BibitemShut {NoStop}%
\bibitem [{\citenamefont {Schindler}\ \emph {et~al.}(2018)\citenamefont {Schindler}, \citenamefont {Cook}, \citenamefont {Vergniory}, \citenamefont {Wang}, \citenamefont {Parkin}, \citenamefont {Bernevig},\ and\ \citenamefont {Neupert}}]{schindler2018higher}%
  \BibitemOpen
  \bibfield  {author} {\bibinfo {author} {\bibfnamefont {F.}~\bibnamefont {Schindler}}, \bibinfo {author} {\bibfnamefont {A.~M.}\ \bibnamefont {Cook}}, \bibinfo {author} {\bibfnamefont {M.~G.}\ \bibnamefont {Vergniory}}, \bibinfo {author} {\bibfnamefont {Z.}~\bibnamefont {Wang}}, \bibinfo {author} {\bibfnamefont {S.~S.}\ \bibnamefont {Parkin}}, \bibinfo {author} {\bibfnamefont {B.~A.}\ \bibnamefont {Bernevig}}, and\ \bibinfo {author} {\bibfnamefont {T.}~\bibnamefont {Neupert}},\ }\bibfield  {title} {\bibinfo {title} {Higher-order topological insulators},\ }\href@noop {} {\bibfield  {journal} {\bibinfo  {journal} {Science advances}\ }\textbf {\bibinfo {volume} {4}},\ \bibinfo {pages} {eaat0346} (\bibinfo {year} {2018})}\BibitemShut {NoStop}%
\bibitem [{\citenamefont {Khalaf}(2018)}]{khalaf2018higher}%
  \BibitemOpen
  \bibfield  {author} {\bibinfo {author} {\bibfnamefont {E.}~\bibnamefont {Khalaf}},\ }\bibfield  {title} {\bibinfo {title} {Higher-order topological insulators and superconductors protected by inversion symmetry},\ }\href@noop {} {\bibfield  {journal} {\bibinfo  {journal} {Physical Review B}\ }\textbf {\bibinfo {volume} {97}},\ \bibinfo {pages} {205136} (\bibinfo {year} {2018})}\BibitemShut {NoStop}%
\bibitem [{\citenamefont {Benalcazar}\ \emph {et~al.}(2019)\citenamefont {Benalcazar}, \citenamefont {Li},\ and\ \citenamefont {Hughes}}]{Benalcazar2019HOTI}%
  \BibitemOpen
  \bibfield  {author} {\bibinfo {author} {\bibfnamefont {W.~A.}\ \bibnamefont {Benalcazar}}, \bibinfo {author} {\bibfnamefont {T.}~\bibnamefont {Li}}, and\ \bibinfo {author} {\bibfnamefont {T.~L.}\ \bibnamefont {Hughes}},\ }\bibfield  {title} {\bibinfo {title} {Quantization of fractional corner charge in ${C}_{n}$-symmetric higher-order topological crystalline insulators},\ }\href {https://doi.org/10.1103/PhysRevB.99.245151} {\bibfield  {journal} {\bibinfo  {journal} {Phys. Rev. B}\ }\textbf {\bibinfo {volume} {99}},\ \bibinfo {pages} {245151} (\bibinfo {year} {2019})}\BibitemShut {NoStop}%
\bibitem [{\citenamefont {Bernevig}(2013)}]{bernevig2013topological}%
  \BibitemOpen
  \bibfield  {author} {\bibinfo {author} {\bibfnamefont {B.~A.}\ \bibnamefont {Bernevig}},\ }\bibfield  {title} {\bibinfo {title} {Topological insulators and topological superconductors},\ }in\ \href@noop {} {\emph {\bibinfo {booktitle} {Topological Insulators and Topological Superconductors}}}\ (\bibinfo  {publisher} {Princeton university press},\ \bibinfo {year} {2013})\BibitemShut {NoStop}%
\bibitem [{\citenamefont {Chiu}\ \emph {et~al.}(2016)\citenamefont {Chiu}, \citenamefont {Teo}, \citenamefont {Schnyder},\ and\ \citenamefont {Ryu}}]{Chiu2016review}%
  \BibitemOpen
  \bibfield  {author} {\bibinfo {author} {\bibfnamefont {C.-K.}\ \bibnamefont {Chiu}}, \bibinfo {author} {\bibfnamefont {J.~C.~Y.}\ \bibnamefont {Teo}}, \bibinfo {author} {\bibfnamefont {A.~P.}\ \bibnamefont {Schnyder}}, and\ \bibinfo {author} {\bibfnamefont {S.}~\bibnamefont {Ryu}},\ }\bibfield  {title} {\bibinfo {title} {Classification of topological quantum matter with symmetries},\ }\href {https://doi.org/10.1103/RevModPhys.88.035005} {\bibfield  {journal} {\bibinfo  {journal} {Rev. Mod. Phys.}\ }\textbf {\bibinfo {volume} {88}},\ \bibinfo {pages} {035005} (\bibinfo {year} {2016})}\BibitemShut {NoStop}%
\bibitem [{\citenamefont {Kruthoff}\ \emph {et~al.}(2017)\citenamefont {Kruthoff}, \citenamefont {de~Boer}, \citenamefont {van Wezel}, \citenamefont {Kane},\ and\ \citenamefont {Slager}}]{Kruthoff2017TIBandComb}%
  \BibitemOpen
  \bibfield  {author} {\bibinfo {author} {\bibfnamefont {J.}~\bibnamefont {Kruthoff}}, \bibinfo {author} {\bibfnamefont {J.}~\bibnamefont {de~Boer}}, \bibinfo {author} {\bibfnamefont {J.}~\bibnamefont {van Wezel}}, \bibinfo {author} {\bibfnamefont {C.~L.}\ \bibnamefont {Kane}}, and\ \bibinfo {author} {\bibfnamefont {R.-J.}\ \bibnamefont {Slager}},\ }\bibfield  {title} {\bibinfo {title} {Topological classification of crystalline insulators through band structure combinatorics},\ }\href {https://doi.org/10.1103/PhysRevX.7.041069} {\bibfield  {journal} {\bibinfo  {journal} {Phys. Rev. X}\ }\textbf {\bibinfo {volume} {7}},\ \bibinfo {pages} {041069} (\bibinfo {year} {2017})}\BibitemShut {NoStop}%
\bibitem [{\citenamefont {Bradlyn}\ \emph {et~al.}(2017)\citenamefont {Bradlyn}, \citenamefont {Elcoro}, \citenamefont {Cano}, \citenamefont {Vergniory}, \citenamefont {Wang}, \citenamefont {Felser}, \citenamefont {Aroyo},\ and\ \citenamefont {Bernevig}}]{Bradlyn2017tqc}%
  \BibitemOpen
  \bibfield  {author} {\bibinfo {author} {\bibfnamefont {B.}~\bibnamefont {Bradlyn}}, \bibinfo {author} {\bibfnamefont {L.}~\bibnamefont {Elcoro}}, \bibinfo {author} {\bibfnamefont {J.}~\bibnamefont {Cano}}, \bibinfo {author} {\bibfnamefont {M.~G.}\ \bibnamefont {Vergniory}}, \bibinfo {author} {\bibfnamefont {Z.}~\bibnamefont {Wang}}, \bibinfo {author} {\bibfnamefont {C.}~\bibnamefont {Felser}}, \bibinfo {author} {\bibfnamefont {M.~I.}\ \bibnamefont {Aroyo}}, and\ \bibinfo {author} {\bibfnamefont {B.~A.}\ \bibnamefont {Bernevig}},\ }\bibfield  {title} {\bibinfo {title} {Topological quantum chemistry},\ }\href@noop {} {\bibfield  {journal} {\bibinfo  {journal} {Nature}\ }\textbf {\bibinfo {volume} {547}},\ \bibinfo {pages} {298} (\bibinfo {year} {2017})}\BibitemShut {NoStop}%
\bibitem [{\citenamefont {Po}\ \emph {et~al.}(2017)\citenamefont {Po}, \citenamefont {Vishwanath},\ and\ \citenamefont {Watanabe}}]{Po2017symmind}%
  \BibitemOpen
  \bibfield  {author} {\bibinfo {author} {\bibfnamefont {H.~C.}\ \bibnamefont {Po}}, \bibinfo {author} {\bibfnamefont {A.}~\bibnamefont {Vishwanath}}, and\ \bibinfo {author} {\bibfnamefont {H.}~\bibnamefont {Watanabe}},\ }\bibfield  {title} {\bibinfo {title} {Symmetry-based indicators of band topology in the 230 space groups},\ }\bibfield  {journal} {\bibinfo  {journal} {Nature Communications}\ }\textbf {\bibinfo {volume} {8}},\ \href {https://doi.org/10.1038/s41467-017-00133-2} {10.1038/s41467-017-00133-2} (\bibinfo {year} {2017})\BibitemShut {NoStop}%
\bibitem [{\citenamefont {Watanabe}\ \emph {et~al.}(2018)\citenamefont {Watanabe}, \citenamefont {Po},\ and\ \citenamefont {Vishwanath}}]{watanabe2018structure}%
  \BibitemOpen
  \bibfield  {author} {\bibinfo {author} {\bibfnamefont {H.}~\bibnamefont {Watanabe}}, \bibinfo {author} {\bibfnamefont {H.~C.}\ \bibnamefont {Po}}, and\ \bibinfo {author} {\bibfnamefont {A.}~\bibnamefont {Vishwanath}},\ }\bibfield  {title} {\bibinfo {title} {Structure and topology of band structures in the 1651 magnetic space groups},\ }\href@noop {} {\bibfield  {journal} {\bibinfo  {journal} {Science advances}\ }\textbf {\bibinfo {volume} {4}},\ \bibinfo {pages} {eaat8685} (\bibinfo {year} {2018})}\BibitemShut {NoStop}%
\bibitem [{\citenamefont {Khalaf}\ \emph {et~al.}(2018)\citenamefont {Khalaf}, \citenamefont {Po}, \citenamefont {Vishwanath},\ and\ \citenamefont {Watanabe}}]{khalaf2018symmetry}%
  \BibitemOpen
  \bibfield  {author} {\bibinfo {author} {\bibfnamefont {E.}~\bibnamefont {Khalaf}}, \bibinfo {author} {\bibfnamefont {H.~C.}\ \bibnamefont {Po}}, \bibinfo {author} {\bibfnamefont {A.}~\bibnamefont {Vishwanath}}, and\ \bibinfo {author} {\bibfnamefont {H.}~\bibnamefont {Watanabe}},\ }\bibfield  {title} {\bibinfo {title} {Symmetry indicators and anomalous surface states of topological crystalline insulators},\ }\href@noop {} {\bibfield  {journal} {\bibinfo  {journal} {Physical Review X}\ }\textbf {\bibinfo {volume} {8}},\ \bibinfo {pages} {031070} (\bibinfo {year} {2018})}\BibitemShut {NoStop}%
\bibitem [{\citenamefont {Tang}\ \emph {et~al.}(2019)\citenamefont {Tang}, \citenamefont {Po}, \citenamefont {Vishwanath},\ and\ \citenamefont {Wan}}]{tang2019comprehensive}%
  \BibitemOpen
  \bibfield  {author} {\bibinfo {author} {\bibfnamefont {F.}~\bibnamefont {Tang}}, \bibinfo {author} {\bibfnamefont {H.~C.}\ \bibnamefont {Po}}, \bibinfo {author} {\bibfnamefont {A.}~\bibnamefont {Vishwanath}}, and\ \bibinfo {author} {\bibfnamefont {X.}~\bibnamefont {Wan}},\ }\bibfield  {title} {\bibinfo {title} {Comprehensive search for topological materials using symmetry indicators},\ }\href@noop {} {\bibfield  {journal} {\bibinfo  {journal} {Nature}\ }\textbf {\bibinfo {volume} {566}},\ \bibinfo {pages} {486} (\bibinfo {year} {2019})}\BibitemShut {NoStop}%
\bibitem [{\citenamefont {Cano}\ and\ \citenamefont {Bradlyn}(2021)}]{Cano_2021}%
  \BibitemOpen
  \bibfield  {author} {\bibinfo {author} {\bibfnamefont {J.}~\bibnamefont {Cano}}and\ \bibinfo {author} {\bibfnamefont {B.}~\bibnamefont {Bradlyn}},\ }\bibfield  {title} {\bibinfo {title} {Band representations and topological quantum chemistry},\ }\href {https://doi.org/10.1146/annurev-conmatphys-041720-124134} {\bibfield  {journal} {\bibinfo  {journal} {Annual Review of Condensed Matter Physics}\ }\textbf {\bibinfo {volume} {12}},\ \bibinfo {pages} {225} (\bibinfo {year} {2021})}\BibitemShut {NoStop}%
\bibitem [{\citenamefont {Elcoro}\ \emph {et~al.}(2021)\citenamefont {Elcoro}, \citenamefont {Wieder}, \citenamefont {Song}, \citenamefont {Xu}, \citenamefont {Bradlyn},\ and\ \citenamefont {Bernevig}}]{Elcoro2021tqc}%
  \BibitemOpen
  \bibfield  {author} {\bibinfo {author} {\bibfnamefont {L.}~\bibnamefont {Elcoro}}, \bibinfo {author} {\bibfnamefont {B.}~\bibnamefont {Wieder}}, \bibinfo {author} {\bibfnamefont {Z.}~\bibnamefont {Song}}, \bibinfo {author} {\bibfnamefont {Y.}~\bibnamefont {Xu}}, \bibinfo {author} {\bibfnamefont {B.}~\bibnamefont {Bradlyn}}, and\ \bibinfo {author} {\bibfnamefont {B.~A.}\ \bibnamefont {Bernevig}},\ }\bibfield  {title} {\bibinfo {title} {Magnetic topological quantum chemistry},\ }\bibfield  {journal} {\bibinfo  {journal} {Nature Communications}\ }\textbf {\bibinfo {volume} {12}},\ \href {https://doi.org/https://doi.org/10.1038/s41467-021-26241-8} {https://doi.org/10.1038/s41467-021-26241-8} (\bibinfo {year} {2021})\BibitemShut {NoStop}%
\bibitem [{\citenamefont {Herzog-Arbeitman}\ \emph {et~al.}(2022)\citenamefont {Herzog-Arbeitman}, \citenamefont {Bernevig},\ and\ \citenamefont {Song}}]{herzogarbeitman2022interacting}%
  \BibitemOpen
  \bibfield  {author} {\bibinfo {author} {\bibfnamefont {J.}~\bibnamefont {Herzog-Arbeitman}}, \bibinfo {author} {\bibfnamefont {B.~A.}\ \bibnamefont {Bernevig}}, and\ \bibinfo {author} {\bibfnamefont {Z.-D.}\ \bibnamefont {Song}},\ }\bibfield  {title} {\bibinfo {title} {Interacting topological quantum chemistry in 2d: Many-body real space invariants},\ }\href@noop {} {\bibfield  {journal} {\bibinfo  {journal} {arXiv preprint arXiv:2212.00030}\ } (\bibinfo {year} {2022})},\ \Eprint {https://arxiv.org/abs/2212.00030} {arXiv:2212.00030 [cond-mat.str-el]} \BibitemShut {NoStop}%
\bibitem [{\citenamefont {Essin}\ and\ \citenamefont {Hermele}(2014)}]{Essin2014spect}%
  \BibitemOpen
  \bibfield  {author} {\bibinfo {author} {\bibfnamefont {A.~M.}\ \bibnamefont {Essin}}and\ \bibinfo {author} {\bibfnamefont {M.}~\bibnamefont {Hermele}},\ }\bibfield  {title} {\bibinfo {title} {Spectroscopic signatures of crystal momentum fractionalization},\ }\href {https://doi.org/10.1103/PhysRevB.90.121102} {\bibfield  {journal} {\bibinfo  {journal} {Phys. Rev. B}\ }\textbf {\bibinfo {volume} {90}},\ \bibinfo {pages} {121102} (\bibinfo {year} {2014})}\BibitemShut {NoStop}%
\bibitem [{\citenamefont {Essin}\ and\ \citenamefont {Hermele}(2013)}]{Essin2013SF}%
  \BibitemOpen
  \bibfield  {author} {\bibinfo {author} {\bibfnamefont {A.~M.}\ \bibnamefont {Essin}}and\ \bibinfo {author} {\bibfnamefont {M.}~\bibnamefont {Hermele}},\ }\bibfield  {title} {\bibinfo {title} {Classifying fractionalization: Symmetry classification of gapped ${\mathbb{z}}_{2}$ spin liquids in two dimensions},\ }\href {https://doi.org/10.1103/PhysRevB.87.104406} {\bibfield  {journal} {\bibinfo  {journal} {Phys. Rev. B}\ }\textbf {\bibinfo {volume} {87}},\ \bibinfo {pages} {104406} (\bibinfo {year} {2013})}\BibitemShut {NoStop}%
\bibitem [{\citenamefont {Qi}\ and\ \citenamefont {Fu}(2015)}]{YangPRL2015}%
  \BibitemOpen
  \bibfield  {author} {\bibinfo {author} {\bibfnamefont {Y.}~\bibnamefont {Qi}}and\ \bibinfo {author} {\bibfnamefont {L.}~\bibnamefont {Fu}},\ }\bibfield  {title} {\bibinfo {title} {Anomalous crystal symmetry fractionalization on the surface of topological crystalline insulators},\ }\href {https://doi.org/10.1103/PhysRevLett.115.236801} {\bibfield  {journal} {\bibinfo  {journal} {Phys. Rev. Lett.}\ }\textbf {\bibinfo {volume} {115}},\ \bibinfo {pages} {236801} (\bibinfo {year} {2015})}\BibitemShut {NoStop}%
\bibitem [{\citenamefont {Hermele}\ and\ \citenamefont {Chen}(2016)}]{hermele2016}%
  \BibitemOpen
  \bibfield  {author} {\bibinfo {author} {\bibfnamefont {M.}~\bibnamefont {Hermele}}and\ \bibinfo {author} {\bibfnamefont {X.}~\bibnamefont {Chen}},\ }\bibfield  {title} {\bibinfo {title} {Flux-fusion anomaly test and bosonic topological crystalline insulators},\ }\href {https://doi.org/10.1103/PhysRevX.6.041006} {\bibfield  {journal} {\bibinfo  {journal} {Phys. Rev. X}\ }\textbf {\bibinfo {volume} {6}},\ \bibinfo {pages} {041006} (\bibinfo {year} {2016})}\BibitemShut {NoStop}%
\bibitem [{\citenamefont {Zaletel}\ \emph {et~al.}(2017)\citenamefont {Zaletel}, \citenamefont {Lu},\ and\ \citenamefont {Vishwanath}}]{zaletel2017}%
  \BibitemOpen
  \bibfield  {author} {\bibinfo {author} {\bibfnamefont {M.~P.}\ \bibnamefont {Zaletel}}, \bibinfo {author} {\bibfnamefont {Y.-M.}\ \bibnamefont {Lu}}, and\ \bibinfo {author} {\bibfnamefont {A.}~\bibnamefont {Vishwanath}},\ }\bibfield  {title} {\bibinfo {title} {Measuring space-group symmetry fractionalization in ${\mathbb{z}}_{2}$ spin liquids},\ }\href {https://doi.org/10.1103/PhysRevB.96.195164} {\bibfield  {journal} {\bibinfo  {journal} {Phys. Rev. B}\ }\textbf {\bibinfo {volume} {96}},\ \bibinfo {pages} {195164} (\bibinfo {year} {2017})}\BibitemShut {NoStop}%
\bibitem [{\citenamefont {Song}\ \emph {et~al.}(2017)\citenamefont {Song}, \citenamefont {Huang}, \citenamefont {Fu},\ and\ \citenamefont {Hermele}}]{song2017}%
  \BibitemOpen
  \bibfield  {author} {\bibinfo {author} {\bibfnamefont {H.}~\bibnamefont {Song}}, \bibinfo {author} {\bibfnamefont {S.-J.}\ \bibnamefont {Huang}}, \bibinfo {author} {\bibfnamefont {L.}~\bibnamefont {Fu}}, and\ \bibinfo {author} {\bibfnamefont {M.}~\bibnamefont {Hermele}},\ }\bibfield  {title} {\bibinfo {title} {Topological phases protected by point group symmetry},\ }\href {https://doi.org/10.1103/PhysRevX.7.011020} {\bibfield  {journal} {\bibinfo  {journal} {Phys. Rev. X}\ }\textbf {\bibinfo {volume} {7}},\ \bibinfo {pages} {011020} (\bibinfo {year} {2017})}\BibitemShut {NoStop}%
\bibitem [{\citenamefont {Huang}\ \emph {et~al.}(2017)\citenamefont {Huang}, \citenamefont {Song}, \citenamefont {Huang},\ and\ \citenamefont {Hermele}}]{Huang2017lowerDimCrysSPT}%
  \BibitemOpen
  \bibfield  {author} {\bibinfo {author} {\bibfnamefont {S.-J.}\ \bibnamefont {Huang}}, \bibinfo {author} {\bibfnamefont {H.}~\bibnamefont {Song}}, \bibinfo {author} {\bibfnamefont {Y.-P.}\ \bibnamefont {Huang}}, and\ \bibinfo {author} {\bibfnamefont {M.}~\bibnamefont {Hermele}},\ }\bibfield  {title} {\bibinfo {title} {Building crystalline topological phases from lower-dimensional states},\ }\href {https://doi.org/10.1103/PhysRevB.96.205106} {\bibfield  {journal} {\bibinfo  {journal} {Phys. Rev. B}\ }\textbf {\bibinfo {volume} {96}},\ \bibinfo {pages} {205106} (\bibinfo {year} {2017})}\BibitemShut {NoStop}%
\bibitem [{\citenamefont {Song}\ \emph {et~al.}(2020{\natexlab{a}})\citenamefont {Song}, \citenamefont {Fang},\ and\ \citenamefont {Qi}}]{Song2020RealSpaceRecupeTopCrystallineState}%
  \BibitemOpen
  \bibfield  {author} {\bibinfo {author} {\bibfnamefont {Z.}~\bibnamefont {Song}}, \bibinfo {author} {\bibfnamefont {C.}~\bibnamefont {Fang}}, and\ \bibinfo {author} {\bibfnamefont {Y.}~\bibnamefont {Qi}},\ }\bibfield  {title} {\bibinfo {title} {Real-space recipes for general topological crystalline states},\ }\href@noop {} {\bibfield  {journal} {\bibinfo  {journal} {Nature communications}\ }\textbf {\bibinfo {volume} {11}},\ \bibinfo {pages} {4197} (\bibinfo {year} {2020}{\natexlab{a}})}\BibitemShut {NoStop}%
\bibitem [{\citenamefont {Thorngren}\ and\ \citenamefont {Else}(2018)}]{Thorngren2018}%
  \BibitemOpen
  \bibfield  {author} {\bibinfo {author} {\bibfnamefont {R.}~\bibnamefont {Thorngren}}and\ \bibinfo {author} {\bibfnamefont {D.~V.}\ \bibnamefont {Else}},\ }\bibfield  {title} {\bibinfo {title} {Gauging spatial symmetries and the classification of topological crystalline phases},\ }\href {https://doi.org/10.1103/PhysRevX.8.011040} {\bibfield  {journal} {\bibinfo  {journal} {Phys. Rev. X}\ }\textbf {\bibinfo {volume} {8}},\ \bibinfo {pages} {011040} (\bibinfo {year} {2018})}\BibitemShut {NoStop}%
\bibitem [{\citenamefont {Manjunath}\ and\ \citenamefont {Barkeshli}(2021)}]{manjunath2021cgt}%
  \BibitemOpen
  \bibfield  {author} {\bibinfo {author} {\bibfnamefont {N.}~\bibnamefont {Manjunath}}and\ \bibinfo {author} {\bibfnamefont {M.}~\bibnamefont {Barkeshli}},\ }\bibfield  {title} {\bibinfo {title} {Crystalline gauge fields and quantized discrete geometric response for abelian topological phases with lattice symmetry},\ }\href {https://doi.org/10.1103/PhysRevResearch.3.013040} {\bibfield  {journal} {\bibinfo  {journal} {Phys. Rev. Research}\ }\textbf {\bibinfo {volume} {3}},\ \bibinfo {pages} {013040} (\bibinfo {year} {2021})}\BibitemShut {NoStop}%
\bibitem [{\citenamefont {Manjunath}\ and\ \citenamefont {Barkeshli}(2020)}]{Manjunath2020fqh}%
  \BibitemOpen
  \bibfield  {author} {\bibinfo {author} {\bibfnamefont {N.}~\bibnamefont {Manjunath}}and\ \bibinfo {author} {\bibfnamefont {M.}~\bibnamefont {Barkeshli}},\ }\bibfield  {title} {\bibinfo {title} {Classification of fractional quantum hall states with spatial symmetries},\ }\href@noop {} {\bibfield  {journal} {\bibinfo  {journal} {arXiv preprint arXiv:2012.11603}\ } (\bibinfo {year} {2020})}\BibitemShut {NoStop}%
\bibitem [{\citenamefont {van Miert}\ and\ \citenamefont {Ortix}(2018)}]{Miert2018dislocationCharge}%
  \BibitemOpen
  \bibfield  {author} {\bibinfo {author} {\bibfnamefont {G.}~\bibnamefont {van Miert}}and\ \bibinfo {author} {\bibfnamefont {C.}~\bibnamefont {Ortix}},\ }\bibfield  {title} {\bibinfo {title} {Dislocation charges reveal two-dimensional topological crystalline invariants},\ }\href {https://doi.org/10.1103/PhysRevB.97.201111} {\bibfield  {journal} {\bibinfo  {journal} {Phys. Rev. B}\ }\textbf {\bibinfo {volume} {97}},\ \bibinfo {pages} {201111} (\bibinfo {year} {2018})}\BibitemShut {NoStop}%
\bibitem [{\citenamefont {Li}\ \emph {et~al.}(2020)\citenamefont {Li}, \citenamefont {Zhu}, \citenamefont {Benalcazar},\ and\ \citenamefont {Hughes}}]{Li2020disc}%
  \BibitemOpen
  \bibfield  {author} {\bibinfo {author} {\bibfnamefont {T.}~\bibnamefont {Li}}, \bibinfo {author} {\bibfnamefont {P.}~\bibnamefont {Zhu}}, \bibinfo {author} {\bibfnamefont {W.~A.}\ \bibnamefont {Benalcazar}}, and\ \bibinfo {author} {\bibfnamefont {T.~L.}\ \bibnamefont {Hughes}},\ }\bibfield  {title} {\bibinfo {title} {Fractional disclination charge in two-dimensional ${C}_{n}$-symmetric topological crystalline insulators},\ }\href {https://doi.org/10.1103/PhysRevB.101.115115} {\bibfield  {journal} {\bibinfo  {journal} {Phys. Rev. B}\ }\textbf {\bibinfo {volume} {101}},\ \bibinfo {pages} {115115} (\bibinfo {year} {2020})}\BibitemShut {NoStop}%
\bibitem [{\citenamefont {Liu}\ \emph {et~al.}(2019)\citenamefont {Liu}, \citenamefont {Vishwanath},\ and\ \citenamefont {Khalaf}}]{Liu2019ShiftIns}%
  \BibitemOpen
  \bibfield  {author} {\bibinfo {author} {\bibfnamefont {S.}~\bibnamefont {Liu}}, \bibinfo {author} {\bibfnamefont {A.}~\bibnamefont {Vishwanath}}, and\ \bibinfo {author} {\bibfnamefont {E.}~\bibnamefont {Khalaf}},\ }\bibfield  {title} {\bibinfo {title} {Shift insulators: Rotation-protected two-dimensional topological crystalline insulators},\ }\href {https://doi.org/10.1103/PhysRevX.9.031003} {\bibfield  {journal} {\bibinfo  {journal} {Phys. Rev. X}\ }\textbf {\bibinfo {volume} {9}},\ \bibinfo {pages} {031003} (\bibinfo {year} {2019})}\BibitemShut {NoStop}%
\bibitem [{\citenamefont {Zhang}\ \emph {et~al.}(2022{\natexlab{a}})\citenamefont {Zhang}, \citenamefont {Manjunath}, \citenamefont {Nambiar},\ and\ \citenamefont {Barkeshli}}]{zhang2022fractional}%
  \BibitemOpen
  \bibfield  {author} {\bibinfo {author} {\bibfnamefont {Y.}~\bibnamefont {Zhang}}, \bibinfo {author} {\bibfnamefont {N.}~\bibnamefont {Manjunath}}, \bibinfo {author} {\bibfnamefont {G.}~\bibnamefont {Nambiar}}, and\ \bibinfo {author} {\bibfnamefont {M.}~\bibnamefont {Barkeshli}},\ }\href@noop {} {\bibinfo {title} {Fractional disclination charge and discrete shift in the hofstadter butterfly}} (\bibinfo {year} {2022}{\natexlab{a}}),\ \Eprint {https://arxiv.org/abs/2204.05320} {arXiv:2204.05320 [cond-mat.str-el]} \BibitemShut {NoStop}%
\bibitem [{\citenamefont {Zhang}\ \emph {et~al.}(2022{\natexlab{b}})\citenamefont {Zhang}, \citenamefont {Manjunath}, \citenamefont {Nambiar},\ and\ \citenamefont {Barkeshli}}]{zhang2022pol}%
  \BibitemOpen
  \bibfield  {author} {\bibinfo {author} {\bibfnamefont {Y.}~\bibnamefont {Zhang}}, \bibinfo {author} {\bibfnamefont {N.}~\bibnamefont {Manjunath}}, \bibinfo {author} {\bibfnamefont {G.}~\bibnamefont {Nambiar}}, and\ \bibinfo {author} {\bibfnamefont {M.}~\bibnamefont {Barkeshli}},\ }\href {https://doi.org/10.48550/ARXIV.2211.09127} {\bibinfo {title} {Quantized charge polarization as a many-body invariant in (2+1)d crystalline topological states and hofstadter butterflies}} (\bibinfo {year} {2022}{\natexlab{b}})\BibitemShut {NoStop}%
\bibitem [{\citenamefont {Zhang}\ \emph {et~al.}(2023)\citenamefont {Zhang}, \citenamefont {Manjunath}, \citenamefont {Kobayashi},\ and\ \citenamefont {Barkeshli}}]{zhang2023complete}%
  \BibitemOpen
  \bibfield  {author} {\bibinfo {author} {\bibfnamefont {Y.}~\bibnamefont {Zhang}}, \bibinfo {author} {\bibfnamefont {N.}~\bibnamefont {Manjunath}}, \bibinfo {author} {\bibfnamefont {R.}~\bibnamefont {Kobayashi}}, and\ \bibinfo {author} {\bibfnamefont {M.}~\bibnamefont {Barkeshli}},\ }\bibfield  {title} {\bibinfo {title} {Complete crystalline topological invariants from partial rotations in $(2+1)\mathrm{D}$ invertible fermionic states and hofstadter's butterfly},\ }\href {https://doi.org/10.1103/PhysRevLett.131.176501} {\bibfield  {journal} {\bibinfo  {journal} {Phys. Rev. Lett.}\ }\textbf {\bibinfo {volume} {131}},\ \bibinfo {pages} {176501} (\bibinfo {year} {2023})}\BibitemShut {NoStop}%
\bibitem [{\citenamefont {Manjunath}\ \emph {et~al.}(2023{\natexlab{a}})\citenamefont {Manjunath}, \citenamefont {Calvera},\ and\ \citenamefont {Barkeshli}}]{manjunath2022mzm}%
  \BibitemOpen
  \bibfield  {author} {\bibinfo {author} {\bibfnamefont {N.}~\bibnamefont {Manjunath}}, \bibinfo {author} {\bibfnamefont {V.}~\bibnamefont {Calvera}}, and\ \bibinfo {author} {\bibfnamefont {M.}~\bibnamefont {Barkeshli}},\ }\href@noop {} {\bibinfo {title} {Non-perturbative constraints from symmetry and chirality on majorana zero modes and defect quantum numbers in (2+1)d}} (\bibinfo {year} {2023}{\natexlab{a}}),\ \Eprint {https://arxiv.org/abs/2210.02452} {arXiv:2210.02452 [cond-mat.str-el]} \BibitemShut {NoStop}%
\bibitem [{\citenamefont {Manjunath}\ \emph {et~al.}(2023{\natexlab{b}})\citenamefont {Manjunath}, \citenamefont {Calvera},\ and\ \citenamefont {Barkeshli}}]{manjunath2023characterization}%
  \BibitemOpen
  \bibfield  {author} {\bibinfo {author} {\bibfnamefont {N.}~\bibnamefont {Manjunath}}, \bibinfo {author} {\bibfnamefont {V.}~\bibnamefont {Calvera}}, and\ \bibinfo {author} {\bibfnamefont {M.}~\bibnamefont {Barkeshli}},\ }\href@noop {} {\bibinfo {title} {Characterization and classification of interacting (2+1)d topological crystalline insulators with orientation-preserving wallpaper groups}} (\bibinfo {year} {2023}{\natexlab{b}}),\ \Eprint {https://arxiv.org/abs/2309.12389} {arXiv:2309.12389 [cond-mat.str-el]} \BibitemShut {NoStop}%
\bibitem [{\citenamefont {Chen}\ \emph {et~al.}(2013)\citenamefont {Chen}, \citenamefont {Gu}, \citenamefont {Liu},\ and\ \citenamefont {Wen}}]{Chen2013SPTGroupCohomology}%
  \BibitemOpen
  \bibfield  {author} {\bibinfo {author} {\bibfnamefont {X.}~\bibnamefont {Chen}}, \bibinfo {author} {\bibfnamefont {Z.-C.}\ \bibnamefont {Gu}}, \bibinfo {author} {\bibfnamefont {Z.-X.}\ \bibnamefont {Liu}}, and\ \bibinfo {author} {\bibfnamefont {X.-G.}\ \bibnamefont {Wen}},\ }\bibfield  {title} {\bibinfo {title} {Symmetry protected topological orders and the group cohomology of their symmetry group},\ }\href {https://doi.org/10.1103/PhysRevB.87.155114} {\bibfield  {journal} {\bibinfo  {journal} {Phys. Rev. B}\ }\textbf {\bibinfo {volume} {87}},\ \bibinfo {pages} {155114} (\bibinfo {year} {2013})}\BibitemShut {NoStop}%
\bibitem [{\citenamefont {Senthil}(2015)}]{Senthil2015SPT}%
  \BibitemOpen
  \bibfield  {author} {\bibinfo {author} {\bibfnamefont {T.}~\bibnamefont {Senthil}},\ }\bibfield  {title} {\bibinfo {title} {{Symmetry Protected Topological phases of Quantum Matter}},\ }\href {https://doi.org/10.1146/annurev-conmatphys-031214-014740} {\bibfield  {journal} {\bibinfo  {journal} {Ann. Rev. Condensed Matter Phys.}\ }\textbf {\bibinfo {volume} {6}},\ \bibinfo {pages} {299} (\bibinfo {year} {2015})},\ \Eprint {https://arxiv.org/abs/1405.4015} {arXiv:1405.4015 [cond-mat.str-el]} \BibitemShut {NoStop}%
\bibitem [{\citenamefont {Gu}\ and\ \citenamefont {Levin}(2014)}]{gu2014}%
  \BibitemOpen
  \bibfield  {author} {\bibinfo {author} {\bibfnamefont {Z.-C.}\ \bibnamefont {Gu}}and\ \bibinfo {author} {\bibfnamefont {M.}~\bibnamefont {Levin}},\ }\bibfield  {title} {\bibinfo {title} {Effect of interactions on two-dimensional fermionic symmetry-protected topological phases withz2symmetry},\ }\bibfield  {journal} {\bibinfo  {journal} {Physical Review B}\ }\textbf {\bibinfo {volume} {89}},\ \href {https://doi.org/10.1103/physrevb.89.201113} {10.1103/physrevb.89.201113} (\bibinfo {year} {2014})\BibitemShut {NoStop}%
\bibitem [{\citenamefont {Kapustin}(2014)}]{kapustin2014SPTbeyond}%
  \BibitemOpen
  \bibfield  {author} {\bibinfo {author} {\bibfnamefont {A.}~\bibnamefont {Kapustin}},\ }\bibfield  {title} {\bibinfo {title} {Symmetry protected topological phases, anomalies, and cobordisms: beyond group cohomology},\ }\href@noop {} {\bibfield  {journal} {\bibinfo  {journal} {arXiv preprint arXiv:1403.1467}\ } (\bibinfo {year} {2014})}\BibitemShut {NoStop}%
\bibitem [{\citenamefont {Kapustin}\ \emph {et~al.}(2015)\citenamefont {Kapustin}, \citenamefont {Thorngren}, \citenamefont {Turzillo},\ and\ \citenamefont {Wang}}]{kapustin2015fSPT}%
  \BibitemOpen
  \bibfield  {author} {\bibinfo {author} {\bibfnamefont {A.}~\bibnamefont {Kapustin}}, \bibinfo {author} {\bibfnamefont {R.}~\bibnamefont {Thorngren}}, \bibinfo {author} {\bibfnamefont {A.}~\bibnamefont {Turzillo}}, and\ \bibinfo {author} {\bibfnamefont {Z.}~\bibnamefont {Wang}},\ }\bibfield  {title} {\bibinfo {title} {Fermionic symmetry protected topological phases and cobordisms},\ }\href@noop {} {\bibfield  {journal} {\bibinfo  {journal} {Journal of High Energy Physics}\ }\textbf {\bibinfo {volume} {2015}},\ \bibinfo {pages} {1} (\bibinfo {year} {2015})}\BibitemShut {NoStop}%
\bibitem [{\citenamefont {Wang}\ and\ \citenamefont {Gu}(2020)}]{Wang2020fSPT}%
  \BibitemOpen
  \bibfield  {author} {\bibinfo {author} {\bibfnamefont {Q.-R.}\ \bibnamefont {Wang}}and\ \bibinfo {author} {\bibfnamefont {Z.-C.}\ \bibnamefont {Gu}},\ }\bibfield  {title} {\bibinfo {title} {Construction and classification of symmetry-protected topological phases in interacting fermion systems},\ }\href {https://doi.org/10.1103/PhysRevX.10.031055} {\bibfield  {journal} {\bibinfo  {journal} {Phys. Rev. X}\ }\textbf {\bibinfo {volume} {10}},\ \bibinfo {pages} {031055} (\bibinfo {year} {2020})}\BibitemShut {NoStop}%
\bibitem [{\citenamefont {Barkeshli}\ \emph {et~al.}(2021)\citenamefont {Barkeshli}, \citenamefont {Chen}, \citenamefont {Hsin},\ and\ \citenamefont {Manjunath}}]{barkeshli2021invertible}%
  \BibitemOpen
  \bibfield  {author} {\bibinfo {author} {\bibfnamefont {M.}~\bibnamefont {Barkeshli}}, \bibinfo {author} {\bibfnamefont {Y.-A.}\ \bibnamefont {Chen}}, \bibinfo {author} {\bibfnamefont {P.-S.}\ \bibnamefont {Hsin}}, and\ \bibinfo {author} {\bibfnamefont {N.}~\bibnamefont {Manjunath}},\ }\href@noop {} {\bibinfo {title} {Classification of (2+1)d invertible fermionic topological phases with symmetry}} (\bibinfo {year} {2021}),\ \Eprint {https://arxiv.org/abs/2109.11039} {arXiv:2109.11039 [cond-mat.str-el]} \BibitemShut {NoStop}%
\bibitem [{\citenamefont {Aasen}\ \emph {et~al.}(2021)\citenamefont {Aasen}, \citenamefont {Bonderson},\ and\ \citenamefont {Knapp}}]{aasen2021characterization}%
  \BibitemOpen
  \bibfield  {author} {\bibinfo {author} {\bibfnamefont {D.}~\bibnamefont {Aasen}}, \bibinfo {author} {\bibfnamefont {P.}~\bibnamefont {Bonderson}}, and\ \bibinfo {author} {\bibfnamefont {C.}~\bibnamefont {Knapp}},\ }\href@noop {} {\bibinfo {title} {Characterization and classification of fermionic symmetry enriched topological phases}} (\bibinfo {year} {2021}),\ \Eprint {https://arxiv.org/abs/2109.10911} {arXiv:2109.10911 [cond-mat.str-el]} \BibitemShut {NoStop}%
\bibitem [{\citenamefont {Barkeshli}\ \emph {et~al.}(2019)\citenamefont {Barkeshli}, \citenamefont {Bonderson}, \citenamefont {Cheng},\ and\ \citenamefont {Wang}}]{Barkeshli2019}%
  \BibitemOpen
  \bibfield  {author} {\bibinfo {author} {\bibfnamefont {M.}~\bibnamefont {Barkeshli}}, \bibinfo {author} {\bibfnamefont {P.}~\bibnamefont {Bonderson}}, \bibinfo {author} {\bibfnamefont {M.}~\bibnamefont {Cheng}}, and\ \bibinfo {author} {\bibfnamefont {Z.}~\bibnamefont {Wang}},\ }\bibfield  {title} {\bibinfo {title} {Symmetry fractionalization, defects, and gauging of topological phases},\ }\href {https://doi.org/10.1103/PhysRevB.100.115147} {\bibfield  {journal} {\bibinfo  {journal} {Phys. Rev. B}\ }\textbf {\bibinfo {volume} {100}},\ \bibinfo {pages} {115147} (\bibinfo {year} {2019})}\BibitemShut {NoStop}%
\bibitem [{\citenamefont {Freed}\ and\ \citenamefont {Hopkins}(2021)}]{freed2016}%
  \BibitemOpen
  \bibfield  {author} {\bibinfo {author} {\bibfnamefont {D.~S.}\ \bibnamefont {Freed}}and\ \bibinfo {author} {\bibfnamefont {M.~J.}\ \bibnamefont {Hopkins}},\ }\bibfield  {title} {\bibinfo {title} {{Reflection positivity and invertible topological phases}},\ }\href {https://doi.org/10.2140/gt.2021.25.1165} {\bibfield  {journal} {\bibinfo  {journal} {Geom. Topol.}\ }\textbf {\bibinfo {volume} {25}},\ \bibinfo {pages} {1165} (\bibinfo {year} {2021})},\ \Eprint {https://arxiv.org/abs/1604.06527} {arXiv:1604.06527 [hep-th]} \BibitemShut {NoStop}%
\bibitem [{\citenamefont {Bulmash}\ and\ \citenamefont {Barkeshli}(2022)}]{bulmashSymmFrac}%
  \BibitemOpen
  \bibfield  {author} {\bibinfo {author} {\bibfnamefont {D.}~\bibnamefont {Bulmash}}and\ \bibinfo {author} {\bibfnamefont {M.}~\bibnamefont {Barkeshli}},\ }\bibfield  {title} {\bibinfo {title} {Fermionic symmetry fractionalization in $(2+1)$ dimensions},\ }\href {https://doi.org/10.1103/PhysRevB.105.125114} {\bibfield  {journal} {\bibinfo  {journal} {Phys. Rev. B}\ }\textbf {\bibinfo {volume} {105}},\ \bibinfo {pages} {125114} (\bibinfo {year} {2022})}\BibitemShut {NoStop}%
\bibitem [{\citenamefont {Song}\ \emph {et~al.}(2020{\natexlab{b}})\citenamefont {Song}, \citenamefont {He}, \citenamefont {Vishwanath},\ and\ \citenamefont {Wang}}]{Song_2020monopole}%
  \BibitemOpen
  \bibfield  {author} {\bibinfo {author} {\bibfnamefont {X.-Y.}\ \bibnamefont {Song}}, \bibinfo {author} {\bibfnamefont {Y.-C.}\ \bibnamefont {He}}, \bibinfo {author} {\bibfnamefont {A.}~\bibnamefont {Vishwanath}}, and\ \bibinfo {author} {\bibfnamefont {C.}~\bibnamefont {Wang}},\ }\bibfield  {title} {\bibinfo {title} {From spinon band topology to the symmetry quantum numbers of monopoles in dirac spin liquids},\ }\bibfield  {journal} {\bibinfo  {journal} {Physical Review X}\ }\textbf {\bibinfo {volume} {10}},\ \href {https://doi.org/10.1103/physrevx.10.011033} {10.1103/physrevx.10.011033} (\bibinfo {year} {2020}{\natexlab{b}})\BibitemShut {NoStop}%
\bibitem [{\citenamefont {You}\ \emph {et~al.}(2020)\citenamefont {You}, \citenamefont {Bibo},\ and\ \citenamefont {Pollmann}}]{You2020hoe}%
  \BibitemOpen
  \bibfield  {author} {\bibinfo {author} {\bibfnamefont {Y.}~\bibnamefont {You}}, \bibinfo {author} {\bibfnamefont {J.}~\bibnamefont {Bibo}}, and\ \bibinfo {author} {\bibfnamefont {F.}~\bibnamefont {Pollmann}},\ }\bibfield  {title} {\bibinfo {title} {Higher-order entanglement and many-body invariants for higher-order topological phases},\ }\href {https://doi.org/10.1103/PhysRevResearch.2.033192} {\bibfield  {journal} {\bibinfo  {journal} {Phys. Rev. Research}\ }\textbf {\bibinfo {volume} {2}},\ \bibinfo {pages} {033192} (\bibinfo {year} {2020})}\BibitemShut {NoStop}%
\bibitem [{\citenamefont {Barkeshli}\ \emph {et~al.}(2025)\citenamefont {Barkeshli}, \citenamefont {Fechisin}, \citenamefont {Komargodski},\ and\ \citenamefont {Zhong}}]{barkeshli2025disclinations}%
  \BibitemOpen
  \bibfield  {author} {\bibinfo {author} {\bibfnamefont {M.}~\bibnamefont {Barkeshli}}, \bibinfo {author} {\bibfnamefont {C.}~\bibnamefont {Fechisin}}, \bibinfo {author} {\bibfnamefont {Z.}~\bibnamefont {Komargodski}}, and\ \bibinfo {author} {\bibfnamefont {S.}~\bibnamefont {Zhong}},\ }\bibfield  {title} {\bibinfo {title} {Disclinations, dislocations, and emanant flux at dirac criticality},\ }\href@noop {} {\bibfield  {journal} {\bibinfo  {journal} {arXiv preprint arXiv:2501.13866}\ } (\bibinfo {year} {2025})}\BibitemShut {NoStop}%
\bibitem [{\citenamefont {Barkeshli}\ \emph {et~al.}(2020)\citenamefont {Barkeshli}, \citenamefont {Bonderson}, \citenamefont {Cheng}, \citenamefont {Jian},\ and\ \citenamefont {Walker}}]{barkeshli2020reflection}%
  \BibitemOpen
  \bibfield  {author} {\bibinfo {author} {\bibfnamefont {M.}~\bibnamefont {Barkeshli}}, \bibinfo {author} {\bibfnamefont {P.}~\bibnamefont {Bonderson}}, \bibinfo {author} {\bibfnamefont {M.}~\bibnamefont {Cheng}}, \bibinfo {author} {\bibfnamefont {C.-M.}\ \bibnamefont {Jian}}, and\ \bibinfo {author} {\bibfnamefont {K.}~\bibnamefont {Walker}},\ }\bibfield  {title} {\bibinfo {title} {Reflection and time reversal symmetry enriched topological phases of matter: path integrals, non-orientable manifolds, and anomalies},\ }\href@noop {} {\bibfield  {journal} {\bibinfo  {journal} {Communications in Mathematical Physics}\ }\textbf {\bibinfo {volume} {374}},\ \bibinfo {pages} {1021} (\bibinfo {year} {2020})}\BibitemShut {NoStop}%
\bibitem [{\citenamefont {Barkeshli}\ and\ \citenamefont {Cheng}(2020)}]{Barkeshli2020Anomaly}%
  \BibitemOpen
  \bibfield  {author} {\bibinfo {author} {\bibfnamefont {M.}~\bibnamefont {Barkeshli}}and\ \bibinfo {author} {\bibfnamefont {M.}~\bibnamefont {Cheng}},\ }\bibfield  {title} {\bibinfo {title} {Relative anomalies in (2+1)d symmetry enriched topological states},\ }\bibfield  {journal} {\bibinfo  {journal} {SciPost Physics}\ }\textbf {\bibinfo {volume} {8}},\ \href {https://doi.org/10.21468/scipostphys.8.2.028} {10.21468/scipostphys.8.2.028} (\bibinfo {year} {2020})\BibitemShut {NoStop}%
\bibitem [{\citenamefont {Shiozaki}\ \emph {et~al.}(2017)\citenamefont {Shiozaki}, \citenamefont {Shapourian},\ and\ \citenamefont {Ryu}}]{shiozaki2017MBIfSPTs}%
  \BibitemOpen
  \bibfield  {author} {\bibinfo {author} {\bibfnamefont {K.}~\bibnamefont {Shiozaki}}, \bibinfo {author} {\bibfnamefont {H.}~\bibnamefont {Shapourian}}, and\ \bibinfo {author} {\bibfnamefont {S.}~\bibnamefont {Ryu}},\ }\bibfield  {title} {\bibinfo {title} {Many-body topological invariants in fermionic symmetry-protected topological phases: Cases of point group symmetries},\ }\href@noop {} {\bibfield  {journal} {\bibinfo  {journal} {Physical Review B}\ }\textbf {\bibinfo {volume} {95}},\ \bibinfo {pages} {205139} (\bibinfo {year} {2017})}\BibitemShut {NoStop}%
\bibitem [{\citenamefont {Kobayashi}\ \emph {et~al.}(2025)\citenamefont {Kobayashi}, \citenamefont {Zhang}, \citenamefont {Manjunath},\ and\ \citenamefont {Barkeshli}}]{kobayashi2025crystalline}%
  \BibitemOpen
  \bibfield  {author} {\bibinfo {author} {\bibfnamefont {R.}~\bibnamefont {Kobayashi}}, \bibinfo {author} {\bibfnamefont {Y.}~\bibnamefont {Zhang}}, \bibinfo {author} {\bibfnamefont {N.}~\bibnamefont {Manjunath}}, and\ \bibinfo {author} {\bibfnamefont {M.}~\bibnamefont {Barkeshli}},\ }\bibfield  {title} {\bibinfo {title} {Crystalline invariants of fractional chern insulators},\ }\href@noop {} {\bibfield  {journal} {\bibinfo  {journal} {Physical Review B}\ }\textbf {\bibinfo {volume} {112}},\ \bibinfo {pages} {035147} (\bibinfo {year} {2025})}\BibitemShut {NoStop}%
\bibitem [{\citenamefont {Levin}\ and\ \citenamefont {Wen}(2006)}]{levin2006}%
  \BibitemOpen
  \bibfield  {author} {\bibinfo {author} {\bibfnamefont {M.}~\bibnamefont {Levin}}and\ \bibinfo {author} {\bibfnamefont {X.-G.}\ \bibnamefont {Wen}},\ }\bibfield  {title} {\bibinfo {title} {Detecting topological order in a ground state wave function},\ }\href {https://doi.org/10.1103/PhysRevLett.96.110405} {\bibfield  {journal} {\bibinfo  {journal} {Phys. Rev. Lett.}\ }\textbf {\bibinfo {volume} {96}},\ \bibinfo {pages} {110405} (\bibinfo {year} {2006})}\BibitemShut {NoStop}%
\bibitem [{\citenamefont {Kitaev}\ and\ \citenamefont {Preskill}(2006)}]{kitaev2006topological}%
  \BibitemOpen
  \bibfield  {author} {\bibinfo {author} {\bibfnamefont {A.}~\bibnamefont {Kitaev}}and\ \bibinfo {author} {\bibfnamefont {J.}~\bibnamefont {Preskill}},\ }\bibfield  {title} {\bibinfo {title} {Topological entanglement entropy},\ }\href@noop {} {\bibfield  {journal} {\bibinfo  {journal} {Physical review letters}\ }\textbf {\bibinfo {volume} {96}},\ \bibinfo {pages} {110404} (\bibinfo {year} {2006})}\BibitemShut {NoStop}%
\bibitem [{\citenamefont {Dehghani}\ \emph {et~al.}(2021)\citenamefont {Dehghani}, \citenamefont {Cian}, \citenamefont {Hafezi},\ and\ \citenamefont {Barkeshli}}]{dehghani2021}%
  \BibitemOpen
  \bibfield  {author} {\bibinfo {author} {\bibfnamefont {H.}~\bibnamefont {Dehghani}}, \bibinfo {author} {\bibfnamefont {Z.-P.}\ \bibnamefont {Cian}}, \bibinfo {author} {\bibfnamefont {M.}~\bibnamefont {Hafezi}}, and\ \bibinfo {author} {\bibfnamefont {M.}~\bibnamefont {Barkeshli}},\ }\bibfield  {title} {\bibinfo {title} {Extraction of the many-body chern number from a single wave function},\ }\href {https://doi.org/10.1103/PhysRevB.103.075102} {\bibfield  {journal} {\bibinfo  {journal} {Phys. Rev. B}\ }\textbf {\bibinfo {volume} {103}},\ \bibinfo {pages} {075102} (\bibinfo {year} {2021})}\BibitemShut {NoStop}%
\bibitem [{\citenamefont {Cian}\ \emph {et~al.}(2021)\citenamefont {Cian}, \citenamefont {Dehghani}, \citenamefont {Elben}, \citenamefont {Vermersch}, \citenamefont {Zhu}, \citenamefont {Barkeshli}, \citenamefont {Zoller},\ and\ \citenamefont {Hafezi}}]{cian2021}%
  \BibitemOpen
  \bibfield  {author} {\bibinfo {author} {\bibfnamefont {Z.-P.}\ \bibnamefont {Cian}}, \bibinfo {author} {\bibfnamefont {H.}~\bibnamefont {Dehghani}}, \bibinfo {author} {\bibfnamefont {A.}~\bibnamefont {Elben}}, \bibinfo {author} {\bibfnamefont {B.}~\bibnamefont {Vermersch}}, \bibinfo {author} {\bibfnamefont {G.}~\bibnamefont {Zhu}}, \bibinfo {author} {\bibfnamefont {M.}~\bibnamefont {Barkeshli}}, \bibinfo {author} {\bibfnamefont {P.}~\bibnamefont {Zoller}}, and\ \bibinfo {author} {\bibfnamefont {M.}~\bibnamefont {Hafezi}},\ }\bibfield  {title} {\bibinfo {title} {Many-body chern number from statistical correlations of randomized measurements},\ }\href {https://doi.org/10.1103/PhysRevLett.126.050501} {\bibfield  {journal} {\bibinfo  {journal} {Phys. Rev. Lett.}\ }\textbf {\bibinfo {volume} {126}},\ \bibinfo {pages} {050501} (\bibinfo {year} {2021})}\BibitemShut {NoStop}%
\bibitem [{\citenamefont {Cian}\ \emph {et~al.}(2022)\citenamefont {Cian}, \citenamefont {Hafezi},\ and\ \citenamefont {Barkeshli}}]{cian2022extracting}%
  \BibitemOpen
  \bibfield  {author} {\bibinfo {author} {\bibfnamefont {Z.-P.}\ \bibnamefont {Cian}}, \bibinfo {author} {\bibfnamefont {M.}~\bibnamefont {Hafezi}}, and\ \bibinfo {author} {\bibfnamefont {M.}~\bibnamefont {Barkeshli}},\ }\href@noop {} {\bibinfo {title} {Extracting wilson loop operators and fractional statistics from a single bulk ground state}} (\bibinfo {year} {2022}),\ \Eprint {https://arxiv.org/abs/2209.14302} {arXiv:2209.14302 [cond-mat.str-el]} \BibitemShut {NoStop}%
\bibitem [{\citenamefont {Kim}\ \emph {et~al.}(2022)\citenamefont {Kim}, \citenamefont {Shi}, \citenamefont {Kato},\ and\ \citenamefont {Albert}}]{Kim2022ccc}%
  \BibitemOpen
  \bibfield  {author} {\bibinfo {author} {\bibfnamefont {I.~H.}\ \bibnamefont {Kim}}, \bibinfo {author} {\bibfnamefont {B.}~\bibnamefont {Shi}}, \bibinfo {author} {\bibfnamefont {K.}~\bibnamefont {Kato}}, and\ \bibinfo {author} {\bibfnamefont {V.~V.}\ \bibnamefont {Albert}},\ }\bibfield  {title} {\bibinfo {title} {Chiral central charge from a single bulk wave function},\ }\href {https://doi.org/10.1103/PhysRevLett.128.176402} {\bibfield  {journal} {\bibinfo  {journal} {Phys. Rev. Lett.}\ }\textbf {\bibinfo {volume} {128}},\ \bibinfo {pages} {176402} (\bibinfo {year} {2022})}\BibitemShut {NoStop}%
\bibitem [{\citenamefont {Fan}\ \emph {et~al.}(2023)\citenamefont {Fan}, \citenamefont {Sahay},\ and\ \citenamefont {Vishwanath}}]{fan2022}%
  \BibitemOpen
  \bibfield  {author} {\bibinfo {author} {\bibfnamefont {R.}~\bibnamefont {Fan}}, \bibinfo {author} {\bibfnamefont {R.}~\bibnamefont {Sahay}}, and\ \bibinfo {author} {\bibfnamefont {A.}~\bibnamefont {Vishwanath}},\ }\bibfield  {title} {\bibinfo {title} {Extracting the quantum hall conductance from a single bulk wave function},\ }\href {https://doi.org/10.1103/PhysRevLett.131.186301} {\bibfield  {journal} {\bibinfo  {journal} {Phys. Rev. Lett.}\ }\textbf {\bibinfo {volume} {131}},\ \bibinfo {pages} {186301} (\bibinfo {year} {2023})}\BibitemShut {NoStop}%
\bibitem [{\citenamefont {Turzillo}\ \emph {et~al.}(2025)\citenamefont {Turzillo}, \citenamefont {Manjunath},\ and\ \citenamefont {Garre-Rubio}}]{turzillo2025}%
  \BibitemOpen
  \bibfield  {author} {\bibinfo {author} {\bibfnamefont {A.}~\bibnamefont {Turzillo}}, \bibinfo {author} {\bibfnamefont {N.}~\bibnamefont {Manjunath}}, and\ \bibinfo {author} {\bibfnamefont {J.}~\bibnamefont {Garre-Rubio}},\ }\href {https://arxiv.org/abs/2503.04510} {\bibinfo {title} {Detection of 2d spt order with partial symmetries}} (\bibinfo {year} {2025}),\ \Eprint {https://arxiv.org/abs/2503.04510} {arXiv:2503.04510 [cond-mat.str-el]} \BibitemShut {NoStop}%
\bibitem [{\citenamefont {Li}\ and\ \citenamefont {Haldane}(2008)}]{Haldane2008entanglement}%
  \BibitemOpen
  \bibfield  {author} {\bibinfo {author} {\bibfnamefont {H.}~\bibnamefont {Li}}and\ \bibinfo {author} {\bibfnamefont {F.~D.~M.}\ \bibnamefont {Haldane}},\ }\bibfield  {title} {\bibinfo {title} {Entanglement spectrum as a generalization of entanglement entropy: Identification of topological order in non-abelian fractional quantum hall effect states},\ }\href {https://doi.org/10.1103/PhysRevLett.101.010504} {\bibfield  {journal} {\bibinfo  {journal} {Phys. Rev. Lett.}\ }\textbf {\bibinfo {volume} {101}},\ \bibinfo {pages} {010504} (\bibinfo {year} {2008})}\BibitemShut {NoStop}%
\bibitem [{\citenamefont {Qi}\ \emph {et~al.}(2012)\citenamefont {Qi}, \citenamefont {Katsura},\ and\ \citenamefont {Ludwig}}]{Qi2012entanglement}%
  \BibitemOpen
  \bibfield  {author} {\bibinfo {author} {\bibfnamefont {X.-L.}\ \bibnamefont {Qi}}, \bibinfo {author} {\bibfnamefont {H.}~\bibnamefont {Katsura}}, and\ \bibinfo {author} {\bibfnamefont {A.~W.~W.}\ \bibnamefont {Ludwig}},\ }\bibfield  {title} {\bibinfo {title} {General relationship between the entanglement spectrum and the edge state spectrum of topological quantum states},\ }\bibfield  {journal} {\bibinfo  {journal} {Physical Review Letters}\ }\textbf {\bibinfo {volume} {108}},\ \href {https://doi.org/10.1103/physrevlett.108.196402} {10.1103/physrevlett.108.196402} (\bibinfo {year} {2012}),\ \Eprint {https://arxiv.org/abs/1103.5437} {arXiv:1103.5437 [cond-mat.mes-hall]} \BibitemShut {NoStop}%
\bibitem [{\citenamefont {Pollmann}\ \emph {et~al.}(2010)\citenamefont {Pollmann}, \citenamefont {Turner}, \citenamefont {Berg},\ and\ \citenamefont {Oshikawa}}]{Pollmann2010EntanglementSpectrumSPT1d}%
  \BibitemOpen
  \bibfield  {author} {\bibinfo {author} {\bibfnamefont {F.}~\bibnamefont {Pollmann}}, \bibinfo {author} {\bibfnamefont {A.~M.}\ \bibnamefont {Turner}}, \bibinfo {author} {\bibfnamefont {E.}~\bibnamefont {Berg}}, and\ \bibinfo {author} {\bibfnamefont {M.}~\bibnamefont {Oshikawa}},\ }\bibfield  {title} {\bibinfo {title} {Entanglement spectrum of a topological phase in one dimension},\ }\href {https://doi.org/10.1103/PhysRevB.81.064439} {\bibfield  {journal} {\bibinfo  {journal} {Phys. Rev. B}\ }\textbf {\bibinfo {volume} {81}},\ \bibinfo {pages} {064439} (\bibinfo {year} {2010})}\BibitemShut {NoStop}%
\bibitem [{\citenamefont {Pollmann}\ \emph {et~al.}(2012)\citenamefont {Pollmann}, \citenamefont {Berg}, \citenamefont {Turner},\ and\ \citenamefont {Oshikawa}}]{pollmann2012symmetry}%
  \BibitemOpen
  \bibfield  {author} {\bibinfo {author} {\bibfnamefont {F.}~\bibnamefont {Pollmann}}, \bibinfo {author} {\bibfnamefont {E.}~\bibnamefont {Berg}}, \bibinfo {author} {\bibfnamefont {A.~M.}\ \bibnamefont {Turner}}, and\ \bibinfo {author} {\bibfnamefont {M.}~\bibnamefont {Oshikawa}},\ }\bibfield  {title} {\bibinfo {title} {Symmetry protection of topological phases in one-dimensional quantum spin systems},\ }\href@noop {} {\bibfield  {journal} {\bibinfo  {journal} {Physical Review B—Condensed Matter and Materials Physics}\ }\textbf {\bibinfo {volume} {85}},\ \bibinfo {pages} {075125} (\bibinfo {year} {2012})}\BibitemShut {NoStop}%
\bibitem [{\citenamefont {Cho}\ \emph {et~al.}(2015)\citenamefont {Cho}, \citenamefont {Hsieh}, \citenamefont {Morimoto},\ and\ \citenamefont {Ryu}}]{cho2015}%
  \BibitemOpen
  \bibfield  {author} {\bibinfo {author} {\bibfnamefont {G.~Y.}\ \bibnamefont {Cho}}, \bibinfo {author} {\bibfnamefont {C.-T.}\ \bibnamefont {Hsieh}}, \bibinfo {author} {\bibfnamefont {T.}~\bibnamefont {Morimoto}}, and\ \bibinfo {author} {\bibfnamefont {S.}~\bibnamefont {Ryu}},\ }\bibfield  {title} {\bibinfo {title} {Topological phases protected by reflection symmetry and cross-cap states},\ }\href {https://doi.org/10.1103/PhysRevB.91.195142} {\bibfield  {journal} {\bibinfo  {journal} {Phys. Rev. B}\ }\textbf {\bibinfo {volume} {91}},\ \bibinfo {pages} {195142} (\bibinfo {year} {2015})}\BibitemShut {NoStop}%
\bibitem [{\citenamefont {Lake}(2016)}]{Lake2016FoldingTrick}%
  \BibitemOpen
  \bibfield  {author} {\bibinfo {author} {\bibfnamefont {E.}~\bibnamefont {Lake}},\ }\bibfield  {title} {\bibinfo {title} {Anomalies and symmetry fractionalization in reflection-symmetric topological order},\ }\href {https://doi.org/10.1103/PhysRevB.94.205149} {\bibfield  {journal} {\bibinfo  {journal} {Phys. Rev. B}\ }\textbf {\bibinfo {volume} {94}},\ \bibinfo {pages} {205149} (\bibinfo {year} {2016})}\BibitemShut {NoStop}%
\bibitem [{\citenamefont {Tantivasadakarn}(2017)}]{NatInvariants}%
  \BibitemOpen
  \bibfield  {author} {\bibinfo {author} {\bibfnamefont {N.}~\bibnamefont {Tantivasadakarn}},\ }\bibfield  {title} {\bibinfo {title} {Dimensional reduction and topological invariants of symmetry-protected topological phases},\ }\href {https://doi.org/10.1103/PhysRevB.96.195101} {\bibfield  {journal} {\bibinfo  {journal} {Phys. Rev. B}\ }\textbf {\bibinfo {volume} {96}},\ \bibinfo {pages} {195101} (\bibinfo {year} {2017})}\BibitemShut {NoStop}%
\bibitem [{\citenamefont {Yoshida}\ \emph {et~al.}(2015)\citenamefont {Yoshida}, \citenamefont {Morimoto},\ and\ \citenamefont {Furusaki}}]{Yoshida2015BSPTsReflection}%
  \BibitemOpen
  \bibfield  {author} {\bibinfo {author} {\bibfnamefont {T.}~\bibnamefont {Yoshida}}, \bibinfo {author} {\bibfnamefont {T.}~\bibnamefont {Morimoto}}, and\ \bibinfo {author} {\bibfnamefont {A.}~\bibnamefont {Furusaki}},\ }\bibfield  {title} {\bibinfo {title} {Bosonic symmetry-protected topological phases with reflection symmetry},\ }\href {https://doi.org/10.1103/PhysRevB.92.245122} {\bibfield  {journal} {\bibinfo  {journal} {Phys. Rev. B}\ }\textbf {\bibinfo {volume} {92}},\ \bibinfo {pages} {245122} (\bibinfo {year} {2015})}\BibitemShut {NoStop}%
\bibitem [{GAP(2024)}]{GAP4}%
  \BibitemOpen
  \href@noop {} {\emph {\bibinfo {title} {{GAP -- Groups, Algorithms, and Programming, Version 4.14.0}}}},\ \bibinfo {organization} {{The GAP~Group}} (\bibinfo {year} {2024}),\ \bibinfo {note} {\url{https://www.gap-system.org}}\BibitemShut {NoStop}%
\bibitem [{\citenamefont {Zhang}\ \emph {et~al.}(2022{\natexlab{c}})\citenamefont {Zhang}, \citenamefont {Yang}, \citenamefont {Qi},\ and\ \citenamefont {Gu}}]{zhang2022realspace}%
  \BibitemOpen
  \bibfield  {author} {\bibinfo {author} {\bibfnamefont {J.-H.}\ \bibnamefont {Zhang}}, \bibinfo {author} {\bibfnamefont {S.}~\bibnamefont {Yang}}, \bibinfo {author} {\bibfnamefont {Y.}~\bibnamefont {Qi}}, and\ \bibinfo {author} {\bibfnamefont {Z.-C.}\ \bibnamefont {Gu}},\ }\bibfield  {title} {\bibinfo {title} {Real-space construction of crystalline topological superconductors and insulators in 2d interacting fermionic systems},\ }\href {https://doi.org/10.1103/PhysRevResearch.4.033081} {\bibfield  {journal} {\bibinfo  {journal} {Phys. Rev. Res.}\ }\textbf {\bibinfo {volume} {4}},\ \bibinfo {pages} {033081} (\bibinfo {year} {2022}{\natexlab{c}})}\BibitemShut {NoStop}%
\bibitem [{\citenamefont {Kobayashi}\ \emph {et~al.}(2024)\citenamefont {Kobayashi}, \citenamefont {Zhang}, \citenamefont {Manjunath},\ and\ \citenamefont {Barkeshli}}]{kobayashi2024FCI}%
  \BibitemOpen
  \bibfield  {author} {\bibinfo {author} {\bibfnamefont {R.}~\bibnamefont {Kobayashi}}, \bibinfo {author} {\bibfnamefont {Y.}~\bibnamefont {Zhang}}, \bibinfo {author} {\bibfnamefont {N.}~\bibnamefont {Manjunath}}, and\ \bibinfo {author} {\bibfnamefont {M.}~\bibnamefont {Barkeshli}},\ }\href {https://arxiv.org/abs/2405.17431} {\bibinfo {title} {Crystalline invariants of fractional chern insulators}} (\bibinfo {year} {2024}),\ \Eprint {https://arxiv.org/abs/2405.17431} {arXiv:2405.17431 [cond-mat.str-el]} \BibitemShut {NoStop}%
\bibitem [{\citenamefont {Handel}(1993)}]{CohomologyDihedral}%
  \BibitemOpen
  \bibfield  {author} {\bibinfo {author} {\bibfnamefont {D.}~\bibnamefont {Handel}},\ }\bibfield  {title} {\bibinfo {title} {On products in the cohomology of the dihedral groups},\ }\href@noop {} {\bibfield  {journal} {\bibinfo  {journal} {Tohoku Mathematical Journal, Second Series}\ }\textbf {\bibinfo {volume} {45}},\ \bibinfo {pages} {13} (\bibinfo {year} {1993})}\BibitemShut {NoStop}%
\bibitem [{\citenamefont {Benini}\ \emph {et~al.}(2019)\citenamefont {Benini}, \citenamefont {Córdova},\ and\ \citenamefont {Hsin}}]{Cordova:2GroupAnomaly}%
  \BibitemOpen
  \bibfield  {author} {\bibinfo {author} {\bibfnamefont {F.}~\bibnamefont {Benini}}, \bibinfo {author} {\bibfnamefont {C.}~\bibnamefont {Córdova}}, and\ \bibinfo {author} {\bibfnamefont {P.-S.}\ \bibnamefont {Hsin}},\ }\bibfield  {title} {\bibinfo {title} {On 2-group global symmetries and their anomalies},\ }\bibfield  {journal} {\bibinfo  {journal} {Journal of High Energy Physics}\ }\textbf {\bibinfo {volume} {2019}},\ \href {https://doi.org/10.1007/jhep03(2019)118} {10.1007/jhep03(2019)118} (\bibinfo {year} {2019})\BibitemShut {NoStop}%
\end{thebibliography}%
\appendix

\newpage

\section{Review of real space construction}
According to Ref.~\cite{Huang2017lowerDimCrysSPT}, crystalline SPTs when $K$ is trivial can be constructed by placing $G_{p}$ charges at $p$ for all maximal Wyckoff positions $p$ for the wallpaper group $\Gwp$.

For a non-trival $K$, Ref.~\cite{Song2020RealSpaceRecupeTopCrystallineState} tells us that the real space construction is done by decorating $n$-simplices of an equivariant decomposition of the unit cell. Each $n$-simplex is decorated by an SPT in $(n+1)$D protected by $K$ and any lattice symmetry that leaves the simplex invariant. There are additional relations and restrictions between the possible decorations. We can read off how many decorations for each $n$ are needed to construct states in each of the different phases from Table III of Ref.~\cite{Song2020RealSpaceRecupeTopCrystallineState}. Using this result, we see that 
\begin{enumerate}
    \item For $K=\U(1)$, we only need to decorate 0-simplices, which corresponds to placing $\U(1)$ charges at the Wyckoff positions. 
    \item For $K=\SO(3)$, we need to decorate 1-simplices. Only reflections can leave 1-simplices invariant and there are no $\ZZ_2\times\SO(3)$ mixed SPT phases in (1+1)D. Then, the decorations correspond to placing $\SO(3)$ SPT states on the 1-simplices. Simple examples of this include singlets made of two $S=1/2$ spins. 
    \item For $K=\ZZ_N$, we need to decorate both 0- and 1- simplices. The decorations on 0 simplices correspond to placing $\Z_N$ charges at different Wyckoff positions. The 1-simplex decorations come from $\Z_N \times \Z_2$ mixed (1+1)D SPTs, which are only present when there are reflections. 
\end{enumerate}

\section{Relation between \texorpdfstring{$\Sigma_{\OO,\LL}$ }{SigmaOL} and TQFT partition function}\label{app:ReflectionInvaraint}

The aim of this appendix is to relate the expectation value which defines $ \Sigma_{\OO,\LL} $ to a TQFT partition function that evaluates to the invariant $k_{2,\OO,\LL}$ in the topological action of $D_{2n}$ (Eq.~\ref{eq:topActionD2n}). This is summarized by the result
\begin{equation}
    \Sigma_{\OO,\LL} = k_{2,\OO,\LL}.
\end{equation}
In App.~\ref{app:Warmup}, we relate the partial double reflection to a particular TQFT partition function, which we evaluate in App.~\ref{app:evalZ}. App.~\ref{app:DressedPartialDoubleReflection} extends these calculations to the dressed partial double reflection $\Sigma_{\OO,\LL}(\kbs,\jbf)$.

\begin{figure*}[t!]
    \centering
    \includegraphics[width=0.85\linewidth]{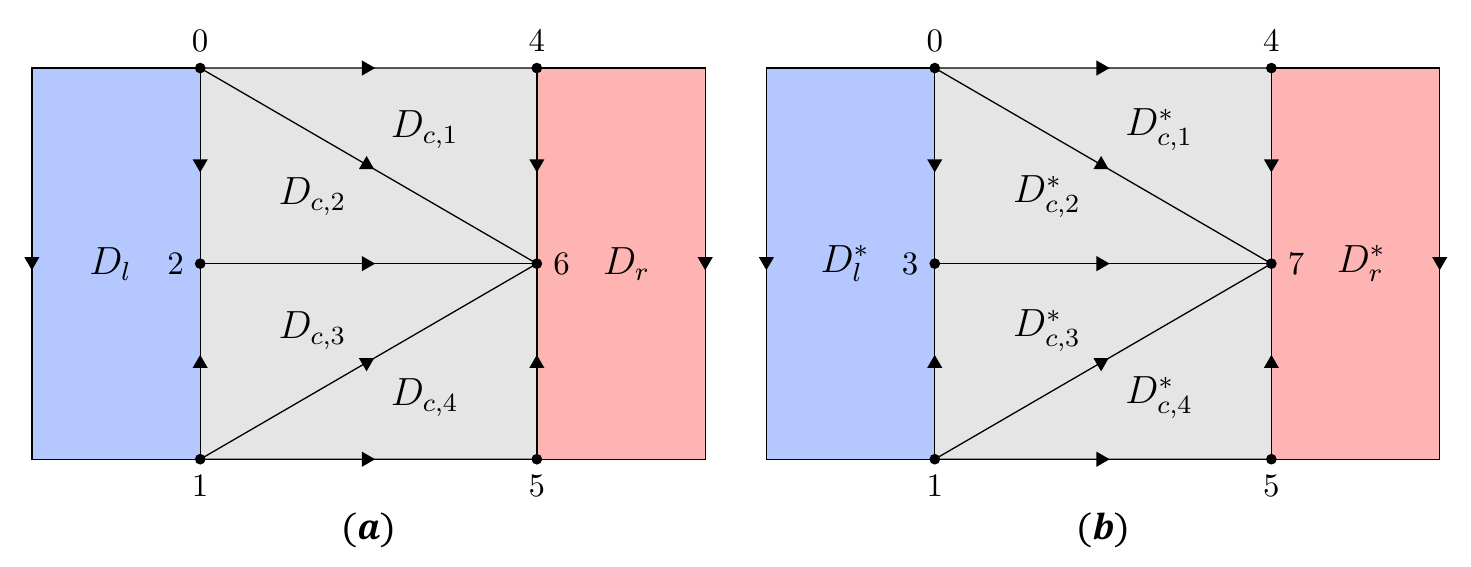}
    \caption{The state on the region $D = D_l \cup D_c \cup D_r$ is represented by a solid sphere, whose boundary is a sphere. The northern hemisphere is the `ket' part of the state, which we triangulate as shown in panel {(a)}. We have broken region $D_c$ into four smaller regions to accommodate the triangulation. Panel {(b)} shows the triangulation of the `bra' part. Note that the boundaries of the regions in panels {(a)} and {(b)} are the same because it corresponds to the equator of the sphere representing the state. }
    \label{fig:SurfaceTriangulation}
\end{figure*}

\subsection{Relation with TQFT}\label{app:Warmup}
We start with the ground state $\Psi$ on a disk and choose a region $D = D_l \cup D_c \cup D_r$ as in Fig.~\ref{fig:DefinitionDForPartialReflection}. Then, the reduced density matrix $\rho = \Tr_{\bar{D}}[\ket{\Psi}\bra{\Psi}]$, representing the state in the region $D$, is represented as a solid sphere. The northern and southern hemispheres correspond to the ket and bra parts of the state, respectively. We then triangulate the bra and ket parts of the surface as shown in panels (a) and (b) of Fig.~\ref{fig:SurfaceTriangulation}, respectively.

Next, we determine the surface gluing prescription by analyzing the partial symmetry operator: $R_{\LL}\eval_{D_c} \cdot (R_{\LL'})\eval_{D_l \cup D_r}$. Recall that $\LL$ and $\LL'$ denote horizontal and vertical reflection lines, respectively (see Fig.~\ref{fig:DefinitionDForPartialReflection}). 

The partial vertical reflection $R_{\LL'}\eval_{D_l\cup D_r}$ acts on the side regions, interchanging the red and blue regions, which dictates the gluing of $D_l \leftrightarrow D_r^*$ and $D_r\leftrightarrow D_l^*$. 

Similarly, the partial horizontal reflection $R_{\LL}\eval_{D_c}$ dictates the following gluing of regions $D_{c,1}\leftrightarrow D_{c,4}^*$, $D_{c,2}\leftrightarrow D_{c,3}^*$, $D_{c,3}\leftrightarrow D_{c,2}^*$, and $D_{c,4}\leftrightarrow D_{c,1}^*$. 

Having established an intuitive understanding of the gluing process based on the symmetries, we now provide a formal description using a specific triangulation of the sphere, which we deform to a cylinder for simplicity. 

\subsection{Evaluation of partition function}\label{app:evalZ}

The analysis below follows the detailed discussion of state-sum constructions in Ref.~\cite{barkeshli2020reflection}. 

We triangulate the cylinder representing the state on region $D$ as shown in Fig.~\ref{fig:triangulation_cylinder}. We used the same vertex labeling as in Fig.~\ref{fig:SurfaceTriangulation}. We denote the $n$-simplices by $\Delta^{n}_{i_1\dots i_n}$, where $i_1 < i_2 < \dots < i_n$ are the vertices. The plane containing the 2-simplex $\Delta_{045}^{2}$ separates the `bra' and `ket' regions of the state. For reference, the faces in Fig.~\ref{fig:SurfaceTriangulation} correspond to 2-simplices $\Delta^2_{ijk}$, where $i,j,k$ are the vertices on the boundary of said face. We now use the previously defined surface gluing prescription to identify 2-simplices. For example, we need to identify $\Delta_{012}^2 \sim \Delta_{457}^2$ and $\Delta^2_{026}\sim \Delta_{137}^2$. 

Since all $0$-simplices and $1$-simplices on the boundary of the 2-simplices also need to be identified, the resulting cellulation has: 2 0-simplices ($\Delta_0^0,\Delta_2^0$), 6 1-simplices ($\Delta^1_{01},\Delta^1_{02},\Delta^1_{03},\Delta^1_{04},\Delta^1_{05},\Delta^1_{26}$), 8 2-simplices, and 6 3-simplices. To simplify the expressions, we define the group element $\cbf \equiv \hbf_{\OO}^{\MO/2}$ and $\rbf \equiv \rbf_{\LL}$. Since the invariant is solely determined by the $D_2= \{\zero,\cbf,\rbf,\cbf\rbf\}$ subgroup of $D_{2n}$, we evaluate the state sum using $G=D_2$. 

Even though we identify the 0-simplices associated to vertices 
$0,1,4,5$ and $2,3,6,7$, the corresponding group elements are 
related by multiplication with the appropriate group element.  
Explicitly, we have
\begin{equation}
\begin{aligned}
    \gbf_1 &= \rbf \gbf_0, &\qquad 
    \gbf_4 &= \rbf\cbf \gbf_0, &\qquad 
    \gbf_5 &= \cbf \gbf_0, \\
    \gbf_3 &= \rbf \gbf_2, &\qquad 
    \gbf_6 &= \cbf \gbf_2, &\qquad 
    \gbf_7 &= \rbf\cbf \gbf_2.
\end{aligned}
\end{equation}

\begin{figure}[h!]
    \centering
    \includegraphics[width=0.5\linewidth]{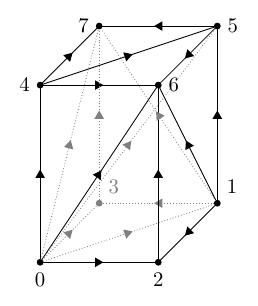}
    \caption{Triangulation of the square prism used to evaluate the partial double reflection as a state sum construction.
    The 2-simplices $\Delta^{2}_{012}$ and $\Delta^{2}_{456}$ are identified with the region $D_r$ and $D_{l}$, respectively, of the `bra' when evaluating the double partial reflection.   }
    \label{fig:triangulation_cylinder}
\end{figure}
To each 3-simplex and configuration of group elements $\{\gbf_j\}$, we assign a complex phase 
\begin{equation}
    \begin{split}
        \Zmc(\Delta_{j_1j_2j_3j_4}^3;\{\gbf_j\}) = \tilde{\nu}_{3}(\gbf_{j_1},\gbf_{j_2},\gbf_{j_3},\gbf_{j_4})^{s(\Delta_{j_1j_2j_3j_4)}^3)},
    \end{split}
\end{equation}
where $\tilde{\nu}_{3}$ is a homogeneous cocycle, satisfying
\begin{equation}
   \begin{split}
        \tilde{\nu}_{3}(\gbf_1,\gbf_2,\gbf_3,\gbf_4) = \tilde{\nu}_{3}(\gbf_0\gbf_1,\gbf_0\gbf_2,\gbf_0\gbf_3,\gbf_0\gbf_4)^{\psf(\gbf_0)}
   \end{split}
\end{equation}
for $\quad \gbf_0,\dots, \gbf_4 \in G$.  $\psf(\gbf_0) = 1$ if $\gbf_0$ is orientation preserving, and $\psf(\gbf_0)= *$ is complex conjugation if $\gbf_0$ is orientation reversing. There is a one to one map between homogeneous and inhomogeneous cocycles:
\begin{equation}
    \nu_{3}(\gbf_1,\gbf_2,\gbf_3)= \tilde{\nu}_{3}(\zero,\gbf_1,\gbf_1\gbf_2,\gbf_1\gbf_2\gbf_3),
\end{equation}
were $\nu_{3}$ is the inhomogeneous cocycle we have used in most of the paper. 

The partition function is
\begin{equation}\label{eq:partitionFx:def}
    \Zmc(\Mmc^3) =\frac{1}{\abs{G}^{N_v}} \sum_{\{\gbf_j\}} \prod_{\Delta^3\in \Imc_3} \Zmc(\Delta^3;\{\gbf_j\})^{s(\Delta^3)},
\end{equation}
where $\Imc_n$ is the set of $n$-simplices of the cellulation of the manifold $\Mmc^3$, and $N_v$ is the number of 0-simplices. $s(\Delta^3)$ is the orientation of the 3-simplex relative to a reference 3-simplex. Then $s(\Delta^3) =1 (*)$ if $\Delta^3$ has the same (opposite) orientation as the reference 3-simplex. As our cellulation is constructed from a triangulation, the orientation is such that any two 3-simplices sharing a 2-simplex must have opposite orientation. In our case, $\Delta_{0126}^3, \Delta_{0456}^3$ and $\Delta_{0157}$ have the same orientation, while $\Delta_{0156}^3, \Delta_{0137}^3$ and $\Delta_{0457}^3$ have orientation opposite to $\Delta_{0126}^3$. Therefore, for a given configuration $\{\gbf_j\}$ (which is specified by $\gbf_0$ and $\gbf_2$ in our case), the summand in Eq.~\ref{eq:partitionFx:def} is then 
\begin{widetext}
    \begin{equation}\label{eq:TopSum:Summand}
    \begin{split}
       e^{i S [\{\gbf_j\}]} &=\frac{\tilde{\nu}_{3}(\gbf_0,\gbf_1,\gbf_2,\gbf_6)\tilde{\nu}_{3}(\gbf_0,\gbf_4,\gbf_5,\gbf_6)\nu_{3}(\gbf_0,\gbf_1,\gbf_5,\gbf_7) }{\tilde{\nu}_{3}(\gbf_0,\gbf_1,\gbf_5,\gbf_6)\nu_{3}(\gbf_0,\gbf_4,\gbf_5,\gbf_7)\nu_{3}(\gbf_0,\gbf_1,\gbf_3,\gbf_7)}\\
        &= \left[\frac{\nu_{3}(\rbf, \rbf\gbf,\cbf)\nu_{3}(\rbf\cbf,\rbf,\gbf)\nu_{3}(\gbf,\rbf\cbf,\rbf\gbf)}{\nu_{3}(\rbf,\cbf \rbf,\gbf)\nu_{3}(\rbf\cbf,\rbf,\rbf\gbf)\nu_{3}(\rbf,\gbf,\cbf)}\right]^{\psf(\gbf_0)}
    \end{split}
\end{equation}
where we have defined $\gbf := \gbf_0^{-1}\gbf_2$. We need to simplify the above action. 

By combining the following expressions
\begin{equation}\label{eq:CohomologyRelationsUsedtoEval}
\begin{split}
    1=\bar{\dd}\nu_{3}(\rbf,\cbf,\rbf,\rbf\gbf)& = \frac{\nu_{3}(\rbf,\cbf\rbf,\rbf\gbf)\nu_{3}(\rbf,\cbf,\rbf)}{\nu_{3}(\cbf,\rbf,\rbf\gbf)\nu_{3}(\cbf\rbf,\rbf,\rbf\gbf)\nu_{3}(\rbf,\cbf,\gbf)};\\
1=     \bar{\dd}\nu_{3}(\rbf,\rbf,\cbf,\gbf) &=\frac{\nu_{3}(\rbf,\rbf\cbf,\gbf)\nu_{3}(\rbf,\rbf,\cbf)}{\nu_{3}(\rbf,\cbf,\gbf)\nu_{3}(\zero,\cbf,\gbf)\nu_{3}(\rbf,\rbf,\cbf\gbf)}\\ 
    1= \bar{\dd}\nu_{3}(\cbf,\rbf,\rbf,\gbf) & =  \frac{\nu_{3}(\rbf,\rbf,\gbf)\nu_{3}(\cbf,\zero,\gbf)\nu_{3}(\cbf,\rbf,\rbf)}{\nu_{3}(\cbf\rbf,\rbf,\gbf)\nu_{3}(\cbf,\rbf,\rbf \gbf)}\\ 
    1= \bar{\dd}\nu_{3}(\rbf,\rbf,\gbf,\cbf) &= \frac{\nu_{3}(\rbf,\rbf \gbf,\cbf)\nu_{3}(\rbf,\rbf,\gbf)}{\nu_{3}(\rbf,\gbf,\cbf)\nu_{3}(\zero,\gbf,\cbf)\nu_{3}(\rbf,\rbf,\gbf \cbf)}; \\
     \iota_{\cbf} \nu_{3}(\rbf,\rbf)&= \frac{\nu_{3}(\cbf,\rbf,\rbf)\nu_{3}(\rbf,\rbf,\cbf)}{\nu_{3}(\rbf,\cbf,\rbf)}; \\
     \iota_{\cbf}\nu_{3}(\zero,\gbf) &= \frac{\nu_{3}(\cbf,\zero,\gbf)\nu_{3}(\zero,\gbf,\cbf)}{\nu_{3}(\zero,\cbf,\gbf)},
\end{split}\end{equation}
\end{widetext}
we obtain 
\begin{equation}\label{eq:TopSum:Summand:Simplified}
    e^{i S[\{\gbf_j\}]} =  [\iota_{\cbf}\nu_{3}(\rbf,\rbf) \iota_{\cbf}\nu_{3}(\zero,\cbf )]^{\psf(\gbf_{0})}.
\end{equation}
Given that $\nu_{3}$ can be taken to be normalized (i.e., $\nu_{3} = 1$ whenever any of its arguments equals $\zero$), we identify
\begin{equation}
    \iota_{\cbf} \nu_{3} (\rbf,\rbf) = \Zmc(\RP^2_{\rbf}\times S^1_{\cbf})
\end{equation}
(see App.~\ref{app:CohInv:D2n}). Since this partition function can only take real values $\pm 1$, we may omit the factor $\psf(\gbf_0)$. Consequently,
\begin{equation}
\begin{split}
    \Zmc(\Mmc^3) 
    &= \frac{1}{4^2} \sum_{\gbf_0,\gbf_2 \in D_2}
\Zmc(\RP^2_{\rbf} \times S^1_{\cbf})\\
&= \Zmc(\RP^2_{\rbf} \times S^1_{\cbf}).
\end{split}
\end{equation}

\subsection{Dressed Partial double reflection}\label{app:DressedPartialDoubleReflection}

The dressed partial double reflection is obtained from the operator defined in Eq.~\ref{eq:dressedPartialDoubleReflection}. The difference compared to the bare partial double reflection is the additional action of $\jbf$, $\kbf$, and $\kbf^{-1}$ on the regions $D_c$, $D_l$, and $D_r$, respectively. The identification of 2-simplicices is the same as for the bare partial double reflection. However, the identification of group elements becomes 
\begin{equation}
\begin{aligned}
    \gbf_1 &= \rbf \jbf \gbf_0, &\qquad 
    \gbf_4 &= \rbf\cbf \kbf \gbf_0, &\qquad 
    \gbf_5 &= \cbf \jbf  \kbf \gbf_0, \\
    \gbf_3 &= \rbf\jbf \gbf_2, &\qquad 
    \gbf_6 &= \cbf\jbf\kbf \gbf_2, &\qquad 
    \gbf_7 &= \rbf\cbf  \kbf\gbf_2.
\end{aligned}
\end{equation}
The group elements now belong to $G=\ZZ_2^{\rbf}\times \ZZ_2^{\cbf} \times \ZZ_2^{\jbf} \times \ZZ_{2n}^{\kbf}$. The summand in the partition function now becomes
\begin{equation}
    e^{i S[\{\gbf_j\}]} = 
    \left[\frac{\nu_{3}(\tilde{\rbf}, \tilde{\rbf}\gbf, \sbf)\nu_{3}(\tilde{\rbf}\sbf,\tilde{\rbf},\gbf)\nu_{3}(\gbf,\tilde{\rbf}\sbf,\tilde{\rbf}\gbf)}{\nu_{3}(\tilde{\rbf},\sbf \tilde{\rbf},\gbf)\nu_{3}(\tilde{\rbf}\sbf,\tilde{\rbf},\tilde{\rbf}\gbf)\nu_{3}(\tilde{\rbf},\gbf,\sbf)}\right]^{\psf(\gbf_0)},
\end{equation}
where $\tilde{\rbf} = \rbf \jbf$ and $\sbf = \cbf \jbf \kbf$.
Using the same steps to get from Eq.~\ref{eq:TopSum:Summand} to Eq.~\ref{eq:TopSum:Summand:Simplified}, we obtain \footnote{To obtain the relations in Eq.~\ref{eq:CohomologyRelationsUsedtoEval} used that $\rbf^2 =\zero$ but never that $\gbf^2 = \cbf^2 = \zero$. Therefore, they remain valid after replacing $(\rbf,\cbf) \to (\tilde{\rbf}, \sbf)$.}
\begin{equation}\label{eq:TopSum:Summand:Simplified:Dressed}
    e^{i S[\{\gbf_j\}]} =  [\iota_{\sbf}\nu_{3}(\tilde{\rbf},\tilde{\rbf} )]^{\psf(\gbf_{0})},
\end{equation}
where we have assumed that $\nu_3$ is a normalized cocycle. From the K\"unneth formula, $\Hmc^3(\ZZ_{2n}\times\ZZ_2\times \ZZ_2 \times \ZZ_2^{\rbf}, \U(1)^{\OR}) = \ZZ_2^{J}$ for some integer $J$. This implies that the invariant $\iota_{\sbf}\nu_3(\tilde{\rbf},\tilde{\rbf})$ can only take the values $+1$ or $-1$. Then, we can remove $\psf(\gbf_0)$ in Eq.~\ref{eq:TopSum:Summand:Simplified} and obtain 
\begin{equation}
    \Zmc(\Mmc^3) = [\iota_{\sbf}\nu_{3}(\tilde{\rbf},\tilde{\rbf} )] ,
\end{equation}
after summing over $\gbf_0$ and $\gbf_2$. Using the same arguments to interpret the bare partial double reflection, we obtain
\begin{equation}
    \Zmc(\Mmc^3) = \Zmc( \RP^2_{\jbf\rbf} \times S^1_{\cbf \jbf \kbf}).
\end{equation}

\section{Details on group cohomology for p4m}\label{app:Coh:p4m}

This appendix contains the explicit cocycles and cohomology invariants we used in Sec.~\ref{sec:p4m}. App.~\ref{app:Cohomology:D2n} focuses on dihedral groups $D_{2n}$, providing the ground work for the full analysis of the wallpaper group \rm{p4m} in App.~\ref{app:Cohomology:p4m}.

\subsection{Group cohomology of \texorpdfstring{$D_{2n}$}{D2n}}\label{app:Cohomology:D2n}

Recall that $D_{2n}$ is the dihedral group with $4n$ elements, generated by a reflection $\rbf$ and $2n$-fold rotation $\hbf$. These elements satisfy the relations 
\begin{equation}
    \rbf^2 = \hbf^{2n} = (\rbf\hbf)^2 = \zero,
\end{equation}
where $\zero$ is the identity. Then, a general element in $D_{2n}$ can be written as $\hbf^{a}\rbf^{b}$ with $a\in \{0,1,\dots,2n-1\}$ and $b \in \{0,1\}$. See Ref.~\cite{CohomologyDihedral} for a calculation of the group cohomology groups we cite below. 

Consider the following cochains $f_0,f_1,f_2 \in C^1(D_{2n},\ZZ)$:
\begin{equation}
    \begin{split}
        f_{0}(\hbf^{a}\rbf^{b}) &= [a]_{2n}; \quad 
        f_{1}(\hbf^{a}\rbf^{b}) = [a]_{2}; \\
        f_{2}(\hbf^{a}\rbf^{b}) &= [b]_{2}.
    \end{split}
\end{equation}
Recall that $[s]_m$ is the residue of $s$ modulo $m$, with $[s]_m$ taking values in $\{0,1, \dots , m-1\}$. Representative cochains for the generators of $\Hmc^2(D_{2n},\ZZ) \cong \ZZ_2 \times \ZZ_2$ are $z_{1} := \frac{\dd f_1}{2}$ and $z_{2} := \frac{\dd f_2}{2}$. More explicitly
\begin{equation}
    \begin{split}
        z_1(\hbf^{a_1}\rbf^{b_1},\hbf^{a_2}\rbf^{b_2}) &= [a_1]_2[a_2]_2;\\
        z_2(\hbf^{a_1}\rbf^{b_1},\hbf^{a_2}\rbf^{b_2}) &= [b_1]_2[b_2]_2.\\
    \end{split}
\end{equation}
Similarly, $\Hmc^2(D_{2n},\ZZ^{\OR})\cong \ZZ_{2n}$ is generated by $\tilde{z}_1:= \frac{\bar{\dd} f_0}{2n}$, or more explicitly:
\begin{widetext}
\begin{equation}
    \tilde{z}_1(\hbf^{a_1}\rbf^{b_1},\hbf^{a_2}\rbf^{b_2}) = \frac{[a_1]_{2n} +(-1)^{b_1}[a_2]_{2n}- [a_1 + (-1)^{b_1}a_2]_{2n}}{2n}.
\end{equation}
According to Ref.~\cite{CohomologyDihedral}, $\Hmc^{4}(D_{2n},\ZZ^{\OR} )$ is isomorphic to $\ZZ_2\times\ZZ_2$ as an Abelian group, and its generators are precisely the cup product of the generators of $\Hmc^2(D_{2n},\ZZ)$ with the generator of $\Hmc^2(D_{2n},\ZZ^{\OR})$. In other words, the generators are $\tilde{Z}_j = z_j \cup \tilde{z}_1$ for $j=1,2$, or more explicitly:
\begin{equation}
    \tilde{Z}_j(\hbf^{a_1}\rbf^{b_1},\hbf^{a_2}\rbf^{b_2},\hbf^{a_3}\rbf^{b_3},\hbf^{a_4}\rbf^{b_4}) = z_j(\hbf^{a_1}\rbf^{b_1},\hbf^{a_2}\rbf^{b_2})(-1)^{b_1+b_2} \tilde{z}_1(\hbf^{a_3}\rbf^{b_3},\hbf^{a_4}\rbf^{b_4}).
\end{equation}
The factor $(-1)^{b_1}$ appears because we are dealing with cocycles with twisted coefficients\footnote{Recall that $\tilde{z}_1$ is a cocycle with values in $\ZZ^{\OR}$.}, e.g. see Appendix A.1 of Ref.~\cite{Cordova:2GroupAnomaly}.
\end{widetext}
Finally, by the Bockstein homorphism one has $\Hmc^3(D_{2n},\U(1)^{\OR}) \cong \Hmc^4(D_{2n},\Z^{\OR})$, and representative cocycles can be obtained by finding cochains $\Phi \in C^3(D_{2n},\RR)$ satisfying $\bar{\dd}\Phi = 2\pi \tilde{Z}$ for $\tilde{Z} \in Z^4(D_{2n}, \ZZ^{\OR})$. The respective cocycle $\zeta\in Z^3(D_{2n},\U(1)^{\OR})$ is $\zeta = e^{i\Phi}$. We can solve for $\Phi$ easily using: (1) $\bar{\dd}^2=0$, and (2) the Leibnitz rule. These conditions give $\bar{\dd}\Phi_j = \tilde{Z}_j$ is $\Phi_j = \pi f_j \cup \tilde{z}_1$ for $j=1,2$. 

Thus a general element $\Hmc^3(D_{2n},\U(1)^{\OR})$ is represented by a cocycle of the form 
\begin{equation}\label{eq:def:nuk1k2}
    \varphi_{3} = k_1 \Phi_1 + k_2\Phi_2; \quad k_1,k_2\in \ZZ_2,
\end{equation}
or, equivalently, 
\begin{equation}
     \nu_3 = \zeta_1^{k_1} \zeta_{2}^{k_2}.
\end{equation}

\subsubsection{Cohomology invariants}\label{app:CohInv:D2n}
To extract the invariants $k_1$ and $k_2$ from a generic cocycle $\nu_{3} \in Z^3(D_{2n},\U(1)^{\OR})$, consider the following quantities :
\begin{equation}\label{eq:InvariantsD2n}
    \begin{split}
        \Imc_1[\nu_{3}] &:= \prod_{j=0}^{2n-1} \nu_{3}(\hbf,\hbf^j,\hbf)\\
        \Imc_2[\nu_{3}] &:= \prod_{j=0}^{1} \frac{\nu_{3}(\hbf^{n},\rbf^j,\rbf)\nu_{3}(\rbf^j,\rbf,\hbf^{n})}{\nu_{3}(\rbf^j,\hbf^{n},\rbf)}.
    \end{split}
\end{equation}
It is a standard result that $\Imc_1$ is coboundary-invariant, e.g. see \cite{NatInvariants}. $\Imc_2$ is an invariant if $\Imc_2[\bar{\dd}\alpha]=1$ for any $\alpha \in C^2(D_{2n},\U(1))$. Note that 
\begin{equation}\label{eq:dbaralpha}
    \bar{\dd}{\alpha}(\gbf_1,\gbf_2,\gbf_3) = \frac{{^{\gbf_1}}\alpha(\gbf_2,\gbf_3) \alpha(\gbf_1,\gbf_2\gbf_3)}{\alpha(\gbf_1\gbf_2,\gbf_3)\alpha(\gbf_1,\gbf_2)}.
\end{equation}
Recall that ${^{\gbf_1}}\alpha(\gbf_2,\gbf_3)=[\alpha(\gbf_2,\gbf_3)]^{(-1)^{b}}$ for $\gbf=\hbf^a\rbf^b$. It is straightforward to show that $\Imc_2[\bar{\dd}\alpha]=1$ by combining Eqs.~\ref{eq:dbaralpha} and~\ref{eq:InvariantsD2n}.

We then evaluate the invariants for $\nu_3$ in Eq.~\ref{eq:def:nuk1k2}:
\begin{equation}
    \Imc_1[\nu_{3}] = (-1)^{k_1};\quad
    \Imc_2[\nu_{3}] = (-1)^{k_2}.
\end{equation}
We have thus shown that $\{\Imc_1,\Imc_2\}$ is an independent and complete set of invariants. 

A well-known interpretation of the invariant $\Imc_1$ is that it evaluates the partition function of the TQFT on a lens space with appropriate fluxes along its non-trivial cycle \cite{NatInvariants}. To interpret $\Imc_2$, we note that
\begin{equation}
    \begin{split}
        \Imc_2[\nu_{3}] &= \Jmc[\iota_{\hbf^n}\nu_{3}];\\
        (\iota_{\gbf_0}\nu_{3})(\gbf_1,\gbf_2) &:=  \frac{\nu_{3}(\gbf_0,\gbf_1,\gbf_2)\nu_{3}(\gbf_1,\gbf_2,\gbf_0)}{\nu_{3}(\gbf_1,\gbf_0,\gbf_2)};\\
        \Jmc[\mu] &:= \mu(\rbf,\rbf)\mu(\zero,\rbf),
    \end{split}
\end{equation}
where $\iota_{\gbf}$ denotes the Slant product and $\Jmc$ is an invariant defined for 2-cochains. On the non-orientable manifold $\RP^2$, the partition function evaluates precisely the invariant $\Jmc$ \cite{barkeshli2020reflection}. The action of the Slant product with a group element $\gbf$ can be interpreted as compactifying one direction into a circle with a $\gbf$-flux threated through it \cite{NatInvariants}. Therefore, $\Imc_2$ is the value of the partition function on the manifold $S^1 \times \RP^2$, with $\hbf^n$ flux through $S^1$.

Note that $D_{2n}$ has $2n$ distinct reflections, given by $\rbf_{a} = \hbf^{a} \rbf$ with $a=0,1,\dots,2n-1$. For each reflection, we can define an invariant $\Imc_{2,a}[\nu_{3}] = \prod_{j=0,1}(\iota_{\hbf^n}\nu_{3})(\rbf_{a}^j,\rbf_l)$. We can evaluate $\Imc_{2,a}[\nu_{3}] = (-1)^{a k_1+k_2}$. In this sense, the $k_2$ invariant depends on a choice of reflection axis. 

\subsubsection{Relation to topological action}

The gauge fields are obtained by pullbacks using the map $B$:
$\omega = \frac{2\pi}{2n}B^*f_0$ and $\sigma = B^*f_2 $. Then $\frac{\bar{\dd}\omega}{2\pi} = B^*\tilde{z}_1 \in \ZZ$ and $n\omega = \pi B^*f_1 \mod 2\pi$. Thus, 
\begin{equation}
    \begin{split}
        B^*\Phi_1 &= n\omega \frac{\bar{\dd}\omega}{2\pi} \mod 2\pi;\\
        \quad
    B^*\Phi_2 &= \pi \sigma \frac{\bar{\dd}\omega}{2\pi} \mod 2\pi.
    \end{split}
\end{equation}
We can recover Eq.~\ref{eq:topActionD2n} by setting $\Lmc = B^*\varphi_3$ and replacing $(k_1,k_2) \to (k_{1,\OO},k_{2,\OO,\LL})$ in Eq.~\ref{eq:def:nuk1k2}. We added the $\LL$ subscript to remark that this coefficient depends on the choice of reflection axis when writing the action. The subscript $\OO$ is redundant at the moment but will be important once we study p4m. 

Another way to see the reflection axis dependence of $k_2$ is to use a different reference reflection to write gauge fields: we factorizing the group elements as $\hbf^{a'}(\hbf \rbf)^{b'}$. This means that the new gauge fields (primed) are related to the old gauge fields (unprimed) by $\omega' \sim \omega + \pi \sigma/n $ and $\sigma' \sim \sigma$. Thus
\begin{equation}
    \begin{split}
        n \omega' \frac{\bar{\dd}\omega'}{2\pi} &= n\omega \frac{\bar{\dd}\omega}{2\pi} + \omega \frac{\bar{\dd}\omega}{2\pi}\\
        \sigma' \frac{\bar{\dd}\omega'}{2\pi}& = \sigma\frac{\bar{\dd}\omega}{2\pi}
    \end{split}
\end{equation}
where we have used that $\bar{\dd}\sigma=0$. Therefore, by changing the reference reflection axis used to define the gauge fields, the coefficients in the action change. If $\LL'$ is defined by $\rbf_{\LL'} = \hbf \rbf_{\LL}$, the above transformations for the gauge fields confirm that 
\begin{equation}
    k_{1,\OO} = k_{1,\OO};\quad k_{2,\OO,\LL'} = k_{2,\OO,\LL} + k_{1,\OO}.
\end{equation}

\subsubsection{More cohomology invariants }\label{app:D2n:moreInv}

For future convenience, we introduce invariants to identify elements in $\Hmc^1(D_{2n},\U(1))$ and $\Hmc^2(D_{2n},\ZZ^{\OR})$. 

Any element in $\Hmc^1(D_{2n},\U(1))$ can be represented by $\Xi_1 \in Z^{1}(D_{2n},\RR/\ZZ)$\footnote{$\RR/\ZZ$ is the additive group of real numbers modulo one which is isomorphic to $\U(1)$ by identifying $x \in \RR/\ZZ$ with $e^{2\pi i x} \in \U(1)$. } where 
\begin{equation}
    \Xi_1 = a_1 w_1 + a_2 w_2, \quad a_1,a_2 \in \ZZ_2;
\end{equation}
here $w_1 = f_1/2$ and $w_2 = f_2/2$ are representative cocycles for generators of $ \H^{1}(D_{2n},\RR/\ZZ)$. $[\Xi_1]$ is fully specified by the invariants $\{\Emc_{\hbf},\Emc_{\rbf}\}$, where
\begin{equation}
    \Emc_{\gbf}(\Xi) := \Xi(\gbf)
\end{equation}
for any normalized cocycle $\Xi \in Z^1(G, \RR/\ZZ)$ and $\gbf \in G$.

Similarly, any class in $\Hmc^2(D_{2n},\ZZ^{\OR})$ can be represented by a cocycle in $\Xi_2 \in Z^{2}(D_{2n},\ZZ^{\OR})$ of the form 
\begin{equation}
    \Xi_2 = \frac{b_1}{2n}\tilde{z}_1;\quad b_1\in \ZZ_{2n}.
\end{equation}
The value of $b_1$ is detected by the invariant $\Fmc_{\hbf}$, where we have defined 
\begin{equation}\label{eq:inv:b:D2n}
    \Fmc_{\gbf}[\Xi_2] := \sum_{j=0}^{d_{\gbf}-1} \Xi_2(\gbf^j,\gbf) \mod d_{\gbf},
\end{equation}
here $\Xi_2 \in Z^2(G,\ZZ)$ and $\gbf \in G$ is an orientation preserving element of order $d_{\gbf}$, {i.e.} $\gbf^{d_{\gbf}-1}\neq \zero$ but $\gbf^{d_{\gbf}}=\zero$. Note that even though $\Xi_2$ is a 2-cocycle on twisted coefficients, its restriction to $C_{2n}$ is a regular 2-cocycle because no element in $C_{2n}$ reverses orientation.

\subsection{Group cohomology of p4m}\label{app:Cohomology:p4m}

Recall that p4m is the symmetry group of the square lattice which is a semidirect product between $D_{4}$ and translations $\ZZ^2$, with generators $\xbf$ and $\ybf$. For the rest of this appendix, we take the origin of rotations as $\alpha$, and the preferred reflection axis as $\lambda_1$ (see Fig.~\ref{fig:p4m+p4g_UnitCell} for unit cell conventions). The group elements satisfy: $\hbf_{\alpha} \xbf = \ybf \hbf_{\alpha} $, 
$\hbf_{\alpha} \ybf = \xbf^{-1} \hbf_{\alpha} $,
$\hbf_{\alpha} \rbf_{\lambda_1} = \rbf_{\lambda_1} \hbf_{\alpha}^3 $,  $\rbf_{\lambda_1} \xbf =  \xbf \rbf_{\lambda_1}$, and 
$\rbf_{\lambda_1} \ybf =  \ybf^{-1} \rbf_{\lambda_1}$. A general element $\gbf\in \rm{p4m}$ can be written as 
\begin{equation}
    \gbf = \xbf^{c_{x}}\ybf^{c_y} \hbf_{\alpha}^{a} \rbf_{\lambda_1}^{b}
\end{equation}
with $c_x,c_y \in \ZZ$, $a\in\{0,1,2,3\}$ and $b \in \{0,1\}$.

\begin{table}[h]
    \centering
    \begin{tabular}{|c|c|c|c|c|}
    \hline
     \multicolumn{5}{|c|}{ Generators of site groups $G_{\PP}$} \\ \hline
        \diagbox[width=1.8cm,height=0.7cm,innerleftsep=0.5em,innerrightsep=0.85em]{}{$\PP$}& $\alpha$ & $\beta$ & $\gamma_1$ & $\gamma_2$ 
          \\\hline
         $\rbf_{\LL_{\PP}}$ & $\rbf_{\lambda_1}$ & $\ybf\rbf_{\lambda_1}$ & $\rbf_{\lambda_1}$ & $\ybf\rbf_{\lambda_1}$\\
         $\hbf_{\PP}$ & $\hbf_{\alpha}$ & $\xbf\hbf_{\alpha}$ & $\xbf\hbf_{\alpha}^{2}$ & $\ybf\hbf_{\alpha}^{2}$\\
         \hline
    \end{tabular}
    \caption{Explicit formulas for the rotation and reflection generators for the site groups in terms of the generators of \text{p4m} with origin $\OO=\alpha$. 
    }
    \label{tab:p4mDef}
\end{table}

To facilitate calculations involving the various site groups ($G_{\PP}$), we fix a preferred reflection line ($\LL_{\PP}$) for each site $\PP \in \{\alpha, \beta, \gamma_1, \gamma_2\}$. We choose $\LL_{\alpha} = \LL_{\gamma_1}=\lambda_1$ and $\LL_{\beta}=\LL_{\gamma_2}=\mu_1$. All $G_{\PP}$ gauge fields are defined with respect to their corresponding reflection axis, $\LL_{\PP}$. The expressions for generators of each $G_{\PP}$ are provided in Table $\text{\ref{tab:p4mDef}}$. Recall that we are computing restrictions to these subgroups to determine relations between different invariants, and to identify their origin or reflection axis dependence.

\subsubsection{Useful cochains}
Consider the following cochains $f_j,f_x,f_y \in C^1(\rm{p4m},\ZZ)$ with $j=0,1,2$:
\begin{equation}\label{eq:app:cocycles}
    \begin{split}
        f_0(\xbf^{c_{x}}\ybf^{c_y} \hbf_{\alpha}^{a} \rbf_{\lambda_1}^{b} ) &= [a]_{4}; \\
        f_1(\xbf^{c_{x}}\ybf^{c_y} \hbf_{\alpha}^{a} \rbf_{\lambda_1}^{b} ) &= [a]_{2};\\
        f_2(\xbf^{c_{x}}\ybf^{c_y} \hbf_{\alpha}^{a} \rbf_{\lambda_1}^{b} ) &= [b]_{2}; \\
        f_x(\xbf^{c_{x}}\ybf^{c_y} \hbf_{\alpha}^{a} \rbf_{\lambda_1}^{b} ) &= c_x;\\
        f_y(\xbf^{c_{x}}\ybf^{c_y} \hbf_{\alpha}^{a} \rbf_{\lambda_1}^{b} ) &= c_y.
    \end{split}
\end{equation}
\begin{widetext}
We construct the area-form cocycle $\Omega \in Z^2(\rm{p4m},\ZZ^{\OR })$ in App.~\ref{app:AreaFormp4m}. The result is
\begin{equation}\label{eq:p4mAreaForm}
            \Omega(\xbf^{c_{x1}}\ybf^{c_{y1}} \hbf_{\alpha}^{a_1} \rbf_{\lambda_1}^{b_1},\xbf^{c_{x2}}\ybf^{c_{y2}} \hbf_{\alpha}^{a_2} \rbf_{\lambda_1}^{b_2} ) := c_{x1} (\sin(\tfrac{\pi}{2} a_1)c_{x2} +\cos(\tfrac{\pi}{2} a_1) (-1)^{b_1}c_{y2}) + [b_1]_2 c_{x2}c_{y2}.
\end{equation}
\end{widetext}

\subsubsection{
\texorpdfstring{$\Hmc^1(\rm{p4m},\RR/\ZZ) $}{H1(p4m,R/Z)},
\texorpdfstring{$\Hmc^2(\rm{p4m},\ZZ) $}{H2(p4m,Z)},
\texorpdfstring{$\Hmc^2(\rm{p4m},\ZZ^{\OR}) $}{H2(p4m,Zor)}
}

A set of representative cocycles of the generators of $\Hmc^1(\rm{p4m},\RR/\ZZ) \cong \ZZ_2^3$ is $(w_1,w_2,w_3)$, where
\begin{equation}
    w_1 := \frac{f_1}{2};
    w_2 := \frac{f_2}{2};
    w_3 := \frac{f_x+f_y}{2} \in \RR.\footnote{We take $w_j$ to be chains valued in $\RR$ and use them as representatives of $\RR/\ZZ$ classes by reducing modulo $1$, which we leave implicit.}
\end{equation}
The coefficients of $\Xi = \sum_{j=1}^3 k_j w_j$ can be obtained as:  $k_1 = \Emc_{\hbf_{\alpha}}[\Xi]$,  $k_2 = \Emc_{\rbf_{\lambda_1}}[\Xi]$, and  $k_3 = \Emc_{\xbf}[\Xi]$. To find the restriction of $[w_j]$ to $G_{\PP}$, it is enough to evaluate $\Emc_{\hbf_{\PP}}$ and $\Emc_{\rbf_{\LL_{\PP}}}$ because these invariants fully determine classes in $\Hmc^1(G_{\PP},\RR/\ZZ) \cong \ZZ_2^2$. Let $\Res_{\PP}$ be shorthand for restriction from p4m to $G_{\PP}$. Let $\vec{w}:=(w_1,w_2,w_3)$, then
\begin{equation}\label{eq:Res:p4m:H1U1}
    \begin{split}
        \Res_{\alpha}\vec{w}&=(w_1,w_2,0)\\
        \Res_{\beta}\vec{w}&=(w_1,w_2,w_1+w_2)\\
        \Res_{\gamma_1}\vec{w}&=(0,w_2,w_1)\\
        \Res_{\gamma_2}\vec{w}&=(0,w_2,w_1+w_2)\\
    \end{split}
\end{equation}
where the above equalities hold as cohomology classes. Expressions for $w_j$ on the RHS are given in App.~\ref{app:D2n:moreInv}. The generators $\Hmc^2(\rm{p4m},\ZZ) \cong \Hmc^1(\rm{p4m},\RR/\ZZ) $ can be taken as $z_j = \dd w_j$, for $j=1,2,3$. Their restrictions can be obtained directly from Eq.~\ref{eq:Res:p4m:H1U1} because $\Res_{\PP}$ and $\dd$ commute.

A set of generators of $\Hmc^2(\rm{p4m},\ZZ^{\OR}) \cong \ZZ_4 \times \ZZ_2 \times \ZZ$ is $(\tilde{z}_1,\tilde{z}_2,\tilde{z}_3)$,\footnote{We always give the list of generators in the same order we present the cohomology group, e.g., in this case $\tilde{z}_1$ generates $\ZZ_4$, $\tilde{z}_2$ generates $\ZZ_2$, and $\tilde{z}_3$ generates $\ZZ$ } where
\begin{equation}
    \tilde{z}_1 : =\frac{\bar{\dd}f_0}{4};\quad  \tilde{z}_2 := \frac{\bar{\dd}(f_x+f_y)}{2};\quad \tilde{z}_3:= \Omega.
\end{equation}
The coefficients of $\Xi = \sum_{j=1}^3 k_j \tilde{z}_j$ can be obtained as:  $k_1 = \Fmc_{\hbf_{\alpha}}[\Xi]$,  $k_2 = \Fmc_{\xbf\hbf_{\alpha}^2}[\Xi]$, and  $k_3 = \Xi(\xbf,\ybf)-\Xi(\ybf,\xbf)$. 

The image of $\vec{\tilde{z}} =(\tilde{z}_{1},\tilde{z}_{2},\tilde{z}_3)$ under $\Res_{\PP}$ is fully determined by $\Fmc_{\hbf_{\PP}}$, which we evaluated. From here, we obtain
\begin{equation}\label{eq:Res:p4m:H2Z}
    \begin{split}
        \Res_{\alpha}\vec{\tilde{z}}&=(\tilde{z}_1,0,0);\\
        \Res_{\beta}\vec{\tilde{z}}&=(\tilde{z}_1,2\tilde{z}_1,\tilde{z}_1);\\
        \Res_{\gamma_1}\vec{\tilde{z}}&=(\tilde{z}_1,\tilde{z}_1,0);\\
        \Res_{\gamma_2}\vec{\tilde{z}}&=(\tilde{z}_1,\tilde{z}_1,0).
    \end{split}
\end{equation}

\subsubsection{\texorpdfstring{$\Hmc^3(\rm{p4m},\U(1)^{\OR}) $}{H3(p4m,U1OR)}}

According to the GAP computation, we find that $\Hmc^4(\mathrm{p4m},\mathbb{Z}^{\OR}) \cong \mathbb{Z}_2^{6}$. Consequently, $\Hmc^4(\mathrm{p4m},\U(1)^{\OR}) \cong \mathbb{Z}_2^{6}$ as well. We propose a basis $\{\zeta_1, \dots, \zeta_6\}$ for the latter group by defining cochains $\Phi_j \in C^3(\mathrm{p4m},\mathbb{R})$ for $j = 1, \dots, 6$, such that $\zeta_j = e^{i \Phi_j}$. The corresponding phases $\Phi_j$ are given by
\begin{equation}
\begin{split}\label{eq:Res:p4m:H3U1}
        {\Phi}_1 &= 2\pi w_1 \cup \tilde{z}_1; \quad
    {\Phi}_2 = 2\pi w_2 \cup \tilde{z}_1;\\
    {\Phi}_3 &= 2\pi w_3 \cup \tilde{z}_1; \quad
    {\Phi}_4 = 2\pi w_2 \cup \tilde{z}_2;\\
    {\Phi}_5 &= 2\pi w_1 \cup \tilde{z}_3; \quad
    {\Phi}_6 = 2\pi w_2 \cup \tilde{z}_3.
\end{split}
\end{equation}
The $\zeta_j$ are clearly co-cycles by construction. To show that they form a basis, consider the invariants in Eq.~\ref{eq:InvariantsD2n} evaluated for different $\OO$ and $\LL$:
\begin{equation}
    \begin{split}
        \Imc_{1,\OO}[\nu_{3}] &:= \prod_{j=0}^{\MO-1} \nu_{3}(\hbf_{\OO},\hbf_{\OO}^{j},\hbf_{\OO});\\
        \Imc_{2,\OO,\LL}[\nu_{3}] &:= \prod_{j=0}^{1} \iota_{\hbf^{\MO/2}_{\OO}}\nu_{3}(\rbf_{\LL}^j,\rbf_{\LL}),
    \end{split}
\end{equation}
where $\OO$ lies on $\LL$. Let
\begin{equation}\label{eq:generalElement:nu3}
    \nu_3 = \prod_{j=1}^{j}\zeta_j^{k_j};\quad k_j \in \{0,1\},
\end{equation}
be a generic element in $\Hmc^3(\rm{p4m},\U(1)^{\OR})$. Then,
\begin{equation}
    \begin{split}
        \Imc_{1,\alpha}[\nu_3] &= (-1)^{k_1} \\
        \Imc_{2,\alpha,\lambda_1}[\nu_3] &= (-1)^{k_2}\\
        \Imc_{1,\beta}[\nu_3] &= (-1)^{k_1+k_3+k_5}\\\Imc_{2,\beta,\mu_1}[\nu_3] &= (-1)^{k_2+k_3+k_6}\\
        \Imc_{1,\gamma_1}[\nu_3] &= (-1)^{k_3}\\
        \Imc_{2,\gamma_1,\lambda_1}[\nu_3] &= (-1)^{k_2+k_4}
    \end{split}
\end{equation}
It is clear that the above invariants are independent. Thus the proposed set indeed gives a basis. 

 \subsubsection{Topological action}

 The gauge fields used in the main text are
\begin{equation}
    \begin{split}
        \omega &= \frac{2\pi}{4} B^* f_0 , \\
    \sigma &= B^* f_2 ,\\
    \vec{R} &= 2\pi [B^* f_x , B^* f_y]^\top.
    \end{split}
\end{equation}
Therefore, the actions in Eq.~\ref{eq:Res:p4m:H3U1} after pulling back by $B$ become
\begin{equation}
    \begin{split}\label{eq:Res:p4m:H3U1:top}
        B^*{\Phi}_1 &= 2\omega \cup \frac{\bar{\dd}\omega}{2\pi};\\
    B^*{\Phi}_2 &= \pi \sigma \cup \frac{\bar{\dd}\omega}{2\pi};\\
    B^*{\Phi}_3 &= (\vec{R}\cdot \Mbs) \cup \frac{\bar{\dd}\omega}{2\pi};\\
    B^*{\Phi}_4 &= \pi \sigma\cup \frac{\bar{\dd} (\vec{R}\cdot \Mbs)}{2\pi};\\
    B^*{\Phi}_5 &= 2\omega \cup \Area; \\
    B^*{\Phi}_6 &=\pi \sigma \cup \Area.
\end{split}
\end{equation}
We can then recover Eq.~\ref{eq:actionp4m} in the main text by pulling back the most general element of $\Hmc^3(\Gwp,\RR/2\pi\ZZ)$:
\begin{widetext}
    \begin{equation}
     \Lmc = B^* (k_{1,\alpha} \Phi_1 + k_{2,\alpha,\lambda_1} \Phi_2 + k_{3,\alpha} \Phi_3 + k_{4,\alpha,\lambda_1} \Phi_4 + k_{5}\Phi_{5} + k_{6} \Phi_6 ).
\end{equation} 
Where we have added a subscript $\alpha$ and $\lambda_1$ to the coefficients used in Eq.~\ref{eq:generalElement:nu3} to indicate dependence on origin and reference reflection line. 
\end{widetext}

\begin{table*}
\centering
\begin{minipage}{0.45\textwidth}
\centering
\begin{tabular}{|c|c|c|c|c|}\hline
      \multicolumn{5}{|c|}{${B^* \Res_{\OO} \Xi_1;\,\, [\Xi_1] \in \Hmc^1(\rm{p4m},\U(1) )}$ } \\ \hline
\diagbox[width=1.8cm,height=0.9cm,innerleftsep=0.5em,innerrightsep=0.85em]{\raisebox{-0.cm}{$\!\! B^*\Xi_1 $}}{   \raisebox{0.cm}{$\OO$}}& $\alpha$ & $\beta$ & $\gamma_1$ & $\gamma_2$ 
\\\hline
        $2\omega$ & $2\omega$ &  $2\omega$ & 0 &0 \\
        $\pi\sigma$ &$\pi\sigma$ & $ \pi\sigma$ & $\pi\sigma$ & $\pi\sigma$\\
        $\Mbs\cdot\vec{R}$ &0 & $2\omega+\pi \sigma$ & $\omega$ & $\omega + \pi \sigma$\\
        \hline
    \end{tabular}
\end{minipage}
\hspace{0.05\textwidth} 
\begin{minipage}{0.45\textwidth}
\centering
    \begin{tabular}{|c|c|c|c|c|}
    \hline
    \multicolumn{5}{|c|}{${B^* \Res_{\OO} \Xi_2;\,\, [\Xi_2] \in \Hmc^2(\rm{p4m},\ZZ^{\OR} )}$ } \\ \hline
         \diagbox[width=1.8cm,height=0.9cm,innerleftsep=0.5em,innerrightsep=0.85em]{\raisebox{0cm}{$\!\! B^*\Xi_2 $}}{   \raisebox{0cm}{$\OO$}}& $\alpha$ & $\beta$ & $\gamma_1$ & $\gamma_2$ 
         \\\hline
        $\frac{\dOR\omega_{\alpha}}{2\pi}$ & $\frac{\dOR\omega_{}}{2\pi}$ &  $\frac{\dOR\omega_{}}{2\pi}$ & $\frac{\dOR\omega_{}}{2\pi}$ & $\frac{\dOR\omega_{}}{2\pi}$ \\
        $\frac{\dOR\Mbs\cdot\vec{R}}{2\pi}$ &$0$ & $2\frac{\dOR\omega}{2\pi}$ & $\frac{\dOR\omega}{2\pi}$ & $\frac{\dOR\omega_{}}{2\pi}$\\
        $\Area$ &0 & $\frac{\dOR\omega_{}}{2\pi}$ & $0$ & $0$\\
        \hline
    \end{tabular}
\end{minipage}
\caption{\label{tab:Resp4m} Entries of the tables are the image under the restriction map ($\Res_{\OO}$) for generators the generators of $\Hmc^1(\rm{p4m},\U(1))$ (left) and $\Hmc^2(\rm{p4m},\ZZ^{\OR})$ (right). }
\end{table*}

To study topological actions of terms protected by p4m and internal symmetries it is useful to use the restrictions in Eqs.~\ref{eq:Res:p4m:H1U1} and ~\ref{eq:Res:p4m:H2Z}. We summarize these restrictions after pulling back by the gauge field in Table~\ref{tab:Resp4m}.

\subsubsection{Area form cocycle}\label{app:AreaFormp4m}

We construct the cocycle corresponding to the area form for p4m. More precisely, we want $ \Omega \in Z^2(\rm{p4m},\ZZ^{\OR})$ such that $\iota_{\ybf}\iota_{\xbf}\Omega=\Omega(\xbf,\ybf)-\Omega(\ybf,\xbf)=1$. We considered the area form separately because its construction was not obvious to us, even though the final expression (Eq.~\ref{eq:p4mAreaForm}) is simple.

The twisted cup product gives an obvious starting point. Consider $\vec{t}(\gbf) = [f_x(\gbf),f_y(\gbf)]^\top \in \ZZ^2$ is the `translation cocycle' that satisfies
\begin{equation}
    {^{\gbf{_1}}}\vec{t}(\gbf_2) - \vec{t}(\gbf_1\gbf_2)+ \vec{t}(\gbf_1) =0 
\end{equation}
for the action
\begin{equation}
    {^{\gbf}} \vec{t} = \begin{pmatrix}
        \cos(\pi a /2) & -\sin(\pi a /2)(-1)^b\\
        \sin(\pi a /2) & \cos(\pi a /2)(-1)^b
    \end{pmatrix} \cdot \vec{t}.
\end{equation}
where $a = f_0(\gbf)$ and $b=f_2(\gbf)$ are the powers of $\hbf_{\OO}$ and $\rbf_{\LL}$ in $\gbf$, respectively. Equivalently, ${^\gbf}\vec{t}(\gbf_1) = \vec{t}(\gbf\gbf_1\gbf^{-1})$. 

Recall that the wedge product  ($\vec{V}\wedge \vec{U}:= V_1U_2 - V_2U_1$) of two $\rm{O}(2)$ vectors ($V,U$) transforms under the sign representation of $\rm{O}(2)$. This still holds upon restriction to any finite subgroup $D_{n}\subset \rm{O}(2)$, in particular the subgroup $D_4$.  
This implies that the twisted cup product 
\begin{equation}
    \tilde{\Omega}(\gbf_1,\gbf_2) = \vec{t}(\gbf_1) \wedge {^{\gbf_1}}\vec{t}(\gbf_2),
\end{equation}
is closed under the twisted cup product, {i.e.} $\bar{\dd}\tilde{\Omega} =0$. However, $\iota_{\ybf}\iota_{\xbf} \tilde{\Omega} = 2$.

Consider now the cochain $\varpi\in C^1(\rm{p4m},\ZZ)$:
\begin{equation}
    \varpi(\xbf^{c_{x}}\ybf^{c_y} \hbf_{\OO}^{a} \rbf_{\LL}^{b}) = c_x c_y,
\end{equation}
which satisfies 
\begin{equation}
    \bar{\dd} \varpi (\gbf_1,\gbf_2) = \tilde{\Omega}(\gbf_1,\gbf_2) \mod 2. 
\end{equation}
 To see this, let $b_1 = f_2(\gbf_1)$, $[x_1,y_1] :=\vec{t}(\gbf_1)^\top$, $[x_2,y_2] :=\vec{t}(\gbf_2)^\top$ and $[x_3,y_3] :=^{\gbf_1}\vec{t}(\gbf_1)^\top$. Then 
\begin{equation} 
     \begin{split}
         \bar{\dd}\varpi(\gbf_1,\gbf_2) &= (-1)^{b_1} x_2 y_2 + x_1 y_1 - (x_1+x_3)(y_1+y_3) \\
          &= - 2[b_1]_2 x_2 y_2  - (x_1y_3 + x_3 y_1) \\
          &= x_1 y_3 - x_3 y_1 \mod 2 \\
          &= \tilde{\Omega}(\gbf_1,\gbf_2) \mod 2.
     \end{split}
 \end{equation}
 Therefore, we can take $\Omega := \frac{\tilde{\Omega} - \bar{\dd}\varpi}{2} \in Z^2(\rm{p4m},\ZZ^{\OR})$. We write the explicit expression for $\Omega$ in Eq.~\ref{eq:p4mAreaForm}, which can be used to check that $\iota_{\ybf}\iota_{\xbf} \Omega = +1$.

\section{Analytical verification of partial symmetry invariants}

In this appendix, we explicitly evaluate the partial symmetry invariants mentioned in the main text for two classes of examples. In App.~\ref{app:CalculationsForSingletCovering} we study a state with zero correlation length, and in App.~\ref{app:AKLT:state} we study a state constructed using AKLT chains.

\subsection{Calculations for the \texorpdfstring{\singletName }{singlet covering} state}\label{app:CalculationsForSingletCovering}

Below we present calculations for the \singletName state with symmetry $G = \SO(3) \times \text{p4m}$. By identifying suitable subgroups of $\SO(3)$ with $\U(1)$ or $\Z_N$, this example allows us to compute and verify all the independent invariant types studied in this paper except type A3. 
We compute a complete set of invariants to characterize the SPT phase of this state and summarize the results in several tables. 

In App.~\ref{app:singletCover:def} we explicitly define the state. 
App.~\ref{app:singletCover:PartialRotation} evaluates the partial rotation ($\Theta_{k,\OO}(\gbf)$, App.~\ref{app:singletCover:2Reflection} evaluates the partial double reflection ($\Sigma_{\OO,\LL}(\gbf,\jbf)$, and App.~\ref{app:singletCover:TwistedReflection} evaluates the phase of the partial reflection with twisted boundary conditions ($\CCtwoSymbol_{\LL}(\gbf;\jbf)$. 

\subsubsection{Definition of state}\label{app:singletCover:def}

\begin{figure}
    \centering
    \includegraphics[width=0.75\linewidth]{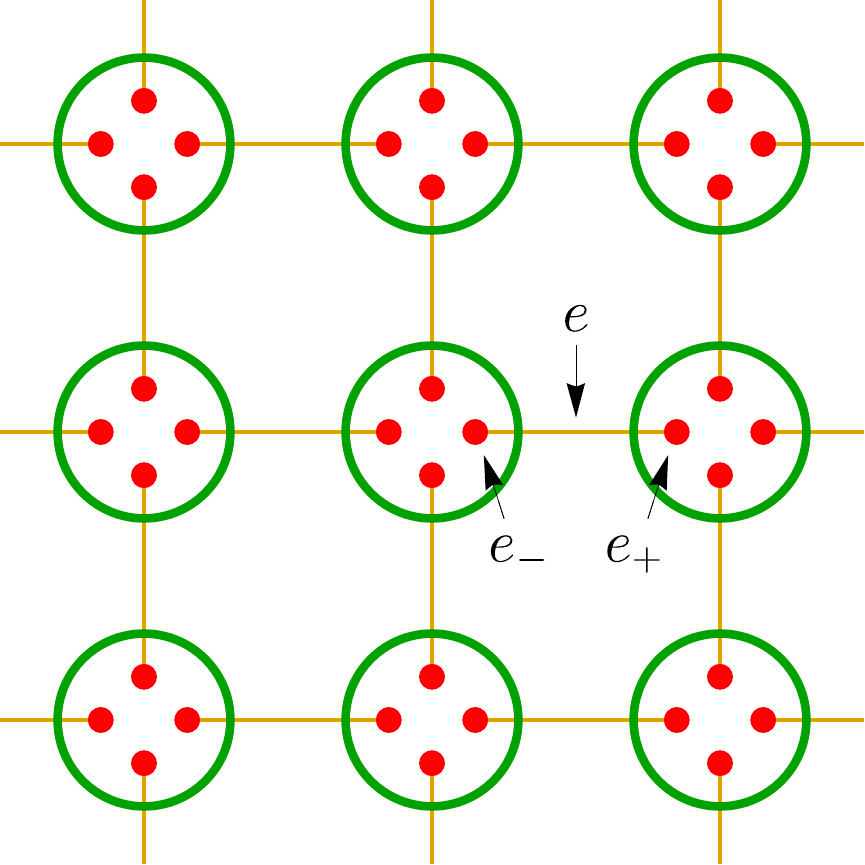}
    \caption{Pictorial representation of \singletName state. Red dots denote a spin $S=1/2$, and green circles represent sites/vertices of the square lattice. Orange lines connect spins that form singlets. Each orange line lies on an edge $e$. The spins lying on the ends of $e$ are denoted by $e_+$ and $e_-$. }
    \label{fig:HilbertSpaceSinglet}
\end{figure}
In this section, we explicitly define the \singletName state ($\ket{\psi_{0}}$). The overall {Hilbert space} is constructed by placing two {$S=1/2$ spins} on the ends of every bond of the square lattice (see Fig.~\ref{fig:HilbertSpaceSinglet}). At any vertex, the {local Hilbert space} is the tensor product of four spins from the four incident edges $\mathcal{H}_{\text{local}} = \mathbb{C}^2 \otimes \mathbb{C}^2 \otimes \mathbb{C}^2 \otimes \mathbb{C}^2 = (\mathbb{C}^2)^{\otimes 4}$. Since this local space involves an {even number} of $S=1/2$ spins, $\mathcal{H}_{\text{local}}$ transforms under a {linear representation of $\SO(3)$} (the group of spatial rotations), rather than the double cover $\SU(2)$. Consider the exactly solvable Hamiltonian 
\begin{equation}\label{eq:Ham:SingletConvering}
    H = \sum_{e} \vec{S}_{e_+} \cdot \vec{S}_{e_-}
\end{equation}
where $\vec{S}_{s}$ is the spin operator for spin $s$. $e_{+}$ and $e_{-}$ denote the two spins living at the ends of the edge $e$ (see Fig.~\ref{fig:HilbertSpaceSinglet}). The \singletName state $\ket{\psi_0}$ is the ground state of the above Hamiltonian, which is the tensor product of singlets
\begin{equation}
    \ket{\psi_0} = \bigotimes_{e}\frac{1}{\sqrt{2}}\left( 
    \ket{\ua}_{e_+}\ket{\da}_{e_-} 
    -
    \ket{\da}_{e_+}
    \ket{\ua}_{e_-}
    \right),
\end{equation}
where $\ket{\ua}_s$ and $\ket{\da}_s$ denote the spin up and spin down states of spin $s$.

This state and Hamiltonian are invariant under: 1) the geometric action of $\Gwp=\text{p4m}$ on the edges of the square lattice\footnote{Geometric action means that it just acts by permuting sites.}; and 2) $\SO(3)$ generated by $\vec{S}_{\text{tot}}=\sum_{e}(\vec{S}_{e_+}+\vec{S}_{e_-})$. In particular, every $\gbf \in \SO(3)$ acts on a spin $S=1/2$ as a $2\times 2$ matrix $V_{\gbf} \in \SU(2)$ that satisfies 
\begin{equation}\label{eq:Vgbf}
    V_{\gbf_1} V_{\gbf_2} = \upmu(\gbf_1,\gbf_2) V_{\gbf_1\gbf_2},
\end{equation}
$\upmu(\gbf_1,\gbf_2) \in \{+1,-1\}$ is a sign that appears because the spins transform as projective representations of $\SO(3)$. In particular, $\upmu$ is a group cocycle representative of the non-trivial class in $\Hmc^2(\SO(3),\U(1))$.

To evaluate the type B invariants we use the subgroup of rotations around the x-axis as the $K=\U(1)$ group. For type C invariants, we identify $K=\ZZ_2$ with the group generated by $\pi$ rotations around the x-axis.

In what follows, we consider the state $\ket{\psi_0}$ where the vertices of the square lattice lie on the Wyckoff position $\alpha$. The region $D$ will be taken to be a rectangle whose boundary crosses $2N_x$ vertical bonds and $2N_y$ horizontal bounds. We denote the `interior' of $D$ by $\mathring{D}$ and its `boundary' by $\partial D$.

The evaluation of the invariants can be simplified by noting that
\begin{equation}\label{eq:Omc:PartialSym}
\Omc_{D} := \mel{\psi_0}{O|_D}{\psi_0} = \Tr[O|_{D}\rho_D],
\end{equation}
where the reduced density matrix decomposes as
\begin{equation}
\rho_{D} = \Tr_{\bar{D}}[\ket{\psi_0}\bra{\psi_0}] = \rho_{\mathring{D}} \otimes \rho_{\partial D}.
\end{equation}
Here, $\rho_{\mathring{D}}$ is the pure state of singlets entirely contained within $D$, while $\rho_{\partial D}$ is a maximally mixed state for spins forming singlets with those outside $D$ (see Fig.~\ref{fig:sketchDecomposition}). Consequently, the calculation factorizes into contributions from the boundary ($\Omc_{\partial D} $) and the interior ($\Omc_{\mathring{D}}$),
\begin{equation}
\Omc_D = \Omc_{\partial D} \cdot \Omc_{\mathring{D}}.
\end{equation}
\begin{figure}[t]
    \centering
    \includegraphics[width=0.42\textwidth]{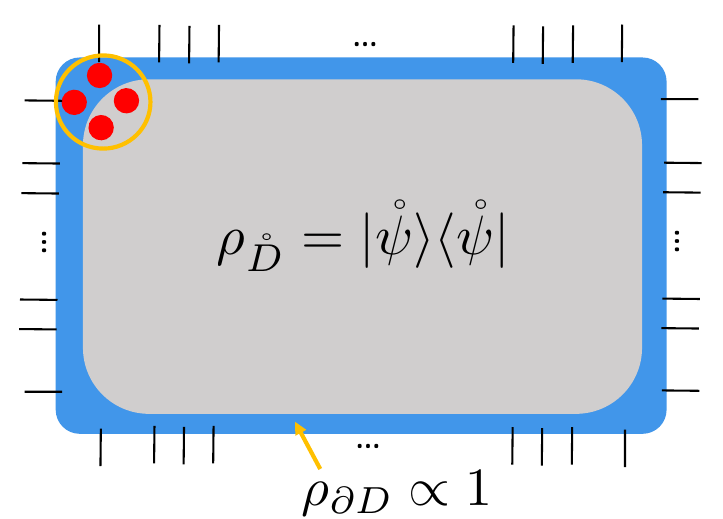}
        \caption{Decomposition of reduced density matrix for the \singletName state. The four red dots surrounded by a circle aim to illustrate the four spins belonging to site closest to the top left corner of region $D$.}
    \label{fig:sketchDecomposition}
\end{figure}

\subsubsection{Partial rotations (type A1/B1/C1/D1)}\label{app:singletCover:PartialRotation}

In this section, we evaluate Eq.~\ref{eq:Omc:PartialSym} with 
\begin{equation}
    \Omc = \tilde{C}_{\MO}^{n} U_{\gbf}.
\end{equation}
for $\gbf \in \SO(3)$. We consider the cases where the order of $\Omc$ is 4 and 2 separately. We show in Tab.~\ref{tab:PartialRotationsPsi0} the real space invariants for $\ket{\psi_0}$.

\paragraph{$\Omc$ is order 4:} This corresponds to $\OO \in {\alpha, \beta}$ with $n=1$. We take the region $D$ to be a square ($N_x = N_y$) centered at $\OO = \alpha$ or $\beta$. Due to the geometry of the square lattice, $N_x$ is odd for $\OO = \alpha$ and even for $\OO = \beta$.

The boundary $\partial D$ is the disjoint union of 4 sides $B_{1}$, $B_2$, $B_3$ and $B_4$ that are permuted cyclically under $C_4$ (see Fig.~\ref{fig:sketchDecompositionBdry}). On each $B_j$, there are $N_x$ boundary spin-1/2 variables.

\begin{figure}
    \centering
    \includegraphics[width=0.42\textwidth]{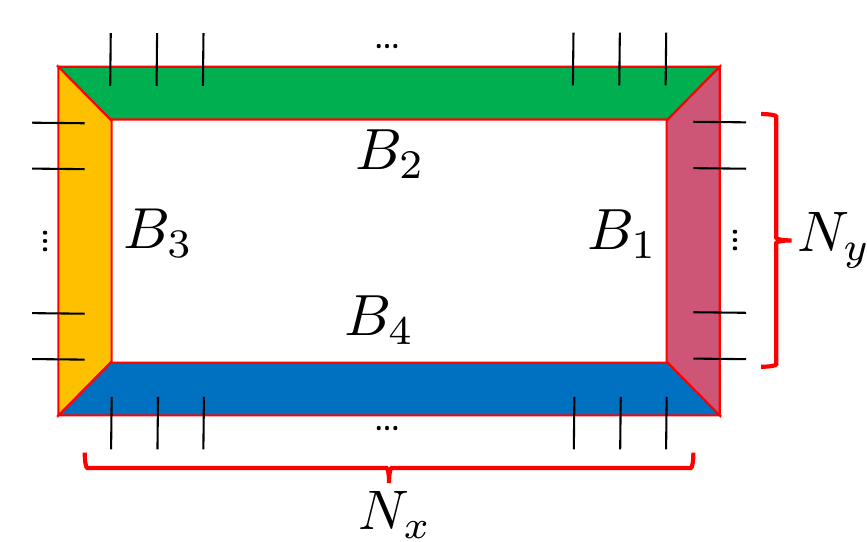}
        \caption{Decomposition used in the calculation of partial symmetry operations.}
    \label{fig:sketchDecompositionBdry}
\end{figure}

The boundary contribution to Eq.~\ref{eq:Omc:PartialSym} is
\begin{equation}
    \Omc_{\partial D} = \frac{\Tr[ \SWAP_{1,2,3,4} \bigotimes_{j\in \partial D }(V_{\gbf})_j]}{2^{2N_x + 2N_y}} 
\end{equation}
where $\SWAP_{1,2,3,4} =\SWAP_{3,4}\SWAP_{2,3}\SWAP_{1,2}$, and $\SWAP_{i,j}$ acts by exchanging the states on subsystems $B_i$ and $B_j$. Recall that $V_{\bf g}$ is the local action of $\SO(3)$ on each spin-1/2 (Eq.~\ref{eq:Vgbf}). $\Omc_{\partial D}$ can be simplified to 
\begin{equation}
    \Omc_{\partial D} = \left(\frac{\Tr[V_{\gbf}^4]}{2^4}\right)^{N_x} = \ee^{\ii N_x\phi_{\gbf} - 3\log(2)N_x  }
\end{equation}
with 
\begin{equation}\label{eq:defVarPhi1}
    \ee^{\ii\phi_{\gbf}} = \prod_{j=0}^{3}\upmu(\gbf,\gbf^j)\in \{+1,-1\}. 
\end{equation}
The sign of the interior contribution, $\Omc_{\mathring{D}}$, can be computed by assigning an orientation to each edge in $\mathring{D}$ and counting the bonds that are flipped under the rotation. Each flipped bond contributes a factor of $-1$, since the spatial wavefunction of a singlet is antisymmetric. The contribution from $U_{\gbf}$ vanishes because the state in $\mathring{D}$ is a product of singlets. By grouping bonds into sets of four that are closed under rotations, one sees that exactly two bonds are flipped in each group, making the overall interior contribution trivial.

The partial rotation evaluates to 
\begin{equation}\label{eq:PartialC4Ug}
    \mel{\psi_0}{(\tilde{C}_{M_{\OO}}U_{\gbf})|_{D}}{\psi_0} = \frac{1}{2^{3N_x}} \begin{cases}
        \ee^{\ii \phi_{\gbf}} \quad &, \OO = \alpha \\
        1 \quad &, \OO = \beta \\
    \end{cases},
\end{equation}
here $\phi_{\gbf}$ is defined in Eq.~\ref{eq:defVarPhi1}.

\paragraph{$\Omc$ is order 2} In this case, $\OO\in\{\a,\b,\g\}$ and $n=(\MO,4)/2$. Now the internal symmetry element $\gbf$ satisfies $\gbf^2=\zero$. From the geometry of the square lattice,
\begin{equation}
    (N_x + N_y) = \begin{cases}
      0 \mod 2 \quad &, \OO \in\{\alpha,\beta\} \\
      1 \mod 2 \quad &, \OO \in\{\gamma\}
    \end{cases}.
\end{equation}

The boundary contribution is 
\begin{equation}
     \Omc_{\partial D} = \frac{\Tr[ \SWAP_{1,3}\SWAP_{2,4} \bigotimes_{j\in \partial D }(V_{\gbf})_j]}{2^{2N_x + 2N_y}}.
\end{equation}
where now $\SWAP_{j,k}$ swaps regions $B_j$ and $B_k$ in Fig.~\ref{fig:sketchDecompositionBdry}. $\Omc_{\partial D}$ simplifies to
\begin{equation}
    \Omc_{\partial D} = \left(\frac{\upmu(\gbf,\gbf)}{2}\right)^{N_x+N_y}.
\end{equation}
The bulk contribution for $\OO \in \{\a,\b\}$ is still trivial. However, for $\OO= \gamma$, $\Omc_{\mathring{D}}=-1$ because of the bond passing through the rotation center.
Putting things together
\begin{equation}\label{eq:PartialC2Ug}
    \mel{\psi_0}{(\tilde{C}_{M_{\OO}}^{\frac{(M_{\OO},4)}{2}}U_{\gbf})|_{D}}{\psi_0} =  \begin{cases}
    \frac{1}{2^{N_x+N_y}} &, \OO =\alpha,\beta \\
       \frac{ -\upmu(\gbf,\gbf)}{2^{N_x+N_y}} &, \OO =\gamma \\
    \end{cases}.
\end{equation}

\paragraph{Evaluation of real space invariants:}
\begin{table}
    \centering
    \begin{tabular}{c|c|c|c}
       \text{Invariant}  & \text{value} & $\gbf$ & $n$\\\hline\hline
        $\Theta_{\a}$   & 0 & $\zero$ & 1\\
        $\Theta_{\b}$  & 0 & $\zero$ & 1\\
        $\Theta_{\g}$  & 1 & $\zero$ & 1\\\hline
        $\Theta_{\a}^{\SO(3)}$ &2 & $[\pi/2,\hat{x}]$ & 1\\
        $\Theta_{\b}^{\SO(3)}$  &0& $[\pi/2,\hat{x}]$ & 1\\
        $\Theta_{\g}^{\SO(3)}$  &0 & $[\pi,\hat{x}]$ & 1\\\hline
        $\Theta_{\a}^{\ZZ_2}$  &0 & $[\pi,\hat{x}]$ & 2\\
        $\Theta_{\b}^{\ZZ_2}$  &0& $[\pi,\hat{x}]$ & 2\\
        $\Theta_{\g}^{\ZZ_2}$  &0 & $[\pi,\hat{x}]$ & 1
    \end{tabular}
    \caption{Dressed partial rotation invariants for the \singletName state $\ket{\psi_0}$ using Eq.~\ref{eq:Omc:PartialSym} with $O = \tilde{C}_{\MO}^n U_\gbf$, $\gbf\in \SO(3)$.}
    \label{tab:PartialRotationsPsi0}
\end{table}

We evaluated the real space invariants using Eq.~\ref{eq:PartialC4Ug} and Eq.~\ref{eq:PartialC2Ug}. In Table~\ref{tab:PartialRotationsPsi0}, we summarize the real space invariant and the group element $\gbf$ used to evaluate them.  Note that $\Theta_{\OO}^{\U(1)}=\Theta_{\OO}^{\SO(3)}$ by our choice of $\U(1)\cong\SO(2)$.

We summarize the type A1, B1, C1 and D1 invariants in Table~\ref{tab:invariants:psi0}, which are calculated by taking differences of the appropriate invariants in Table~\ref{tab:PartialRotationsPsi0}. 

\begin{table}
    \centering
    \begin{tabular}{c|c|c|c|c}
  \diagbox[width=1.3cm]{$\OO$}{} & $\Theta_{\OO}^{}$ & $\so^{\U(1)}$ & $\so^{\ZZ_2}$ & $\so^{\SO(3)}$  \\\hline
   $\a$ & 0& 2&0& 2\\
   $\b$ & 0& 0&0& 0\\
   $\g$ & 1& 1&1& 1\\
    \end{tabular}
    \caption{Type A1, B1, C1 and D1 invariants for the state $\ket{\psi_0}$.}
    \label{tab:invariants:psi0}
\end{table}

\subsubsection{Partial double reflections (Type A2/C2/D2)}\label{app:singletCover:2Reflection}

The main result of this section is Eq.~\eqref{eq:psi_0:A2}, which evaluates $\Sigma_{\OO,\LL}(\gbf,\jbf)$, and the numerical result for the singlet covering state, which is summarized in Table~\ref{tab:PartialReflectionPsi0}. We decompose the region $D$ into three subregions $D_{l}, D_{c}, D_{r}$ as in Fig.~\ref{fig:DefinitionDForPartialReflection} in the main text. 

\paragraph{$\LL$ is horizontal:} we orient $\LL$ parallel to the $x$ axis. We take the boundary of the region $D$ to cross $2N_y$ horizontal bonds and $2N_x$ vertical bonds. Let $2N_1$ and $2N_2$ be the number of vertical bonds the boundary of regions $D_l$ and $D_c$ cross, respectively. The group elements $\gbf,\jbf\in \SO(3)$ satisfy $\gbf  \jbf=\jbf \gbf$ and $\jbf^2=\zero$. 

The expectation value $\Omc$ now decomposes as
\begin{equation}
    \Omc = \Omc_{B_{lr}}  \Omc_{B_v} \Omc_{B_h} \Omc_{\mathring{D}_{lr}}
    \Omc_{\mathring{D}_c}.
\end{equation}
The subscripts correspond to the regions in Fig.~\ref{fig:sketch regions partial reflection} and $B_{lr} = B_l \cup B_r$ and $\mathring{D}_{lr} = \mathring{D}_l \cup \mathring{D}_r$. 

The contribution from $\mathring{D}_{lr}$ ($B_{lr}$) is $2^{-N}$, where $N$ is the number of singlets on $\mathring{D}_{l}$ ($B_{l}$), because we can choose the orientation of singlets in $\mathring{D}_{l}$ ($B_l$) to be the mirror image of those in $\mathring{D}_{r}$ ($B_r$). For the same reason, only singlets lying on or crossing $\LL$ can contribute a non-trivial phase to ${\mathring{\Omc}_{D_c}}$. A direct calculation shows that only the singlets crossing $\LL$ can contribute a phase.

Let's consider the contribution to $\Omc_{B_{v}}$ from the two singlets lying on $\LL$. Their wavefunction is\footnote{we identify $\ket{\ua}=\ket{0}$ and $\ket{\da}=\ket{1}$.}
\begin{equation}
    \ket{\phi} = \frac{1}{2}\sum_{s_1,s_2,s_3,s_4= 0,1} \varepsilon_{s_1s_2}\varepsilon_{s_3s_4}\ket{s_1s_2s_3s_4},
\end{equation}
where we have numbered the spins from left to right ( {e.g.}, spin 2 is in region $D_c$, near its left boundary), and $\varepsilon_{ss'}$ is the Levi-Civita tensor. 

Their contribution to $\Omc_{B_v}$ is
\begin{equation}
    A_{\gbf,\jbf}:=\mel{\phi}{\SWAP_{1,4}\left(V_{\gbf}\otimes V_{\jbf}\otimes V_{\jbf}\otimes V_{\gbf}^{\dagger}\right)}{\phi},
\end{equation}
because, to calculate $\Sigma_{\OO,\LL}(\gbf,\jbf)$, we act with $\gbf$ on $D_{l}$ (where spin~1 lies), with $\gbf^{-1}$ on $D_{r}$ (where spin~4 lies), and with $\jbf$ on $D_{c}$ (where spins~2 and~3 are located).

Using the fact that for any $\gbf\in \SO(3)$, $V_{\gbf}\otimes V_{\gbf} \ket{\varphi_0} = \ket{\varphi_0}$, where $\ket{\varphi_0}$ is the spin-singlet state, we can rewrite
\begin{equation}
    A_{\gbf,\jbf}=\mel{\phi}{\SWAP_{1,4}\left(\Id\otimes V_{\jbf}V_{\gbf}^{\dagger}\otimes V_{\jbf}V_{\gbf}^{}\otimes \Id \right)}{\phi}.
\end{equation}
Using the relation $\sum_{s}\varepsilon_{s_1 s}\varepsilon_{s_2 s} = \delta_{s_1 s_2}$, $A$ simplifies to
\begin{equation}
\begin{split}
A_{\gbf,\jbf}= \frac{1}{4}\Tr[V_{\gbf} V_{\jbf} V_{\gbf}^{\dagger} V_{\jbf}] = \frac{\upmu(\jbf,\jbf)}{2} \frac{\upmu(\gbf,\jbf)}{\upmu(\jbf,\gbf)}.
\end{split}
\end{equation}
The ratio $\upmu(\gbf,\jbf)/\upmu(\jbf,\gbf)$ arises when commuting $V_{\gbf}$ past $V_{\jbf}$, while the factor $\upmu(\jbf,\jbf)$ comes from the product $V_{\jbf}^2$. 

The remaining singlets on $B_v$ appear in groups of four, each forming an orbit under the action of $D_2$, generated by $\rbf_{\LL}$ and $\rbf_{\LL'}$. Their contribution can be rewritten as $A_{\gbf,\jbf}^2$. The singlets in $B_h$ come in pairs, related by $\LL$. Each pair contributes $A_{\zero,\jbf}$.

Putting everything together, we obtain
\begin{equation}\label{eq:psi_0:A2}
    \Omc = \frac{1}{2^{N_x+2N_y}}    \begin{cases}
        \frac{\upmu(\gbf,\jbf)}{\upmu(\jbf,\gbf)} \quad &, \OO =\alpha\\
    \upmu(\jbf,\jbf)\times\frac{\upmu(\gbf,\jbf)}{\upmu(\jbf,\gbf)} \quad &, \OO=\gamma_1\\
        -\upmu(\jbf,\jbf) \quad &, \OO=\gamma_2\\
        1 \quad &, \OO =\beta
    \end{cases}.
\end{equation}
We see that non-trivial phases are only contributed by degrees of freedom on $\LL$ and $\LL'$.

\begin{figure}[b]
    \centering
\includegraphics[width=0.45\textwidth]{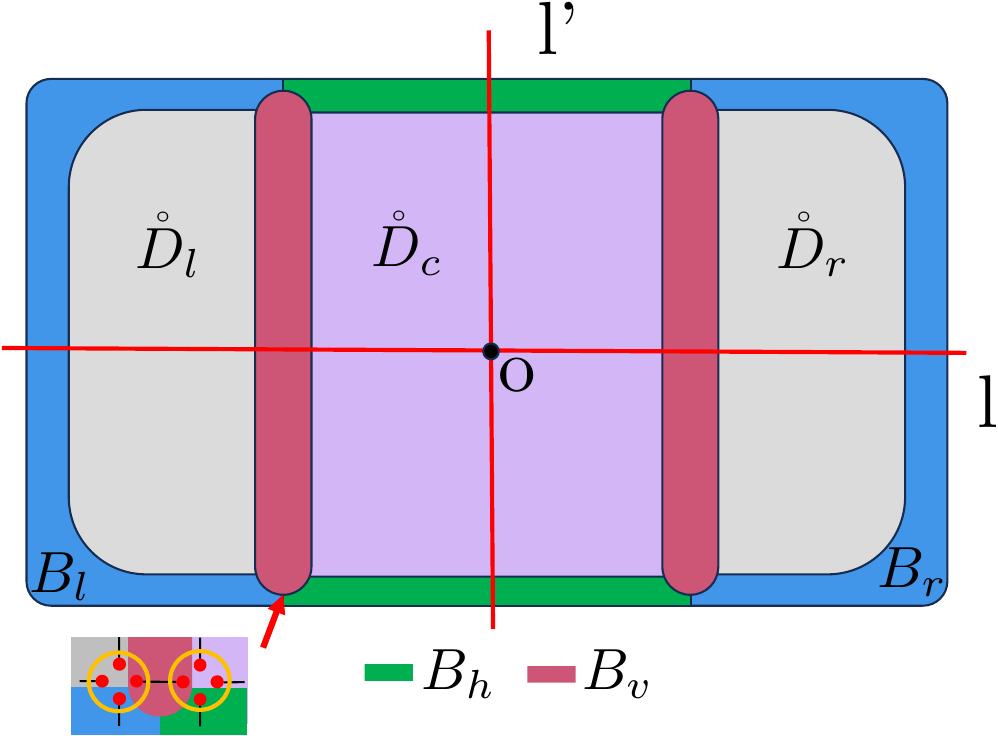}
    \caption{Decomposition of the regions used in the evaluation of $\Sigma_{\OO,\LL}$. The inset illustrates how spins located at the corners of the regions $D_{l}$ and $D_{r}$ are assigned to the regions used to evaluate the partial symmetry.  }
    \label{fig:sketch regions partial reflection}
\end{figure}

\paragraph{$\LL$ is diagonal:} now $\LL$ is parallel to the line $y=x$, and $\OO = \alpha$ or $\beta$. The calculation can be done as in the previous case, but we always find that the phase of $\Omc$ is trivial. 

\paragraph{Evaluation of real space invariants:} The A2 and D2 invariants calculated using geometric reflection are tabulated in Table~\ref{tab:PartialReflectionPsi0}. In the same table, we tabulated C2 invariants with reflections given by combining geometric reflections with $\Xbf$, and $\Sbf = [2\pi/ N , \hat{z}]$.

\begin{table}
    \centering
    \begin{tabular}{c|c}
        \text{Invariant}  & \text{value} \\\hline\hline
               $\Sigma_{\a,\lambda_1}$   & 0 \\
        $\Sigma_{\g_1,\lambda_1}$   & 0 \\
        $\Sigma_{\b,\mu_1}$   & 0 \\ \hline
          $\tilde\Sigma_{\a,\lambda_1}^{\SO(3)}$   & 1 \\
        $\tilde\Sigma_{\b,\mu_1}^{\SO(3)}$   & 0 \\
        $\tilde\Sigma_{\a,\nu_1}^{\SO(3)}$   & 0 \\ \hline
        $\tilde\Sigma_{\a,\lambda_1}^{\ZZ_N}$   & 1 \\
        $\tilde\Sigma_{\b,\mu_1}^{\ZZ_N}$   & 0 \\
        $\tilde\Sigma_{\a,\nu_1}^{\ZZ_N}$   & 0 \\ \hline
    \end{tabular}
    \caption{Type-A2 ($\Sigma_{\OO,\LL}$), type-D2 ($\tilde\Sigma_{\OO,\LL}^{\SO(3)}$) and type-C2 ($\tilde\Sigma_{\OO,\LL}^{\ZZ_{2m}}$) invariants for the \singletName state. Type A2 and D2 are calculated with geometric reflections, while type C2 are calculated by combining the geometric reflection with $\Xbf\in \SO(3)$.}
    \label{tab:PartialReflectionPsi0}
\end{table}
\subsubsection{Partial reflection with twisted boundary conditions}\label{app:singletCover:TwistedReflection}

For the Hamiltonian in Eq.~\ref{eq:Ham:SingletConvering}, $\gbf$-twisted boundary conditions amount to replacing $\vec{S}_{e_-}\to U_{\gbf}\vec{S}_{e_-}U_{\gbf}^{\dagger}$ for all bonds $e$ that cross one vertical line $\LL_*$  that lies between two vertical axes of the square lattice. We are taking $e_-$ to be the spin lying to the right of $\LL_*$. The new ground state is given by 
\begin{equation}\label{eq:psi0g}
    \ket{\psi_{0,\gbf}} = \prod_{e\in \Lmc} U_{\gbf}|_{e_-} \ket{\psi_0}.
\end{equation}

Since we are after the evaluation of type C4 and D4 invariants, we need the \textit{relative phase} of the expectation value $\expval{(\rbf_{\LL} U_{\jbf})|_D}$ between $\ket{\psi_0}$ and $\ket{\psi_{0,\gbf}}$. It is clear that this relative phase will come from the spins where $\gbf$ acts in Eq.~\ref{eq:psi0g}.

By doing manipulations similar to the ones done for $\Sigma_{\OO,\LL}(\gbf,\jbf)$, we arrive at 
\begin{equation}
    (-1)^{\CCtwoSymbol_{\LL}(\gbf;\jbf)} = \left(\frac{\upmu(\gbf,\jbf)}{\upmu(\jbf,\gbf)}\right)^{N_y},
\end{equation}
where $N_y$ is the number horizontal bonds inside $D$ that intersect $\LL_*$. Therefore, $N_y$ is odd (even) when $\LL = [\alpha,\gamma_1] ([\beta,\gamma_2])$.

\subsection{Calculations for stacked AKLT chains}\label{app:AKLT:state}

Here we briefly outline the calculation for a state with finite correlation length that lies in the same phase as the \singletName. This state is constructed by placing Affleck–Kennedy–Lieb–Tasaki (AKLT) chains along each line of the square lattice. At every vertex of the square lattice, we place two $S = 1$ spins. The Hamiltonian for each line is
\begin{equation}
    H_{\text{AKLT}} = \sum_{i} \vec{S}_i\cdot\vec{S}_{i+1}+\frac{1}{3}\left(\vec{S}_i\cdot\vec{S}_{i+1}\right)^2,
\end{equation}
where the sum is over sites on the line.

It is well known that the ground state of $H_{\text{AKLT}}$ can be understood by decomposing each $S=1$ spin into two virtual $S = 1/2$ spins, which form singlets with neighboring sites. This state admits an explicit representation as a matrix product state (MPS) with finite correlation length.

Using the explicit MPS representation, the evaluation of partial symmetries can be simplified considerably when the system and the size of $D$ are both large. This is because the correlation functions for the MPS are dominated by a single eigenvector of the transfer matrix. To leading order in system and $D$ size, the results agree with the \singletName state. For partial rotations, we analytically checked that expectation value $\mel{\psi}{(\Cmo)|_D}{\psi}$ behaves as \ref{ThetaOExplicit} for a rectangular region $D$.

\section{Relation between invariants}\label{app:RelationBetweenInvariants}
\subsection{Relations between type-A invariants}\label{app:RelationBetweenTypeAInvariants}

As in Ref.~\cite{Huang2017lowerDimCrysSPT}, we assume that any short-range state can be deformed to an `atomic limit' while preserving $\Gwp$. We derive the relations assuming that $\ket{\Psi}$ is already in its atomic limit. 

First notice that $\Theta_{\OO}$ and $\Sigma_{\OO,\LL}$, when they exist, satisfy
\begin{equation}\label{eq:AtomicLimit}
    \begin{split}
        \Cmo|_{\{\OO\}}\ket{\Psi} &= \ee^{2\pi \ii \Theta_{\OO}/M_{\OO} }\ket{\Psi}\\
        {R}_{\LL}|_{\{\OO\}}\ket{\Psi} &= (-1)^{\Sigma_{\OO,\LL}} \ket{\Psi}
    \end{split}
\end{equation}
where $\hbf_{\OO}$ is anticlockwise rotation by $2\pi/M_\OO$ radians around $\OO$ and $\rbf_{\LL}$ is reflection along $\LL$. To see that Eq.~\ref{eq:AtomicLimit} is true, note that we choose the regions $D$ in the definitions of $\Theta_\OO$ and $\Sigma_{\OO,\LL}$ such that the contributions away from $\OO$ come in  groups of $M_{\OO}$ and $2$ for $\Theta_\OO$ and $\Sigma_{\OO,\LL}$, respectively. Therefore, all contributions away from the origin cancel out.

Consider the case when $D_2\subset G_{\OO}$, then if we define $\LL'$ by $\rbf_{\LL'} = \hbf_{\OO}\rbf_{\LL}$, we can write 
\begin{equation}
   R_{{\LL'}}|_{\{\OO\}}\ket{\Psi} = \Cmo|_{\{\OO\}}R_{\LL}|_{\{\OO\}}\ket{\Psi}
\end{equation}
and use Eq.~\ref{eq:AtomicLimit} to show that 
\begin{equation}
    \Sigma_{\OO,\LL'} = \frac{2}{M_{\OO}}\Theta_{\OO} + \Sigma_{\OO,\LL} \mod 2.
\end{equation}

Similarly, consider the case when there are two maximal Wyckoff positions $\OO$ and $\OO'$ lying on a line $\LL$, whose site group includes a $D_2$ subgroup ({i.e.} $M_{\OO}$ and $M_{\OO'}$ are even). $\Lambda_{\LL}$ also needs a translation parallel to $\LL$, which we denote as $\ZZ^{(\LL)}$. Let $n_{\OO}(n_{\OO'})$ be the number of sites in the MWP $\OO$($\OO'$) appearing in the unit cell of $\ZZ^{(\LL)}$. Then from the definition 
\begin{equation}
         \frac{\mel{\Psi(L_x+1)}{R_{\LL}}{\Psi(L_x+1)}}{\mel{\Psi(L_x)}{R_{\LL}}{\Psi(L_x)}}\propto (-1)^{\Lambda_{\LL}},
\end{equation}
where $\ket{\Psi(L_x)}$ is the state on a cylinder (or torus) with $L_x$ unit cells along the direction parallel to $\LL$. 

From Eq.~\ref{eq:AtomicLimit},
\begin{equation}
\begin{split}
    R_{\LL}\ket{\Psi(L_x)} &= (-1)^{[n_{\OO}\Sigma_{\OO',\LL} + n_{\OO'} \Sigma_{\OO',\LL}] L_x}\ket{\Psi(L_x)}\\
    \Rightarrow\Lambda_{\LL} &=n_{\OO}\Sigma_{\OO,\LL} + n_{\OO'}\Sigma_{\OO',\LL} \mod 2.
\end{split}
\end{equation}

\subsection{Relations between type-D invariants}\label{app:RelationBetweenTypeDInvariants}

\subsubsection{p4m}\label{app:RelationBetweenTypeDInvariants:p4m}

In this section we derive Eq.~\ref{eq:RelationBetweenD1&D2} using group cohomology methods. Recall that for $\gbf,\jbf \in K$ such that $\gbf\jbf=\jbf\gbf$, we get
\begin{equation}\label{eq:app:Upsilon:g:j}
           (-1)^{\Upsilon_{\gbf;\jbf}} = \Zmc(S^1_{\gbf}\times \RP^2_{\jbf\rbf_{\LL}}) = \iota_{\gbf}\nu_3(\jbf\rbf_{\LL},\jbf\rbf_{\LL}),
\end{equation}
where $\nu_3$ is a 3-cochain, such that $B^*\nu_3 = e^{i \Lmc}$, where $\Lmc$ is the Lagrangian of the topological action.

When we evaluate the type D invariants, we evaluate suitable differences that get rid of any contribution from pure invariants. Therefore, we can express $\Upsilon_{\LL}^{\SO(3)}$ solely in terms of the coefficients in Eq.~\ref{eq:p4m_mixed_SO3}. A choice of $\varphi_3\in Z^3(\rm{p4m}\times \SO(3),\U(1)^{\sigma} )$, such that $B^*\varphi_3 $ returns Eq.~\ref{eq:p4m_mixed_SO3}, is
\begin{equation}\label{eq:app:CocycleGeneralPhi3MixedSO3:p4m}
\varphi_3 = 2\pi{\ww_{2}(u_{1,\a,\lambda_1}  \frac{f_2}{2}+u_{2,\a}\frac{f_1}{2} + u_{3}\frac{f_x+f_y}{2} )}.
\end{equation}
Here we used the cocycles from App.~\ref{app:Coh:p4m}, and $\ww_2$ is a cocycle representing the Stiefel-Whitney class of the $\SO(3)$ bundle. Then by applying Eq.~\eqref{eq:app:Upsilon:g:j}, we find
\begin{equation}
    \begin{split}
        \CCtwoSymbol^{\SO(3)}_{\lambda_1}  &= u_{1,\a,\lambda_1}\\
        \CCtwoSymbol^{\SO(3)}_{\nu_1} &= u_{1,\a,\lambda_1} + u_{2,\a} \\
        \CCtwoSymbol^{\SO(3)}_{\mu_1} &= u_{1,\a,\lambda_1} + u_{3} \\
        \Rightarrow \CCtwoSymbol^{\SO(3)}_{\lambda_1} + \CCtwoSymbol^{\SO(3)}_{\nu_1} &= u_{2,\a} \mod 2,\\
        \Rightarrow \CCtwoSymbol^{\SO(3)}_{\lambda_1} + \CCtwoSymbol^{\SO(3)}_{\mu_1} &= u_{3} \mod 2.
    \end{split}
\end{equation}
To derive Eq.~\ref{eq:RelationBetweenD1&D2} we finally note that $u_{2,\alpha} = \mathscr{S}_{\alpha}^{\SO(3)}$.

\subsubsection{pmm}\label{app:RelationBetweenTypeDInvariants:pmm}
The analogous equation to Eq.~\ref{eq:app:CocycleGeneralPhi3MixedSO3:p4m} for \text{pmm} is
\begin{equation}\label{eq:pmm_mixed_SO3:app}
\varphi_3 = \pi{\ww_{2}(u_{1,\a,\lambda} {f_2}+u_{2,\alpha}f_1 + u_{3, x}f_x + u_{3,y} f_y )},
\end{equation}
where the cocycles $f_{\dots}$ are the same as those in Eq.~\ref{eq:app:cocycles}, with the only modification that the variable $a$ is now defined modulo 2.

The various reflections (see Fig.~\ref{fig:UnitCell_pmm_6}) are 
\begin{equation}
    \rbf_{\mu} = \xbf\rbf_{\lambda};\,\,
    \rbf_{\nu} = \hbf_{\alpha}\rbf_{\lambda};\,\,
    \rbf_{\kappa} = \ybf\hbf_{\alpha}\rbf_{\lambda}.
\end{equation}
Then using Eq.~\ref{eq:app:Upsilon:g:j}, we obtain 
\begin{equation}
    \begin{split}
        \Upsilon_{\lambda}^{\SO(3)} &= u_{1,\alpha,\lambda};\\
    \Upsilon_{\mu}^{\SO(3)} &= u_{1,\alpha,\lambda}+u_{3,x};\\
    \Upsilon_{\nu}^{\SO(3)} &= u_{1,\alpha,\lambda}+u_{2,\alpha};\\
    \Upsilon_{\kappa}^{\SO(3)} &= u_{1,\alpha,\lambda}+u_{2,\alpha}+u_{3,y}.
    \end{split}
\end{equation}

The various rotations are 
\begin{equation}
    \hbf_{\beta}= \xbf\ybf\hbf_{\alpha};\,\,
    \hbf_{\gamma}= \ybf\hbf_{\alpha};\,\,
    \hbf_{\delta}= \xbf\hbf_{\alpha}.
\end{equation}
Recall also that for $\gbf$ of order two,
\begin{equation}
    (-1)^{\Theta_{\OO}(\gbf)} = \nu_3(\gbf\hbf_{\OO},\gbf\hbf_{\OO},\gbf\hbf_{\OO}).
\end{equation}
Therefore, the type D1 invariants are
\begin{equation}
\begin{split}
        \mathscr{S}_{\alpha}^{\SO(3)} &= u_{2,\alpha,\lambda};\\
    \mathscr{S}_{\beta}^{\SO(3)} &= u_{2,\alpha,\lambda}+ u_{3,x}+u_{3,y}\\
    \mathscr{S}_{\gamma}^{\SO(3)} &= u_{2,\alpha,\gamma}+u_{3,y}\\
    \mathscr{S}_{\delta}^{\SO(3)} &= u_{2,\alpha,\lambda}+u_{3,x}.
\end{split}
\end{equation}

We thus see that only 3 of the 4 D1 invariants are independent. To fully determine the invariants appearing in Eq.~\ref{eq:pmm_mixed_SO3:app}, we need to include at least one type D2 invariant.

\onecolumngrid
\section{Group cohomology tables}\label{app:GroupCohTables}

\def\nothing{$\ZZ_1$}
In Table~\ref{tab:GrCoho_untwisted} we present the group cohomology with $\ZZ$ coefficients for every wallpaper group up to degree 5.
\begin{table}[h!]
    \centering
 \begin{tabular}{r|c||c|c|c|c|c}
 \multicolumn{7}{c}{\textbf{Group cohomology $\Hmc^n(\Gwp,\ZZ)$}} \\ 
        \hline
        $\#$ & 
        \diagbox[height=2em,width=6em]{$\!\!G_{\text{space}}$}{$n$}
        & 1 & 2 & 3 & 4 & 5\\ \hline\hline
        1 & p1  & $\Z^2$ & $\Z$ & \nothing & \nothing & \nothing \\
        2 & p2    & \nothing & $\Z\times \Z_2^3$ & \nothing &$\Z_2^4$ & \nothing\\
        3 & pm  & $\Z$ & $\Z_2^2$ & $\Z_2^2$ & $\Z_2^2$ & $\Z_2^2$\\
        4 & pg   & $\Z$ & $\Z_2$ & \nothing & \nothing & \nothing\\
        5 & cm  & $\Z$ & $\Z_2$ & $\Z_2$  & $\Z_2$ & $\Z_2$ \\
        6 & pmm   & \nothing  & $\Z_2^4$ & $\Z_2^4$ & $\Z_2^8$ & $\Z_2^8$\\
        7 & pmg  & \nothing & $\Z_2^3$ & $\Z_2$  & $\Z_2^3$ & $\Z_2$\\
        8 & pgg  & \nothing & $\Z_4\times\Z_2$ & \nothing & $\Z_2^2$ & \nothing\\
        9 & cmm  & \nothing & $\Z_2^3$ & $\Z_2^2$ & $\Z_2^5$ & $\Z_2^4$ \\
        10 & p4  & \nothing & $\Z\times\Z_4\times\Z_2$ & \nothing & $\Z_4^2\times\Z_2$ & \nothing \\
        11 & p4m & \nothing & $\Z_2^3$ & $\Z_2^3$ & $\Z_4^2\times\Z_2^4$ & $\Z_2^6$\\
        12 & p4g  & \nothing & $\Z_4\times\Z_2$ & $\Z_2$  & $\Z_4\times\Z_2^2$ & $\Z_2^2$\\
        13 & p3  & \nothing & $\Z\times\Z_3^2$ & \nothing & $\Z_3^3$ & \nothing\\
        14 & p3m1 &\nothing & $\Z_2$ & $\Z_2$ & $\Z_3^2\times\Z_6$ & $\Z_2$\\
        15 & p31m & \nothing & $\Z_6$ & $\Z_2$ & $\Z_3\times\Z_6$ & $\Z_2$\\
        16 & p6  & \nothing & $\Z\times\Z_6$ &  \nothing & $\Z_6^2$ & \nothing\\
        17 & p6m  & \nothing & $\Z_2^2$ & $\Z_2^2$ & $\Z_2^2\times\Z_6^2$ & $\Z_2^4$
    \end{tabular}
    \caption{Group cohomology $\H^n(\Gwp,\Z)$ for $n=1,2,3,4,5$, where $\Gwp$ acts trivially on the $\Z$ coefficients. Note that $\H^0(\Gwp,\Z) \cong \Z$. $\ZZ_1$ denotes the trivial group.}
    \label{tab:GrCoho_untwisted}
\end{table}
\newpage
\begin{table}[t]
    \centering
    \begin{tabular}{r|c||c|c|c|c|c|c}
    \multicolumn{7}{c}{\textbf{Group cohomology twisted by orientation: $\Hmc^n(\Gwp,\ZZ^{\OR})$}} \\  \hline 
    $\#$ & 
        \diagbox[height=2em,width=6em]{$\!\!G_{\text{space}}$}{$n$}
        & 1 & 2 & 3 & 4 & 5\\ \hline
        \hline
         1 &p1  & $\Z^2$ & $\Z$ & \nothing & \nothing& \nothing\\
         2&p2    & \nothing & $\Z\times \Z_2^3$ & \nothing & $\Z_2^4$ & \nothing\\
         3&pm  & $\Z\times\Z_2$ & $\Z\times\Z_2$ & $\Z_2^2$  & $\Z_2^2$ & $\Z_2^2$\\
         4&pg   & $\Z\times\Z_2$ & $\Z$ & \nothing & \nothing & \nothing\\
         5&cm    & $\Z\times\Z_2$ & $\Z$ & $\Z_2$  & $\Z_2$ & $\Z_2$\\
         6&pmm   &  $\Z_2$ & $\Z\times\Z_2^3$ & $\Z_2^4$  & $\Z_2^8$ & $\Z_2^8$\\
         7 &pmg  &$\Z_2$ & $\Z\times\Z_2^2$ & $\Z_2$  & $\Z_2^3$ & $\Z_2$\\
         8 &pgg  & $\Z_2$ & $\Z\times \Z_2$ & \nothing & $\Z_2^2$ & \nothing\\
         9 &cmm  & $\Z_2$& $\Z\times \Z_2^2$ & $\Z_2^2$ & $\Z_2^5$ & $\Z_2^4$\\
         10 &p4  & \nothing & $\Z\times\Z_4\times\Z_2$ & \nothing & $\Z_4^2\times\Z_2$ & \nothing\\
         11 &p4m & $\Z_2$ & $\Z\times\Z_4\times\Z_2$ & $\Z_2^3$  & $\Z_2^6$ & $\Z_2^6$\\
         12 &p4g  & $\Z_2$ & $\Z\times\Z_4$ & $\Z_2$ & $\Z_4\times\Z_2^2$ & $\Z_2^2$\\
         13 &p3 & \nothing & $\Z\times\Z_3^2$ & \nothing & $\Z_3^3$ & \nothing\\
         14 &p3m1 & $\Z_2$ & $\Z\times\Z_3^2$ & $\Z_2$ & $\Z_2$ & $\Z_2$\\
         15 &p31m  & $\Z_2$ & $\Z\times\Z_3$ & $\Z_2$ & $\Z_6$ & $\Z_2$\\
         16 &p6  & \nothing & $\Z\times \Z_6$ & \nothing & $\Z_6^2$ & \nothing \\
         17 &p6m  & $\Z_2$ & $\Z\times\Z_6$ & $\Z_2^2$ & $\Z_2^4$ & $\Z_2^4$
    \end{tabular}
    \caption{Group cohomology $\H^n(\Gwp,\Z^{\OR})$ for $n=1,2,3,4,5$, in which reflections act by changing the sign of the coefficient. Note that $\H^0(\Gwp,\Z^{\OR}) \cong \Z$ if $\Gwp$ is orientation preserving, and is the trivial group. {{ Furthermore, for the groups without reflections (p1,p2,p3, p4 and p6), $\Hmc^n(\Gwp,\ZZ) \cong \Hmc^n(\Gwp,\ZZ^{\OR}) $. $\ZZ_1$ denotes the trivial group}}}
    \label{tab:GrCoho_twisted}
\end{table}

\clearpage
\section{Crystallography concepts}\label{app:CrystallographyConcepts}

This appendix gives a summary of our conventions for wallpaper groups. In App.~\ref{app:DefWPGrps}, we give a concrete definition of the 17 wallpaper groups in terms of generators. App.~\ref{app:UnitCellConventions} contains our conventions for the high symmetry points and lines used to label the invariants in the main text. 

\subsection{Definition of wallpaper groups}\label{app:DefWPGrps}

Recall that wallpaper groups are extensions of a finite group ($G_0$) by translations $\ZZ^2$. The extension is specified by $\rho$, an automorphism of $\ZZ^2$ enacted by $G_0$, and a 2-cocycle $m\in \Zmc^2_{\rho}(G_0,\ZZ^2)$. $\rho$ specifies the point group, while $m$ specifies if any point group symmetry squares to a translation. $m$ is trivial for symmorphic groups. 

Two-dimensional point groups have the general form $G_0 = \ZZ_M$ or $\ZZ_M\rtimes \ZZ_2$ with $M=1,2,3,4,6$. For the abelian groups $\ZZ_M$, $\hbf$ denotes a generator. For the dihedral groups $\ZZ_M\rtimes \ZZ_2$ we denote by $\hbf$ the generator of the $\ZZ_M$ subgroup,  and by $\rbf$ the generator of $\ZZ_2$. Since $\rho$ is a group homomorphism, it is enough to specify $\rho(\rbf)$ and $\rho(\hbf)$ as elements of $\GL_2(\ZZ)$. 

\begin{table}[]
    \centering
    \begin{tabular}{c|c|c|c}
       $\#$ & $G_{\rm{space}}$ & $G_0$ & $\rho(\hbf)$\\\hline
         2&p2 & $\ZZ_2$ & {\scriptsize$\begin{pmatrix}
           -1 & 0\\
           0  & -1
         \end{pmatrix}$}\\[0.3cm]
         10 &p4 & $\ZZ_4$ & {\scriptsize$\begin{pmatrix}
           0 &  -1\\
           +1  & 0
         \end{pmatrix}$}\\[0.3cm]
         13&p3 & $\ZZ_3$ & {\scriptsize$\begin{pmatrix}
           -1 &  -1\\
           +1  & 0
         \end{pmatrix}$}\\[0.3cm]
         16&p6 & $\ZZ_6$ & {\scriptsize$\begin{pmatrix}
           0 &  -1\\
           +1  & 1
         \end{pmatrix}$}
    \end{tabular}
    \caption{\textbf{Orientation preserving wallpaper groups:} As abstract groups $\Gwp = \Z^2 \rtimes_{\rho} G_0$. The point groups corresponding to $G_0=\ZZ_M$ are denoted by $C_M$.}
    \label{tab:ChiralWP}
\end{table}

\begin{table}[]
    \centering
    \begin{tabular}{c|c|c|c|c}
        $\#$&$G_{\rm{space}}$ & $G_0$ & $\rho(\hbf)$ & $\rho(\rbf)$\\\hline
        3 & pm & $\ZZ_2$ & - & {\scriptsize$\begin{pmatrix}
           -1 & 0\\
           0  & 1
         \end{pmatrix}$}\\[0.3cm]
        5 & cm & $\ZZ_2$ & - & {\scriptsize$\begin{pmatrix}
           0 & 1\\
           1  & 0
         \end{pmatrix}$}\\[0.3cm]
         6 & pmm & $\ZZ_2^2$ & {\scriptsize$\begin{pmatrix}
           -1 & 0\\
           0  & -1
         \end{pmatrix}$}& {\scriptsize$\begin{pmatrix}
           -1 & 0\\
           0  & 1
         \end{pmatrix}$}\\[0.3cm]
          9 & cmm & $\ZZ_2^2$ & {\scriptsize$\begin{pmatrix}
           -1 & 0\\
           0  & -1
         \end{pmatrix}$}& {\scriptsize$\begin{pmatrix}
           0 & 1\\
           1  & 0
         \end{pmatrix}$}\\[0.3cm]
        11 & p4m & $\ZZ_4\rtimes\ZZ_2$ & {\scriptsize$\begin{pmatrix}
           0 &  -1\\
           +1  & 0
         \end{pmatrix}$}& {\scriptsize$\begin{pmatrix}
           -1 & 0\\
           0  & 1
         \end{pmatrix}$}\\[0.3cm]
        14 &  p3m1 & $\ZZ_3\rtimes\ZZ_2$ & {\scriptsize$\begin{pmatrix}
           -1 &  -1\\
           +1  & 0
         \end{pmatrix}$} & {\scriptsize$\begin{pmatrix}
           -1 & -1\\
           0  & 1
         \end{pmatrix}$}\\[0.3cm]
         15 & p31m & $\ZZ_3\rtimes\ZZ_2$ & {\scriptsize$\begin{pmatrix}
           -1 &  -1\\
           +1  & 0
         \end{pmatrix}$} & {\scriptsize$\begin{pmatrix}
           1 & +1\\
           0  & -1
         \end{pmatrix}$}\\[0.3cm]
        17 & p6m & $\ZZ_6\rtimes\ZZ_2$ & {\scriptsize$\begin{pmatrix}
           0 &  -1\\
           +1  & 1
         \end{pmatrix}$}& {\scriptsize$\begin{pmatrix}
           -1 & 0\\
           0  & 1
         \end{pmatrix}$}
    \end{tabular}
     \caption{\textbf{Orientation reversing symmorphic groups wallpaper groups:} As abstract groups $\Gwp = \Z^2 \rtimes_{\rho} G_0$. the reflection in $G_0$ is $\rbf$, while the generator of the rotation subgroup is $\hbf$. The point groups corresponding to $G_0=\ZZ_M \rtimes\ZZ_ 2$ are denoted by $D_M$.}
    \label{tab:nonChiral}
\end{table}

The non-symmorphic groups are defined as follows. The group pg is simply $\ZZ\rtimes \ZZ$; in this case, $G_0 = \ZZ_1\rtimes \ZZ_2$ with $\rho(\rbf) = \diag(+1 , -1)$ and $m(\rbf,\rbf)= [1,0]^\top$: this indicates that there is a glide symmetry along the $x$ axis. The groups pmg and pgg both have $G_0 = \ZZ_2\rtimes\ZZ_2 \cong \ZZ_2\times\ZZ_2$ and are of the form $\rm{pg}\rtimes\ZZ_2^{\hbf} = (\ZZ\rtimes\ZZ )\rtimes \ZZ_2$. Let $\gbf$ be the horizontal glide and $\ybf$ the vertical translation. For pmg, we have $\hbf \gbf = \gbf^{-1} \hbf$ and $\hbf \ybf= \ybf^{-1} \hbf$. While for pgg, $\hbf \gbf = \ybf^{-1}\gbf^{-1} \hbf$ and $\hbf \ybf= \ybf^{-1} \hbf$. Finally, p4g$\cong \rm{p4}\rtimes\ZZ_2^{\rbf}$ but now $\rbf\xbf = \ybf^{-1}\rbf$, $\rbf\ybf = \xbf^{-1}\rbf$ and $\rbf\hbf =\ybf \hbf^{-1}\rbf$.

Explicitly, we can think of the wallpaper groups as subgroups of $\GL_3(\QQ)$ (3-by-3 matrices with rational entries). Translations are represented as 
\begin{equation}
    \xbf \mapsto \begin{pmatrix}
        1 & 0 & 1\\
        0 & 1 & 0\\
        0 & 0 &1
    \end{pmatrix};\quad 
     \ybf \mapsto \begin{pmatrix}
        1 & 0 & 0\\
        0 & 1 & 1\\
        0 & 0 &1
    \end{pmatrix}.
\end{equation}
For symmorphic groups, rotations and reflections are mapped to
\begin{equation}
    \hbf \mapsto \begin{pmatrix}
        \rho(\hbf) & 0\\
        0 &1
    \end{pmatrix};\quad 
     \rbf \mapsto \begin{pmatrix}
        \rho(\rbf) & 0\\
        0 &1
    \end{pmatrix}.
\end{equation}

The non-symmorphic group pg has point group $D_1$. The glide $\gbf$ is represented by $$\gbf\mapsto \begin{pmatrix}
    1 & 0 & 1/2\\
    0 & -1 &0\\
    0 & 0 & 1
\end{pmatrix}$$
For pmg and pgg, the point groups is $D_2$. However, they differ in how the $C_2$ rotation is represented:
$$\hbf_{\rm{pmg}}\mapsto \begin{pmatrix}
    -1 & 0 & 0\\
    0 & -1 &0\\
    0 & 0 & 1
\end{pmatrix};\quad 
\hbf_{\rm{pgg}}\mapsto \begin{pmatrix}
    -1 & 0 & 1/2\\
    0 & -1 & 1/2\\
    0 & 0 & 1
\end{pmatrix}.
$$
Finally, the point group of p4g is $D_4$. The reflection and rotation are represented as 
\begin{equation*}
    \rbf_{\rm{p4g}} \mapsto \begin{pmatrix}
        0 & -1 & 1/2\\
        -1 & 0 & 1/2\\
        0 & 0 & 1
    \end{pmatrix}; 
    \quad 
    \hbf_{\rm{p4g}} \mapsto \begin{pmatrix}
        0 & -1 & 0\\
        1 & 0 & 0\\
        0 & 0 & 1
    \end{pmatrix}
\end{equation*}

\subsection{Unit cells for  Wallpaper groups}\label{app:UnitCellConventions}
The conventions used for all the wallpaper groups are shown in  Figs.~\ref{fig:UnitCells1}, \ref{fig:UnitCells2} and \ref{fig:UnitCells3}. We omit wallpaper group p1 because there is nothing to label. 

\begin{figure*}
    \centering
    \includegraphics[width=0.7\textwidth]{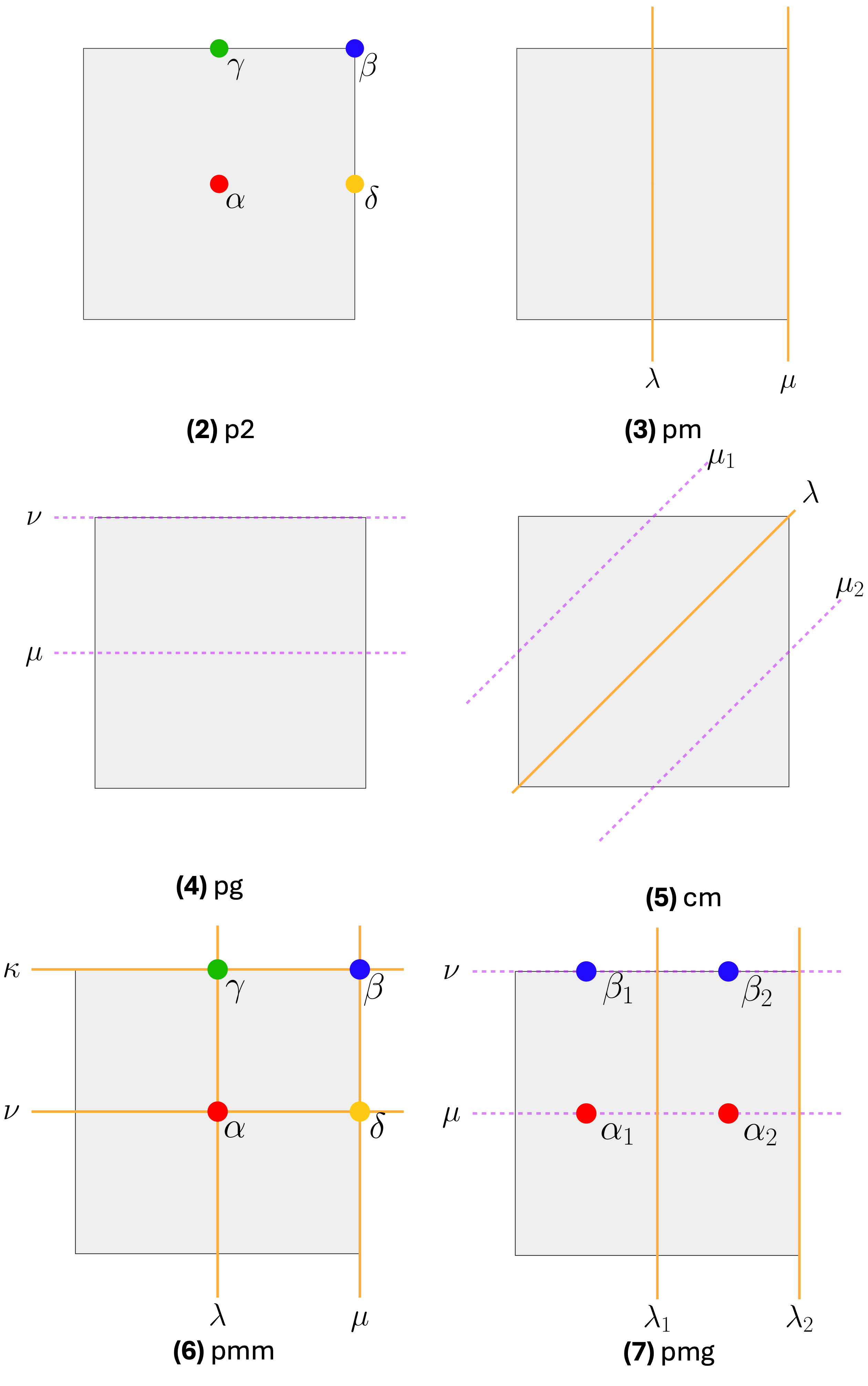}
    \caption{Conventions for the unit cell, Wyckoff positions and symmetry lines for wallpaper groups p2, pm, pg, cm, pmm, and pmg. }
    \label{fig:UnitCells1}
\end{figure*}

\begin{figure*}
    \centering
    \includegraphics[width=0.7\textwidth]{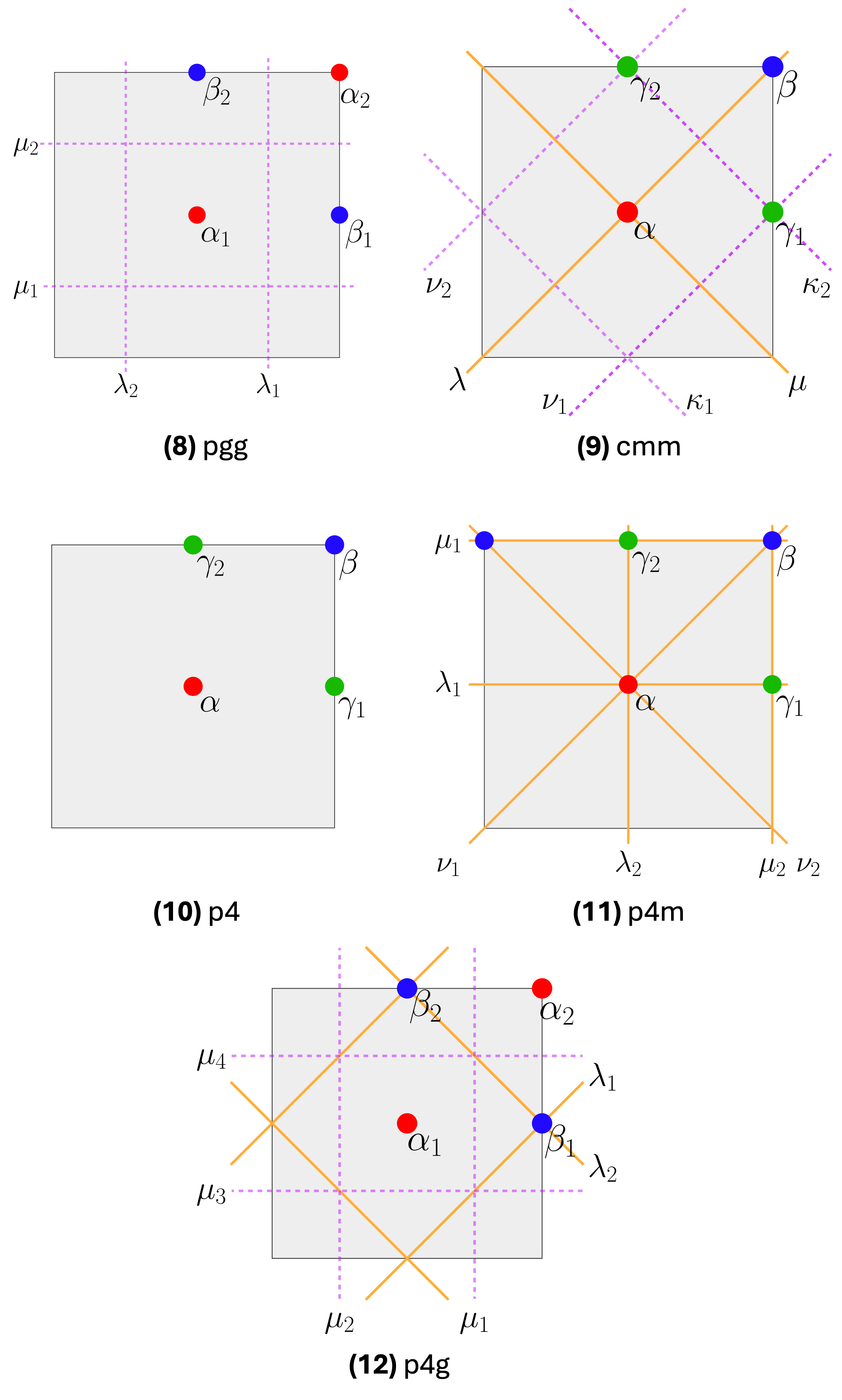}
    \caption{Conventions for the unit cell, Wyckoff positions and symmetry lines for wallpaper groups pgg, cmm, p4, p4m, and p4g. }
    \label{fig:UnitCells2}
\end{figure*}

\begin{figure*}
    \centering
    \includegraphics[width=0.7\textwidth]{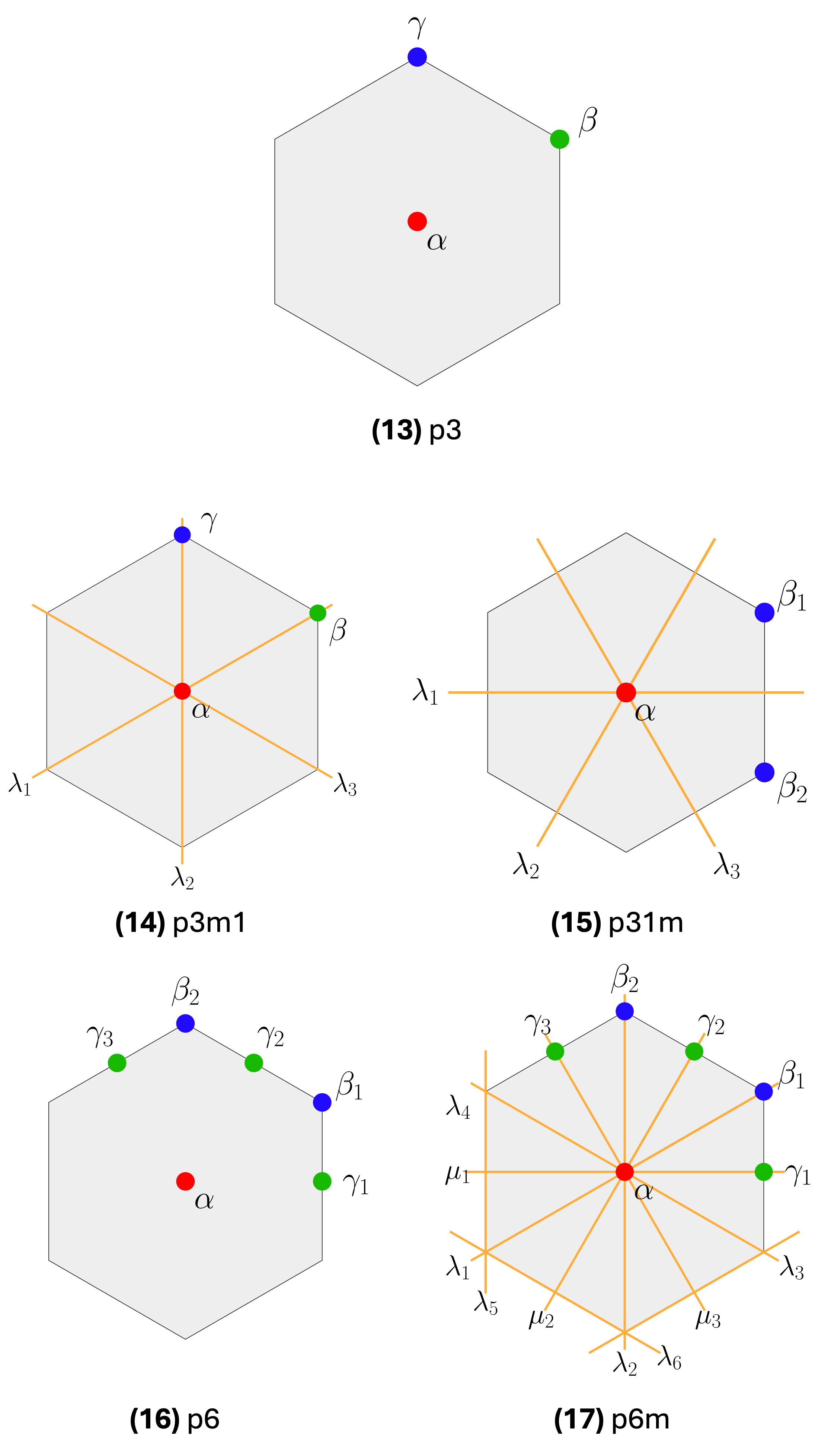}
    \caption{Conventions for the unit cell, Wyckoff positions and symmetry lines for wallpaper groups p3, p3m1, p31m, p6, and p6m.}
    \label{fig:UnitCells3}
\end{figure*}

\end{document}